\documentstyle[12pt,a4p,epsfig,rotating,]{article}

\begin{document}
\def\mrm{\mathrm}
\def\mbf{\mathbf}

\newcommand{\sqrts}{\mbox{$\protect \sqrt{s}$}}
\newcommand{\sqrtsp}{\mbox{$\sqrt{s'}$}}
\newcommand{\epm} {\mbox{$\mathrm{e}^+ \mathrm{e}^-$}}
\newcommand{\mpm} {\mbox{$\mu^+ \mu^-$}}
\newcommand{\nprode}{\mbox{$N^{\mathrm{e}^+ \mathrm{e}^-}_{prod}$}}
\newcommand{\nprodm}{\mbox{$N^{\mu^+ \mu^-}_{prod}$}}
\newcommand{\nexpe}{\mbox{$N_{exp}^{\mathrm{e}^+ \mathrm{e}^-}$}}
\newcommand{\nexpm}{\mbox{$N_{exp}^{\mu^+ \mu^-}$}}
\newcommand{\nexpt}{\mbox{$N_{exp}^{total}$}}
\newcommand{\Zo}{\mbox{$\protect {\rm Z}^0$}}
\newcommand{\bZo}{{\bf \mbox{$\rm Z}^0$}}
\newcommand{\Zs}{\mbox{${\mathrm{Z}}^{*}$}}
\newcommand{\Zgs}{\mbox{$\mathrm{(Z/\gamma)}^{*}$}}
\newcommand{\ho}{\mbox{$\mathrm{h}^{0}$}}
\newcommand{\Ho}{\mbox{$\mathrm{H}^{0}$}}
\newcommand{\Hosm}{\mbox{$\mathrm{H}^{0}_{\mathrm{SM}}$}}
\newcommand{\Ao}{\mbox{$\mathrm{A}^{0}$}}
\newcommand{\Wpm}{\mbox{$\mathrm{W}^{\pm}$}}
\newcommand{\Wp}{\mbox{$\mathrm{W}^+$}}
\newcommand{\Wm}{\mbox{$\mathrm{W}^-$}}
\newcommand{\Hpm}{\mbox{$\mathrm{H}^{\pm}$}}
\newcommand{\Hp}{\mbox{$\mathrm{H}^+$}}
\newcommand{\Hm}{\mbox{$\mathrm{H}^-$}}
\newcommand {\HH}{\Hp\Hm}
\newcommand{\WW}{\mbox{$\mathrm{W}^{+}\mathrm{W}^{-}$}}
\newcommand{\ZZ}{\mbox{$\mathrm{Z}^{0}{\mathrm{Z}^{0}}^{*}$}}
\newcommand{\sctop}{\mbox{$\tilde{\mathrm{t}}$}}
\newcommand{\msctop}{\mbox{$m_{\tilde{\mathrm{t}}}$}}
\newcommand{\ko}{\mbox{${\tilde{\chi_1}^0}$}}
\newcommand{\koko}{\mbox{${\tilde{\chi_1}^0}{{\tilde{\chi_1}^0}}$}}
\newcommand{\bho}{\mbox{$\boldmath{\mathrm{H}^{0}}$}}
\newcommand{\leplep}{\mbox{$\ell^+\ell^-$}}
\newcommand{\tautau}{\mbox{$\tau^+\tau^-$}}
\newcommand{\ee}{\mbox{$\mathrm{e}^{+}\mathrm{e}^{-}$}}
\newcommand{\bee}{\mbox{$\boldmath {\mathrm{e}^{+}\mathrm{e}^{-}} $}}
\newcommand{\mm}{\mbox{$\mu^{+}\mu^{-}$}}
\newcommand{\ellell}{\mbox{$\ell^{+}\ell^{-}$}}
\newcommand{\bmm}{\mbox{$\boldmath {\mu^{+}\mu^{-}} $}}
\newcommand{\nn}{\mbox{$\nu \bar{\nu}$}}
\newcommand{\bnn}{\mbox{$\boldmath {\nu \bar{\nu}} $}}
\newcommand{\qq}{\mbox{$\protect {\rm q} \protect \bar{{\rm q}}$}}
\newcommand{\ff}{\mbox{$\mathrm{f} \bar{\mathrm{f}}$}}
\newcommand{\bqq}{\mbox{$\boldmath {\mathrm{q} \bar{\mathrm{q}}} $}}
\newcommand{\qqp}{\mbox{$\mathrm{q\overline{q}^\prime}$}}
\newcommand{\qpq}{\mbox{$\mathrm{q^\prime\overline{q}}$}}
\newcommand{\qppqppp}{
            \mbox{$\mathrm{q^{\prime\prime}\overline{q}^{\prime\prime\prime}}$}}
\newcommand{\tnt}{\mbox{${\tau\nu_{\tau}}$}}
\newcommand{\tpnu}{\mbox{${\tau^+\nu_{\tau}}$}}
\newcommand{\tmnu}{\mbox{${\tau^-{\bar{\nu}}_{\tau}}$}}
\newcommand{\lnu}{\mbox{$\ell\nu$}}
\newcommand{\tnu}{\mbox{$\tau\nu$}}
\newcommand{\pb}{\mbox{$\protect{\rm pb}^{-1}$}}
\newcommand{\ra}{\mbox{$\rightarrow$}}
\newcommand{\br}{\mbox{$\boldmath {\rightarrow}$}}
\newcommand{\erh}{\mbox{$\mathrm{e}^+\mathrm{e}^-\rightarrow\mathrm{hadrons}$}}
\newcommand{\tpm}{\mbox{$\tau^{\pm}$}}
\newcommand{\pzvis}{\mbox{$\protect P^z_{\rm vis}$}}
\newcommand{\evisn}{\mbox{$\protect E_{\rm vis}$/$\protect \sqrt{s}$}}
\newcommand{\rvis}{\mbox{$R_{\mrm{vis}}$}}
\newcommand{\zzqqll}{\mbox{$\Zo\Zo^{(*)}\ra\qq\ellell$}}
\newcommand{\wwln}{\mbox{$\mrm{W}^{+}\mrm{W}^{-}
            \ra\mathrm{q}\bar{{\mathrm{q}}^{\prime}}\lnu$}}
\newcommand{\gamgam}{\mbox{$\gamma\gamma$}}
\newcommand{\uu}{\mbox{$\mathrm{u} \bar{\mathrm{u}}$}}
\newcommand{\dd}{\mbox{$\mathrm{d} \bar{\mathrm{d}}$}}
\newcommand{\bb}{\mbox{$\mathrm{b} \bar{\mathrm{b}}$}}
\newcommand{\cc}{\mbox{$\mathrm{c} \bar{\mathrm{c}}$}}
\newcommand{\csbar}{\mbox{$\mathrm{c} \bar{\mathrm{s}}$}}
\newcommand{\cbars}{\mbox{$\bar{\mathrm{c}} \mathrm{s}$}}
\newcommand{\nunu}{\mbox{$\nu \bar{\nu}$}}
\newcommand{\mZ}{\mbox{$m_{\mathrm{Z}^{0}}$}}
\newcommand{\mH}{\mbox{$m_{\mathrm{H}^{0}}$}}
\newcommand{\mh}{\mbox{$m_{\mathrm{h}^{0}}$}}
\newcommand{\mhs}{\mbox{$m^2_{\mathrm{h}^{0}}$}}
\newcommand{\mA}{\mbox{$m_{\mathrm{A}^{0}}$}}
\newcommand{\mHpm}{\mbox{$m_{\mathrm{H}^{\pm}}$}}
\newcommand{\mHp}{\mbox{$m_{\mathrm{H}^+}$}}
\newcommand{\mHm}{\mbox{$m_{\mathrm{H}^-}$}}
\newcommand{\mW}{\mbox{$m_{\mathrm{W}^{\pm}}$}}
\newcommand{\mtop}{\mbox{$m_{\mathrm{t}}$}}
\newcommand{\mstop}{\mbox{$m_{\tilde{\mathrm{t}}}$}}
\newcommand{\mstopL}{\mbox{$m_{\tilde{\mathrm{t}}_{\mathrm{L}}}$}}
\newcommand{\mstopR}{\mbox{$m_{\tilde{\mathrm{t}}_{\mathrm{R}}}$}}
\newcommand{\mb}{\mbox{$m_{\mathrm{b}}$}}
\newcommand{\lpm}{\mbox{$\ell ^+ \ell^-$}}
\newcommand{\G}{\mbox{$\mathrm{GeV}$}}
\newcommand{\Gc}{\mbox{$\mathrm{GeV}$}}
\newcommand{\Gcs}{\mbox{$\mathrm{GeV}$}}
\newcommand{\epsnn}{\mbox{$\epsilon^{\nu\bar{\nu}}$(\%)}}
\newcommand{\Nnn}{\mbox{$N^{\nu \bar{\nu}}_{exp}$}}
\newcommand{\epsll}{\mbox{$\epsilon^{\ell^{+}\ell^{-}}$(\%)}}
\newcommand{\Nll}{\mbox{$N^{\ell^+\ell^-}_{exp}$}}
\newcommand{\Nexp}{\mbox{$N^{total}_{exp}$}}
\newcommand{\kl}{\mbox{$\mathrm{K_{L}}$}}
\newcommand{\dedx}{\mbox{d$E$/d$x$}}
\newcommand{\etal}{\mbox{$et$ $al.$}}
\newcommand{\ie}{\mbox{$i.e.$}}
\newcommand{\sba}{\mbox{$\sin ^2 (\beta -\alpha)$}}
\newcommand{\cba}{\mbox{$\cos ^2 (\beta -\alpha)$}}
\newcommand{\tanb}{\mbox{$\tan \beta$}}
\newcommand{\czcz}{\mbox{$\chi^0_1\chi^0_1$}}
\newcommand{\cz}{\mbox{$\chi^{0}$}}
\newcommand{\co}{\mbox{$\chi^0_1$}}
\newcommand{\ct}{\mbox{$\chi^0_2$}}
\newcommand{\coct}{\mbox{$\chi^0_1\chi^0_2$}}
\newcommand{\ctcoz}{\mbox{$\chi^0_2\ra\chi^0_1$\Zs}}
\newcommand{\ctcog}{\mbox{$\chi^0_2\ra\chi^0_1\gamma$}}
\newcommand{\hczcz}{\mbox{$h\ra\czcz$}}
\newcommand{\hcoct}{\mbox{$h\ra\coct$}}
\newcommand{\PhysLett}  {Phys.~Lett.}
\newcommand{\PRL} {Phys.~Rev.\ Lett.}
\newcommand{\PhysRep}   {Phys.~Rep.}
\newcommand{\PhysRev}   {Phys.~Rev.}
\newcommand{\NPhys}  {Nucl.~Phys.}
\newcommand{\NIM} {Nucl.~Instr.\ Meth.}
\newcommand{\CPC} {Comp.~Phys.\ Comm.}
\newcommand{\ZPhys}  {Z.~Phys.}
\newcommand{\IEEENS} {IEEE Trans.\ Nucl.~Sci.}
\newcommand{\thelimit}  {84.2}
\newcommand{\Dbi}{\mbox{${\cal B}_i$}}
\newcommand{\Dbj}{\mbox{${\cal B}_{\mathrm{jet}}$}}
\begin{titlepage}
\begin{center}{\large   EUROPEAN LABORATORY FOR PARTICLE PHYSICS
}\end{center}\bigskip
\begin{flushright}
       CERN-EP/98-173 \\ 27th October 1998
\end{flushright}
\bigskip\bigskip\bigskip\bigskip\bigskip
\begin{center}{\huge\bf\boldmath
      Search for Higgs Bosons \\ 
      in \ee\ Collisions at 183~GeV
}\end{center}\bigskip\bigskip
\begin{center}{\LARGE The OPAL Collaboration
}\end{center}\bigskip\bigskip
\bigskip\begin{center}{\large  Abstract}\end{center}
The data collected by the OPAL experiment at $\sqrts=183$~GeV were used to
search for Higgs bosons which are predicted by
the Standard Model and various extensions, such as general models with 
two Higgs field doublets and the Minimal Supersymmetric Standard Model (MSSM).
The data correspond to an integrated luminosity of approximately 54~\pb.
None of the searches for neutral and charged Higgs bosons have revealed
an excess of events beyond the expected background. This negative outcome,
in combination with similar results from searches at lower energies, leads
to new limits for the Higgs boson masses and other model parameters.
In particular, the 95\% confidence level lower limit for the mass of
the Standard Model Higgs boson is 88.3~GeV.
Charged Higgs bosons can be excluded for masses up to 59.5~GeV.
In the MSSM, $\mh > 70.5$~GeV and $\mA > 72.0$~GeV are obtained
for $\tan{\beta}>1$, no and maximal scalar top mixing and
soft SUSY-breaking masses of 1~TeV.
The range $0.8 < \tanb < 1.9$ is excluded for minimal scalar
top mixing and $m_{\mathrm{top}} \le 175$ GeV.
More general scans of the MSSM parameter space are also considered.
\bigskip\bigskip\bigskip\bigskip
\bigskip\bigskip
\begin{center}{\large
(Submitted to the European Physical Journal C)
}\end{center}

\end{titlepage}
\begin{center}{\Large        The OPAL Collaboration
}\end{center}\bigskip
\begin{center}{
G.\thinspace Abbiendi$^{  2}$,
K.\thinspace Ackerstaff$^{  8}$,
G.\thinspace Alexander$^{ 23}$,
J.\thinspace Allison$^{ 16}$,
N.\thinspace Altekamp$^{  5}$,
K.J.\thinspace Anderson$^{  9}$,
S.\thinspace Anderson$^{ 12}$,
S.\thinspace Arcelli$^{ 17}$,
S.\thinspace Asai$^{ 24}$,
S.F.\thinspace Ashby$^{  1}$
% ...
}\end{center}\bigskip
\bigskip
$^{  1}$School of Physics and Astronomy, University of Birmingham,
Birmingham B15 2TT, UK
\newline
$^{  2}$Dipartimento di Fisica dell' Universit\`a di Bologna and INFN,
I-40126 Bologna, Italy
\newline
$^{  3}$Physikalisches Institut, Universit\"at Bonn,
D-53115 Bonn, Germany
\newline
$^{  4}$Department of Physics, University of California,
Riverside CA 92521, USA
\newline
$^{  5}$Cavendish Laboratory, Cambridge CB3 0HE, UK
\newline
$^{  6}$Ottawa-Carleton Institute for Physics,
Department of Physics, Carleton University,
Ottawa, Ontario K1S 5B6, Canada
\newline
$^{  7}$Centre for Research in Particle Physics,
Carleton University, Ottawa, Ontario K1S 5B6, Canada
\newline
$^{  8}$CERN, European Organisation for Particle Physics,
CH-1211 Geneva 23, Switzerland
\newline
$^{  9}$Enrico Fermi Institute and Department of Physics,
University of Chicago, Chicago IL 60637, USA
\newline
$^{ 10}$Fakult\"at f\"ur Physik, Albert Ludwigs Universit\"at,
D-79104 Freiburg, Germany
\newline
$^{ 11}$Physikalisches Institut, Universit\"at
Heidelberg, D-69120 Heidelberg, Germany
\newline
$^{ 12}$Indiana University, Department of Physics,
Swain Hall West 117, Bloomington IN 47405, USA
\newline
$^{ 13}$Queen Mary and Westfield College, University of London,
London E1 4NS, UK
\newline
$^{ 14}$Technische Hochschule Aachen, III Physikalisches Institut,
Sommerfeldstrasse 26-28, D-52056 Aachen, Germany
\newline
$^{ 15}$University College London, London WC1E 6BT, UK
\newline
$^{ 16}$Department of Physics, Schuster Laboratory, The University,
Manchester M13 9PL, UK
\newline
$^{ 17}$Department of Physics, University of Maryland,
College Park, MD 20742, USA
\newline
$^{ 18}$Laboratoire de Physique Nucl\'eaire, Universit\'e de Montr\'eal,
Montr\'eal, Quebec H3C 3J7, Canada
\newline
$^{ 19}$University of Oregon, Department of Physics, Eugene
OR 97403, USA
\newline
$^{ 20}$CLRC Rutherford Appleton Laboratory, Chilton,
Didcot, Oxfordshire OX11 0QX, UK
\newline
$^{ 22}$Department of Physics, Technion-Israel Institute of
Technology, Haifa 32000, Israel
\newline
$^{ 23}$Department of Physics and Astronomy, Tel Aviv University,
Tel Aviv 69978, Israel
\newline
$^{ 24}$International Centre for Elementary Particle Physics and
Department of Physics, University of Tokyo, Tokyo 113-0033, and
Kobe University, Kobe 657-8501, Japan
\newline
$^{ 25}$Institute of Physical and Environmental Sciences,
Brunel University, Uxbridge, Middlesex UB8 3PH, UK
\newline
$^{ 26}$Particle Physics Department, Weizmann Institute of Science,
Rehovot 76100, Israel
\newline
$^{ 27}$Universit\"at Hamburg/DESY, II Institut f\"ur Experimental
Physik, Notkestrasse 85, D-22607 Hamburg, Germany
\newline
$^{ 28}$University of Victoria, Department of Physics, P O Box 3055,
Victoria BC V8W 3P6, Canada
\newline
$^{ 29}$University of British Columbia, Department of Physics,
Vancouver BC V6T 1Z1, Canada
\newline
$^{ 30}$University of Alberta,  Department of Physics,
Edmonton AB T6G 2J1, Canada
\newline
$^{ 31}$Duke University, Dept of Physics,
Durham, NC 27708-0305, USA
\newline
$^{ 32}$Research Institute for Particle and Nuclear Physics,
H-1525 Budapest, P O  Box 49, Hungary
\newline
$^{ 33}$Institute of Nuclear Research,
H-4001 Debrecen, P O  Box 51, Hungary
\newline
$^{ 34}$Ludwigs-Maximilians-Universit\"at M\"unchen,
Sektion Physik, Am Coulombwall 1, D-85748 Garching, Germany
\newline
\bigskip\newline
$^{  a}$ and at TRIUMF, Vancouver, Canada V6T 2A3
\newline
$^{  b}$ and Royal Society University Research Fellow
\newline
$^{  c}$ and Institute of Nuclear Research, Debrecen, Hungary
\newline
$^{  d}$ and Department of Experimental Physics, Lajos Kossuth
University, Debrecen, Hungary
\newline
$^{  e}$ on leave of absence from the University of Freiburg
\newline

%
\section{Introduction}\label{sect:intro}

The OPAL detector at LEP collected in 1997 approximately 54~\pb\ of integrated 
luminosity at a centre-of-mass energy in the vicinity of $183$~GeV.
These data are used to search for neutral and charged 
Higgs bosons within the framework of the Standard Model (SM)~\cite{sm}, 
extensions with two Higgs field doublets (2HDM)~\cite{higgshunter},
and the Minimal Supersymmetric extension of the 
Standard Model (MSSM)~\cite{mssm}.

In the SM one Higgs boson, \Hosm, is predicted with unspecified 
mass~\cite{higgs}.
In \ee\ collisions at centre-of-mass energies accessible by
LEP2, the \Hosm\ boson is expected to be produced predominantly
via the ``Higgs-strahlung" process \ee\ra\Hosm\Zo .
Contributions from the
\WW\ and \Zo\Zo\ fusion processes account for a small part of the total
production, except close to the kinematic limit of the
\ee\ra\Hosm\Zo\ process.

In any 2HDM,
the Higgs sector comprises five physical Higgs bosons: two neutral CP-even
scalars \ho\ and \Ho\ (with masses satisfying $\mh<\mH$ by definition), one
CP-odd scalar \Ao\ and two charged scalars \Hpm. Our search is interpreted
within the Type II Two Higgs Doublet Model without
extra particles besides those of the SM and the two scalar doublets.
In this model,
the Higgs fields couple separately to up-type quarks for the first doublet,
and to down-type quarks and charged leptons for the second doublet.
At the current \ee\ centre-of-mass energies (\sqrts) accessed by
LEP, the \ho\ and \Ao\  
bosons are expected to be produced predominantly via two processes: 
the ``Higgs-strahlung"
process \ee\ra\ho\Zo\ (as for \Hosm )
and the ``pair production" process \ee\ra\ho\Ao.
For these two processes, the cross-sections 
$\sigma_{\mathrm{hZ}}$ and $\sigma_{\mathrm{hA}}$
are related at tree-level 
to the SM cross-sections~\cite{higgshunter}: 
\begin{eqnarray}
\ee\ra\ho\Zo\;:&&
\sigma_{\mathrm{hZ}}=\sin^2(\beta -\alpha)~\sigma^{\mathrm{SM}}_{\mathrm{HZ}},
\label{equation:xsec_zh} \\
\ee\ra\ho\Ao\;:&&
\sigma_{\mathrm{hA}}=
\cos^2(\beta-\alpha)~\bar{\lambda}~\sigma^{\mathrm{SM}}_{\nn},
\label{equation:xsec_ah}
\end{eqnarray} 
where $\sigma^{\mathrm{SM}}_{\mathrm{HZ}}$ and
$\sigma^{\mathrm{SM}}_{\nn}$ are the cross-sections for the SM processes
\ee\ra\Hosm\Zo\ and \ee\ra\nn, and
$\bar{\lambda}$ is a kinematic factor, depending 
on \mh, \mA\ and \sqrts, typically having values between 0.5 and 0.7 for the
centre-of-mass energies under consideration.
The angle $\beta$ is defined in terms of the
vacuum expectation values $v_1$ and $v_2$ of the two scalar fields, 
$\tanb=v_2/v_1$, and
$\alpha$ is the mixing angle of the two CP-even fields.
The coefficients \sba\ and \cba\ provide complementarity
of the cross-sections for the two processes,
a feature which is exploited in deriving bounds for
Higgs boson masses and other model parameters.
The MSSM is a model with two Higgs field doublets with precise predictions
for the production cross-sections and Higgs boson decay branching ratios
for a given set of MSSM model parameters.

Charged Higgs bosons are expected to be pair-produced in the reaction
\ee\ra\Hp\Hm. The cross-section for this reaction in the 2HDM
is completely determined by SM parameters  for a given charged Higgs mass.
However, the \Hpm\ branching ratio is model-dependent.
While in the MSSM, even with radiative 
corrections included~\cite{carena}, $\mHpm<\mW$ is barely possible, there
are non-minimal models, e.g.~with $R$-parity violation~\cite{hpmrpv}, which
allow the charged Higgs boson to be lighter than the W-boson. 

In this search, the dominant decays for neutral Higgs bosons, 
\Ho\ra\bb\ and \Ho\ra\tautau\ are considered.
In the MSSM, the decay \ho\ra\Ao\Ao\ is also searched for where it is
kinematically allowed. Higgs boson decays into SUSY particles are 
not searched for in this paper.
For charged Higgs bosons, both the decay into \qqp\ and into \tnt\ are
considered.

The OPAL search for \Hosm\ at centre-of-mass energies ranging from 
\mZ\ to 172~GeV has resulted in a lower bound on its mass of $\mH>69.4$~GeV
at the 95\% confidence level (CL)~\cite{smpaper172}.
Previous OPAL searches for neutral Higgs bosons
in 2HDM and the MSSM for $\sqrts\leq 172$~GeV have been reported
in~\cite{mssmpaper172}.
For charged Higgs bosons,  the
published OPAL limit for $\sqrts\leq 172$~GeV is $\mHpm>52$~GeV
at 95\% CL~\cite{hpmpaper172}.

Recent searches performed by the other LEP collaborations
are listed in~\cite{higgsalllep} for neutral Higgs bosons and 
in~\cite{higgschlep} for charged Higgs bosons.
The CLEO and CDF collaborations have set more stringent limits on
the mass of the charged Higgs bosons~\cite{chcleo,chcdf} which are
valid under certain model assumptions.
The combined mass limit for 
the SM Higgs boson using data taken at $\sqrts \le 172$~GeV
by the four LEP experiments is reported in~\cite{lephiggs172}.

Section~\ref{sect:detector} contains 
a short description of the OPAL detector, the data 
samples used, and the various Monte Carlo
simulations. Section~\ref{sect:btag}
gives a description of the procedure for tagging b-flavoured jets.
The event selections for \Hosm\Zo, \ho\Zo, \ho\Ao, and \Hp\Hm\
are described in Sections~\ref{sect:zhsearches},
\ref{sect:hasearches}, and \ref{sect:hpmsearches}.
The interpretation of the searches within the SM, 2HDM, and MSSM
is presented in Section~\ref{sect:interpret}. 
Here also a model-independent interpretation
of the neutral Higgs boson searches is given.
In many cases, the results are combined with earlier search 
results~\cite{mssmpaper172,hpmpaper172}.

\section{Experimental Considerations} \label{sect:detector}
The present analysis is based on data collected with the OPAL 
detector~\cite{detector} during 1997 at an average luminosity
weighted centre-of-mass energy of 182.7~GeV
corresponding to an integrated luminosity of approximately\footnote{
Due to different requirements on the operation of the OPAL subdetectors
the precise integrated luminosity differs from one search channel
to the other.} 54~\pb.
 
The OPAL experiment has
nearly complete solid angle coverage and excellent hermeticity.
The central tracking detector consists of a high-resolution
silicon microstrip vertex detector ($\mu$VTX)~\cite{simvtx}
with polar angle\footnote{
OPAL uses a right-handed
coordinate system where the $+z$ direction is along the electron beam and
where $+x$ points to the centre of the LEP ring.  
The polar angle, $\theta$, is
defined with respect to the $+z$ direction and the azimuthal angle, $\phi$,
with respect to the horizontal, $+x$ direction.}
coverage $|\cos\theta|<0.9$, which immediately
surrounds the beam-pipe. It is followed by a high-precision 
vertex drift chamber,
a large-volume jet chamber, and $z$--chambers to measure the $z$ coordinate
of tracks, all in a uniform
0.435~T axial magnetic field. The lead-glass electromagnetic calorimeter
with presampler
is located outside the magnet coil which provides, in combination with
the forward calorimeter, gamma catcher, MIP plug~\cite{llpaper},
and silicon-tungsten luminometer~\cite{sw}, a geometrical acceptance
down to 33~mrad from the beam direction.  The silicon-tungsten luminometer
serves to measure the integrated luminosity using small-angle Bhabha
scattering events~\cite{lumino}.
The magnet return yoke is instrumented with streamer tubes and thin gap
chambers for hadron calorimetry; it is surrounded by several layers 
of muon chambers.

Events are reconstructed from charged-particle tracks and
energy deposits (``clusters") in the electromagnetic and hadron calorimeters.
The tracks and clusters must pass a set of quality requirements
similar to those used in
previous OPAL Higgs boson searches~\cite{higgsold}.
In calculating the total visible energies and momenta, $E_{\rm vis}$
and $\vec{P}_{\rm vis}$, of events and
individual jets, corrections are applied to prevent 
double-counting of energy in the case of tracks and associated
clusters~\cite{lep2neutralino}. 

The signal detection efficiencies and accepted background cross-sections
are estimated using a variety of Monte Carlo samples. 
The HZHA generator~\cite{hzha} is used to simulate Higgs boson
production processes. The detection efficiencies are determined at 
fixed values of Higgs boson masses using sample sizes varying between 500
and 10,000 events. Efficiencies at arbitrary masses are evaluated using
spline fits in \mH , \mHpm\ or in the $(\mh,\mA)$ plane.
The background processes are simulated primarily by
the following event generators:
PYTHIA~\cite{pythia} (\Zgs\ra\qq($\gamma$)), 
EXCALIBUR~\cite{excalibur} and grc4f~\cite{grc4f} (four-fermion processes 
(4f)),
BHWIDE~\cite{bhwide} (\ee$(\gamma)$),
KORALZ~\cite{koralz} (\mm$(\gamma)$ and \tautau$(\gamma)$),
and PHOJET~\cite{phojet}, HERWIG~\cite{herwig}, and
Vermaseren~\cite{vermaseren} (hadronic and leptonic two-photon processes
($\gamma\gamma$)).
The generated partons are hadronised using JETSET~\cite{pythia} with parameters
described in~\cite{opaltune}.
For systematic studies, cluster fragmentation implemented in
HERWIG is also used.
The resulting particles are
processed through a full simulation of the OPAL detector~\cite{gopal}.

\section{Tagging of b-jets}\label{sect:btag}
  Since neutral Higgs bosons decay 
  preferentially to b\=b pairs, 
  the tagging of jets originating from b-quarks plays an important role
  in Higgs boson searches.
  A jet-wise b-tagging algorithm
  has been developed
  using three independent b-tagging methods:
  (1) lifetime tag, (2) high-$p_{\mathrm{t}}$ lepton tag, and (3) jet shape tag.
  These three methods, described below, are combined using an 
  unbinned likelihood method to form a single discriminating variable 
  for each jet. 

\begin{figure}[htbp]
  \begin{center}
    \epsfig{file=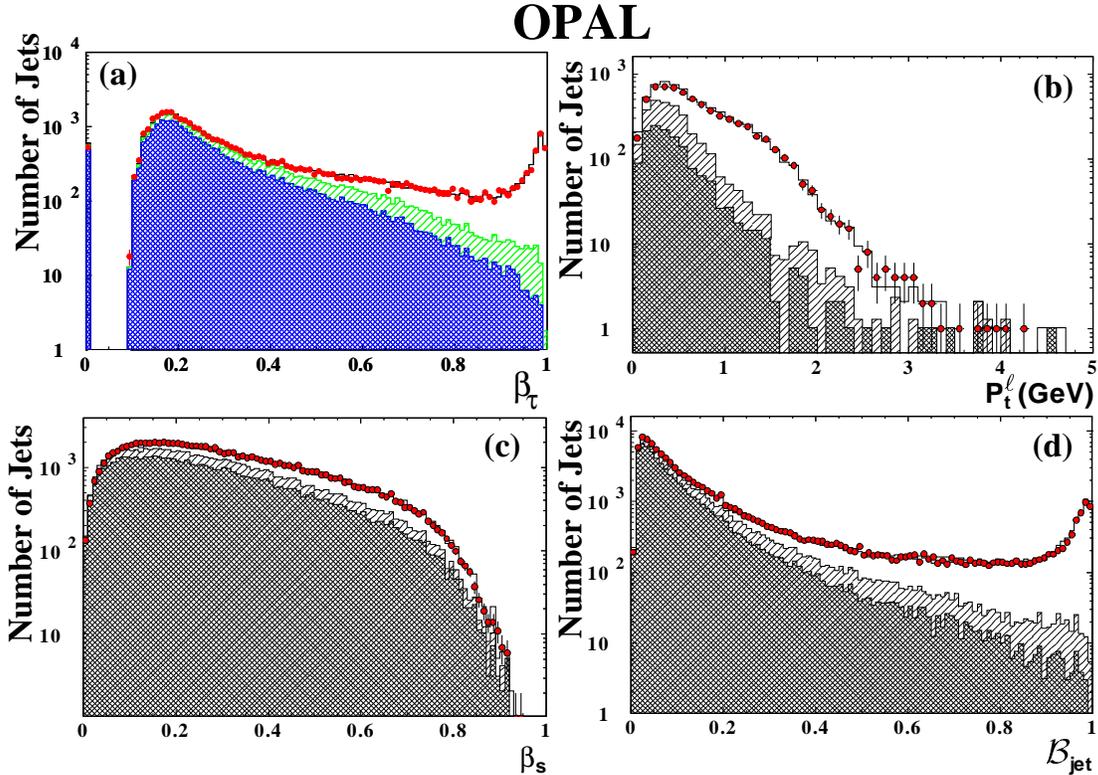,width=.85\textwidth}
\caption[]{\label{fig:btag}\sl
           (a) Lifetime tag $\beta_{\tau}$ (ANN output) for central detector
               region $|\mathrm{cos}\theta_{jet}| \leq 0.75$;
           (b) $p_t^\ell$ -distribution for events with tagged leptons;
           (c) distribution of boosted sphericity, $\beta_s$;
           (d) distribution of \Dbj\ .
           The histograms are Monte Carlo simulations of $\Zo\rightarrow
           \qq$ decays
           for different flavours (cross-hatched: uds flavour;
           hatched: c flavour; open: b-flavour);
           the dots with error bars are OPAL data taken in 1997 at
           $\sqrts=\mZ$.
            }
  \end{center}
\end{figure}

\begin{itemize}
\item[(1)]
  The lifetime tag exploits the relatively long lifetime, high
  decay multiplicity and high mass of the b-flavoured hadrons.
  Five quantities are calculated
  from the tracks and clusters assigned to a given jet.
  These five quantities are input to an artificial neural network (ANN)
  to form a lifetime tag, $\beta_\tau$, for each jet considered.
  Figure~\ref{fig:btag}(a) shows the distribution of $\beta_\tau$ in the 
  central detector region for \Zo\ra\qq\ events 
  for OPAL data at $\sqrts = \mZ$ together with the Monte Carlo simulation.
  Details are given in the Appendix.

\item[(2)]
  Semileptonic b-decays are identified using electron and muon selections,
  rejecting electrons from $\gamma$ conversions
  as described in~\cite{conv}.  
  The transverse momentum $p_{t}^{\ell}$ of the lepton, 
  calculated with respect to the
  direction of the sub-jet (see Appendix) 
  which includes the lepton track, is used as 
  a b-tag variable.
  Figure~\ref{fig:btag}(b) shows the $p_t^{\ell}$ spectrum of 
  the tagged leptons for \Zo\ra\qq\ events
  for OPAL data at $\sqrts = \mZ$ together with the Monte Carlo simulation.

\item[(3)]
  The larger decay multiplicity and higher
  mass of the b-flavoured hadrons tend to result 
  in a more spherical shape for
  b-jets compared to lighter flavour jets.
  As a measure of the jet shape, 
  the boosted sphericity $\beta_s$, defined as the sphericity of the jet 
  calculated in its rest frame, is used as a jet shape tag. The distribution
  of $\beta_s$ is shown in Figure~\ref{fig:btag}(c).
\end{itemize}

  Since the three quantities described above, $\beta_{\tau}$, $ p_t^\ell$, 
  and $\beta_s$, exploit different properties of b-flavoured hadron decays
  which
  are almost uncorrelated, they are combined using an unbinned likelihood 
  method.
  The final b-tagging discriminant \Dbj ,   
  defined as
\begin{equation}
  \Dbj =\frac{w_{\mrm{b}}\cdot f_{\mrm{b}}^{\tau}\cdot f_{\mrm{b}}^{\ell} 
\cdot f_{\mrm{b}}^{s}}{
{w_{\mrm{b}}\cdot f_{\mrm{b}}^{\tau}\cdot f_{\mrm{b}}^{\ell} 
\cdot f_{\mrm{b}}^{s}} + 
{w_{\mrm{c}}\cdot f_{\mrm{c}}^{\tau}\cdot f_{\mrm{c}}^{\ell} 
\cdot f_{\mrm{c}}^{s}} + 
{w_{\mrm{uds}}\cdot f_{\mrm{uds}}^{\tau}\cdot f_{\mrm{uds}}^{\ell} 
\cdot f_{\mrm{uds}}^{s}} },
\label{fun-lhc}
\end{equation} 
is calculated for each jet.
Here, $w_{\mrm{b}}$, $w_{\mrm{c}}$, and $w_{\mrm{uds}}$ 
are weight factors to accommodate different flavour compositions
of the background in different search channels. It has been found, however,
that the sensitivity does not strongly depend on the choice of 
these weight factors.
The functions $f_{\mrm{q}}^{t}$ are the probability density functions
for flavour q $=$ b,c,uds for the tagging
method $t =\tau,\ell,s$, determined from Monte Carlo.
The distribution of the final b-tagging discriminant ${\cal B}_{jet}$
is shown in Figure~\ref{fig:btag}(d).
Good agreement can be seen between data and Monte Carlo simulation.
The agreement has also been checked
using data and Monte Carlo samples of
\ee\ra\Zo$\gamma$ events at $\sqrts = 183$ GeV. The efficiency of
the algorithm has been verified from identified
\Zo\ra\bb\ events at $\sqrts = \mZ$ using the double tagging method 
described in~\cite{tcrit}.

\section{Searches for ${\protect \boldmath \ee\ra\Zo\Ho}$}
\label{sect:zhsearches}

The process \ee\ra\Zo\Ho\ is searched for in the following final
states: \Zo\Ho\ra\qq\bb\ (four jet channel), \Zo\Ho\ra\nn\bb\ (missing
energy channel), \Zo\Ho\ra\tautau\bb\ and \Zo\Ho\ra\qq\tautau\
(tau channels), \Zo\Ho\ra\ee\bb\ and \Zo\Ho\ra\mm\bb\ (electron and
muon channels). Throughout this section \Ho\ denotes a ``generic''
neutral Higgs boson, i.e.~\Hosm\ in the SM and \ho\ in the 2HDM and MSSM.
A search for the process \Zo\ho\ra\Zo\Ao\Ao\ which is possible 
only in 2HDM and in the
MSSM is also described in this section.

\subsection{\label{sect:sm4jet} The Four Jet Channel}
\def\lumi4j{54.1}
The process \ee\ra\Zo\Ho\ra\qq\bb\ accounts for approximately 60\% of the
SM Higgs boson production cross-section. 
It is characterised by four energetic hadronic
jets, large visible energy and the presence of b-hadron decays.
The backgrounds are \Zgs\ra\qq\ with and without initial state 
radiation and hard gluon emission, as well as
four-fermion processes, in particular, \WW\ra\qqp\qqp.
The suppression of these backgrounds relies on the
kinematic reconstruction of the \Zo\ boson and on
the identification of b-quarks from the Higgs boson decay.
The tagging of jets containing b-flavoured hadrons proceeds as explained in
Section~\ref{sect:btag}.

The selection of candidate events is done in two steps.
A preselection using cuts is first applied to retain  
only four-jet-like events.
The preselection requires: (1) a hadronic final state~\cite{l2mh}, (2)
an effective centre-of-mass energy~\cite{l2mh}, 
$\sqrtsp$, in excess of 150 GeV,
(3) the jet resolution parameter in the Durham scheme~\cite{durham},
$y_{34}$, larger than 0.003,
(4) the event shape $C$--parameter~\cite{cpar} larger than 0.25,
(5) at least two charged particle tracks in each of the four jets, and
(6) the 4-C fit (requiring energy and momentum conservation)
 and the 5-C fit (additionally constraining 
two jets to have an invariant mass of $m_{{\mathrm Z}^0}$),
as described in~\cite{smpaper172}, must each yield 
a $\chi^2$ probability larger than 10$^{-5}$.

Table \ref{fourj_t1} shows the number of events selected for the data and 
the Monte 
Carlo simulations of the various background processes at each stage of the cuts.
\begin{table}[htbp]
\begin{center}
\begin{tabular}{|l||r||r||r|r|r||c|} \hline 
\multicolumn{1}{|c||}{Cut} & 
\multicolumn{1}{c||}{Data} & 
\multicolumn{1}{c||}{Total bkg.} & 
\multicolumn{1}{c|}{\qq($\gamma$)} & 
\multicolumn{1}{c|}{4f} & 
\multicolumn{1}{c||}{$\gamma\gamma$} &
\multicolumn{1}{c|}{Efficiency (\%)} \\
    & 183 \Gc\  &   &  &  &  & $\mH = 85$ GeV     \\
\hline
(1) &6131 &  6153.3 &  5095.8 &   949.5 &  108.0 & 99.8 \\
(2) &1956 &  1958.5 &  1404.6 &   548.7 &    5.2 & 94.5 \\
(3) & 711 &  677.2 &   254.1 &   421.0 &     2.1 & 91.7 \\
(4) & 683 &  656.1 &   234.1 &   420.0 &     2.0 & 91.4 \\
(5) & 576 &  563.8 &   192.5 &   369.9 &     1.4 & 88.2 \\
(6) & 514 &  498.2 &   159.4 &   338.4 &     0.4 & 85.6 \\ \hline
${\cal L}^{HZ} > \mbox{0.96} $ &
         7 &  4.95$\pm$0.23  & 1.8  & 3.1   &     - &  39.2 \\ \hline
\end{tabular}
\end{center}
\caption[]{\label{fourj_t1}\sl
        The number of events after each cut of the selection
        for the data at $\sqrt{s}$ = 183 \Gc\, 
        and the expected background in the four jet channel.
        The background estimates are normalised to the integrated luminosity
        corresponding to the data, \lumi4j\ pb$^{-1}$. The quoted error
        on the total background estimate is statistical. 
        The last column shows the selection efficiencies for the
        \Zo\Ho\ra\qq\bb\ final state for a Higgs boson mass of 85~GeV.}
\end{table}

After the preselection, a likelihood technique~\cite{smpaper172}
is employed to classify the remaining events as 
\Zgs\ra\qq\ , four-fermion processes,
or \Zo\Ho\ra\qq\bb. 
To select signal events with low background, 
eight quantities are used.
The first six variables exploit the different kinematics
of the background and signal events:
(1)~the logarithm of $y_{34}$ in the Durham scheme;
(2)~the $C$-parameter;
(3)~the logarithm of the probability of the 5-C fit
in which the two jets
with the smallest b-tagging discriminants \Dbi\ 
(see Section~\ref{sect:btag}) 
are constrained to have an invariant mass of \mZ ;
(4)~the logarithm of the probability of the best kinematic fit requiring 
energy and momentum conservation and both di-jet masses to be equal to the
nominal W mass;
(5)~the difference between the largest and smallest jet energies; 
(6) $\beta_{\mathrm{min}}$: the minimum of
$\beta_{\mathrm{di-jet1}} + \beta_{\mathrm{di-jet2}}$ for
each of the three possible di-jet combinations, where $\beta_{\mathrm{di-jet}(i)}$
is the ratio of di-jet momentum and energy after the 4-C fit.

\begin{figure}[htbp]
\centerline{\epsfig{file=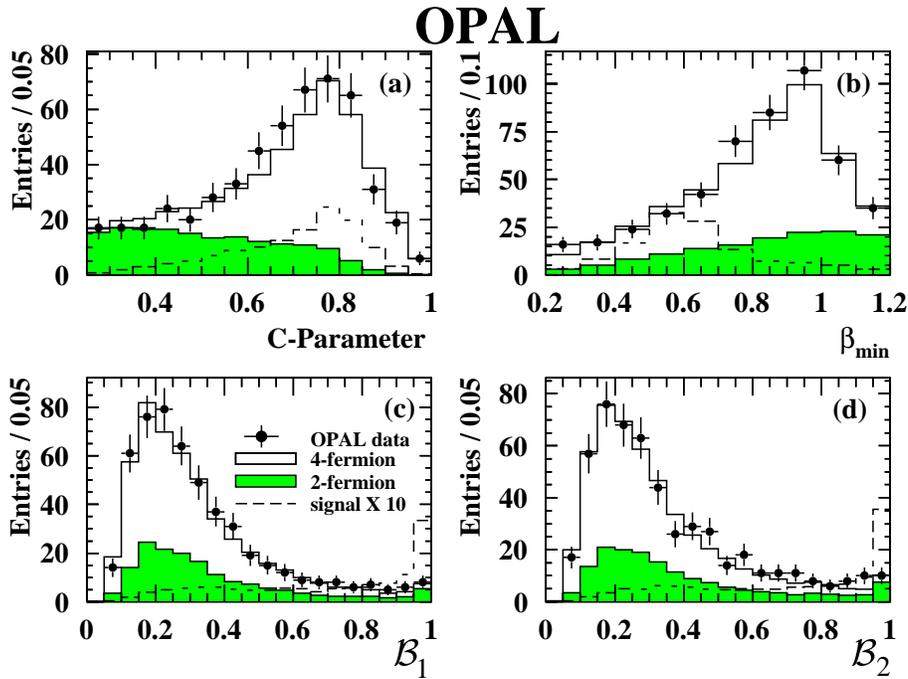,width=12cm}}
\caption[]{\label{fig1-zhqb}\sl
           Four jet channel: distribution of likelihood input variables
           (as described in the text) for data compared to Monte Carlo
           expectations.
           The points with error bars are OPAL data, the shaded (open)
           histogram is the simulation of \Zgs\ra\qq\ (four-fermion) events,
           normalised to the recorded luminosity. The dashed line
           is a simulated signal (\mH=85~GeV) scaled by a factor of 10 for
           better visibility.
}
\end{figure}

To tag jets with b-flavoured hadrons, the two largest
b-tagging discriminants \Dbi\ complete
the set of input variables ((7) and (8)) to the likelihood selection.
The two b-tagging discriminants \Dbi\ are ordered by decreasing energy of the 
jets.
In the calculation of \Dbi, the weight factors have been set to
$w_{\mathrm{b}} = w_{\mathrm{c}} = w_{\mathrm{uds}} = 1$ 
(see Section~\ref{sect:btag}).
The distributions for four of the eight 
input quantities 
are shown in Figure~\ref{fig1-zhqb}.

\begin{figure}[htbp]
\centerline{\epsfig{file=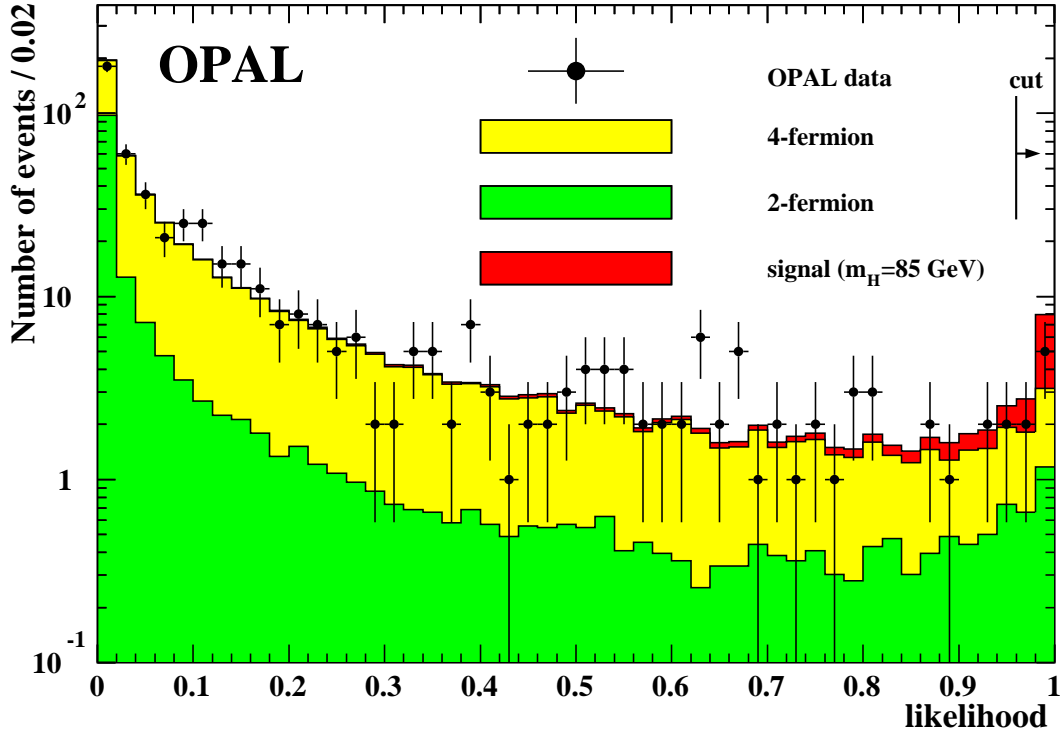,width=14cm}}
\caption[]{\label{fig2-zhqb}\sl
           Four jet channel: signal likelihood.
           The points with error bars are OPAL data, the light grey (dark grey)
           histogram is the simulation of \Zgs\ra\qq\ (four-fermion) events,
           normalised to the recorded luminosity. The black histogram
           represents a simulated signal (\mH=85~GeV) added to the expected
           background. The arrow indicates the position of the cut.
}
\end{figure}

Figure~\ref{fig2-zhqb} shows the distributions of 
the signal likelihood, ${\cal L}^{HZ}$, for the preselected events.
It can be seen that the expected signal is concentrated at large values 
of the likelihood. Candidate events are selected by 
requiring ${\cal L}^{HZ}>0.96$.
The efficiency for $\mH = \mbox{85 \Gc\ }$ is
39.2$\pm$0.2(stat.)$\pm$1.2(syst.)\,\%. 
The signal selection efficiencies as a function of the \Ho\ mass are
given in Table~\ref{tab:smsummary}.
The expected background is 1.8$\pm$0.2 events from \Zgs\ and 
3.1$\pm$0.2 events from four-fermion processes. Other sources of background 
are negligible. Seven candidate events are selected, consistent
with a total expected background of 5.0$\pm$0.2(stat.)$\pm$0.6(syst.) events. 
Their likelihood values and candidate Higgs masses are listed in 
Table~\ref{fourj_t2}.

The results of the 5-C fit are used as a measure of \mH .
For each candidate event there are 3 ways to combine the four
final state jets into a pair of di-jets. For a given di-jet
combination, each pair, in turn, is constrained to \mZ\ while the
other pair is taken as a measure of \mH . The pair yielding the fit with
the largest $\chi^2$ probability is used. According to the signal Monte Carlo,
the correct di-jet pairing is obtained 
in 70\% of the selected events. The fraction of
times the correct di-jet assignments to \mZ\ and \mH\ are made is a
strong function of \mH ; for $\mH =$~85~GeV, the correct assignment is
made in 43\% of the selected events.


The signal selection efficiencies (background estimates) are
affected by the following uncertainties expressed in relative
percentages:
description of the kinematic 
variables used in the preselection
and in the likelihood selection, 0.6\% (2.3\%);
modelling of the kinematic fit probabilities, 1.4\% (3.2\%);
tracking resolution modelling, 0.6\% (7.8\%);
b-hadron charged decay multiplicity uncertainty~\cite{bmul}, 1.7\% (6.3\%);
uncertainty in the b-quark fragmentation function~\cite{bfrag}, 1.9\% (5.2\%).
Different Monte Carlo generators have been used to evaluate
the background from \Zgs -events (HERWIG instead of PYTHIA) and four-fermion
events (EXCALIBUR instead of grc4f), yielding an uncertainty of 4.3\% on the
background estimates.
Adding the above sources in quadrature yields a $\pm$3.0\% ($\pm$12.7\%)
systematic error on the selection efficiency (background estimate).
The additional error from Monte Carlo statistics is 1.8\% (4.6\%).

\begin{table}[hbtp]
\begin{center}
\begin{tabular}{|l|r|r|r|r|r|r|r|} \hline
Event& 1 & 2 & 3 & 4 & 5 & 6 & 7 \\ \hline
${\cal L}^{HZ}$ &
0.960 &
0.999 &
0.993 &
0.987 &
0.967 &
0.997 &
0.989 \\ \hline
${\rm m_{candidate}}$ (GeV) & 
52.5&
67.9&
72.4&
75.6&
78.9&
82.9&
89.1 \\ \hline
\end{tabular}
\end{center}
\caption[]{\label{fourj_t2}\sl
           The likelihood value and reconstructed mass of accepted candidates
           in the four-jet channel.}
\end{table}

\label{sect:smann}
As a cross check, an ANN selection for the
four jet channel has been
performed~\cite{DavisMSc}. It proceeds through a preselection similar
to the one used in the main analysis. Then a set of discriminating variables
is input to an ANN.
The sensitivity is similar to the main analysis.
As an example, for $\mH =$  85 \Gc\  the efficiency of this 
analysis is 37.2\% with an expected background
of 6.2$\pm$0.5 events. 
Of the selected simulated
signal events, 80\% are in common for both analyses. 
Of the accepted background 
cross-section for the main analysis, approximately 60\% is also 
accepted by the ANN analysis.
This is consistent with the observation that five of the six selected
candidate events of the ANN analysis are shared with the likelihood
analysis.

\subsection{The Missing Energy Channel}
\label{sect:sm-miss}
The \ee\ra\nn\Ho\ra\nn\bb\ process accounts for approximately 
18\% of the 
SM Higgs boson  production cross-section
with a small contribution (1.3\% (relative) for \mH=85~GeV) 
from the \WW\ fusion process.
These events are characterised by large missing momentum and two
energetic, acoplanar, b-flavoured jets.
The dominant backgrounds are mis-measured \Zgs\ra\qq\ events,
four-fermion processes with final state neutrinos such as
\ZZ\ra\nn\qq, 
\WW\ra$\ell^{\pm}\nu$\qq, {\Wpm}e$^{\mp}\nu$\ra{\qq}e$^{\mp}\nu$
with the charged lepton escaping detection and, in general, 
events in which 
particles go undetected down the beam pipe such as \ee\ra\Zo$\gamma$
and two-photon events. For the latter backgrounds, the missing momentum
vector points close to the beam direction, while  signal events tend
to have missing momentum in the transverse plane. The rest of the above 
mentioned backgrounds are largely reduced via b-tagging. 
The process \ZZ\ra\nn\bb\ remains an irreducible background.

\begin{figure}[htbp]
\centerline{\epsfig{file=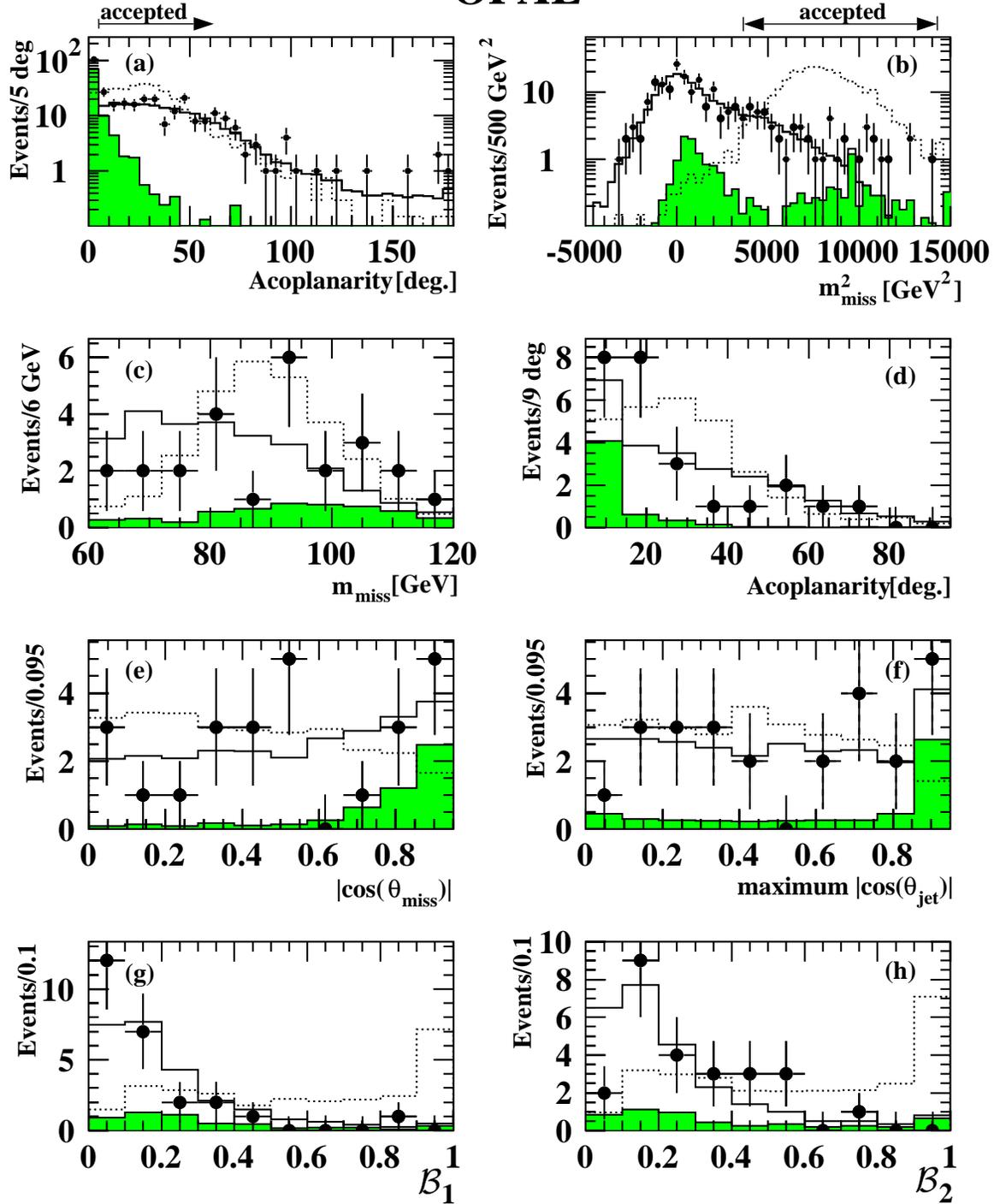,width=0.90\textwidth}}
\caption[]{\label{fig:miss2}\sl
           Missing energy channel: 
           (a) distributions of the
           acoplanarity angle after cut (3) and
           (b) missing mass squared after cut (4);
           (c)-(h)
           distribution of likelihood input variables after
           cuts (1)-(6).
           The points with error bars are OPAL data, the shaded (open)
           histogram represents the simulation 
           of \Zgs\ra\qq\ (four-fermion) events
           normalised to the recorded luminosity. The dotted line
           represents a simulated signal 
           (\mH=85~GeV) scaled by a factor of 100  
           in (a)--(b) and by a factor of 10 in (c)--(h) for better visibility.
}
\end{figure}

The preselection requires: (1)~the number of selected tracks~\cite{higgsold}
to be at least seven and at least 20\% of the total number of tracks;
no significant energy in the forward detectors
as described in~\cite{smpaper172}; the fraction of energy in the region
$|\cos{\theta}| > 0.90$ must not exceed 50\% of the total 
visible energy, $E_{\mathrm{vis}}$;
the total transverse momentum, $P^T_{\mathrm{vis}}$, must be greater than 8~GeV;
the visible mass and energy must satisfy $m_{\mathrm{vis}}>4$~GeV 
and $E_{\mathrm{vis}}/\sqrts<0.80$;
(2)~the polar angle, $\theta_{\mathrm{miss}}$, of the missing momentum
($\vec{P}_{\mathrm{miss}} = -\vec{P}_{\mathrm{vis}}$) 
must satisfy $|\cos\theta_{\rm miss}|<0.95$ and
the $z$-component of the visible momentum, $P^z_{\mathrm{vis}}$, is required to 
be less than 35~GeV;
(3)~the events are reconstructed as two-jet events
using the Durham algorithm; both jet polar angles are required to satisfy
$|\cos\theta_{\mathrm{jet}}|<0.95$;
(4)~the acoplanarity angle $\phi_{\mathrm{acop}} = $ 180$^{\circ} - \phi_{jj}$
($\phi_{jj}$ is the angle between the two jets in the  plane 
perpendicular to the beam axis) must be larger than $5^{\circ}$;
(5)~the missing mass $m_{\mathrm{miss}}$ is required to be consistent
with \mZ :
$(60~\mathrm{GeV})^2< m_{\mathrm{miss}}^2 <(120$~GeV)$^2$.
(6)~the event is required to have no isolated leptons as defined 
in~\cite{smpaper172}.
The distributions of $\phi_{\mathrm{acop}}$ and $m_{\mathrm{miss}}^2$ are shown
in Figs.~\ref{fig:miss2}(a) and (b).

\begin{figure}[htbp]
\centerline{\epsfig{file=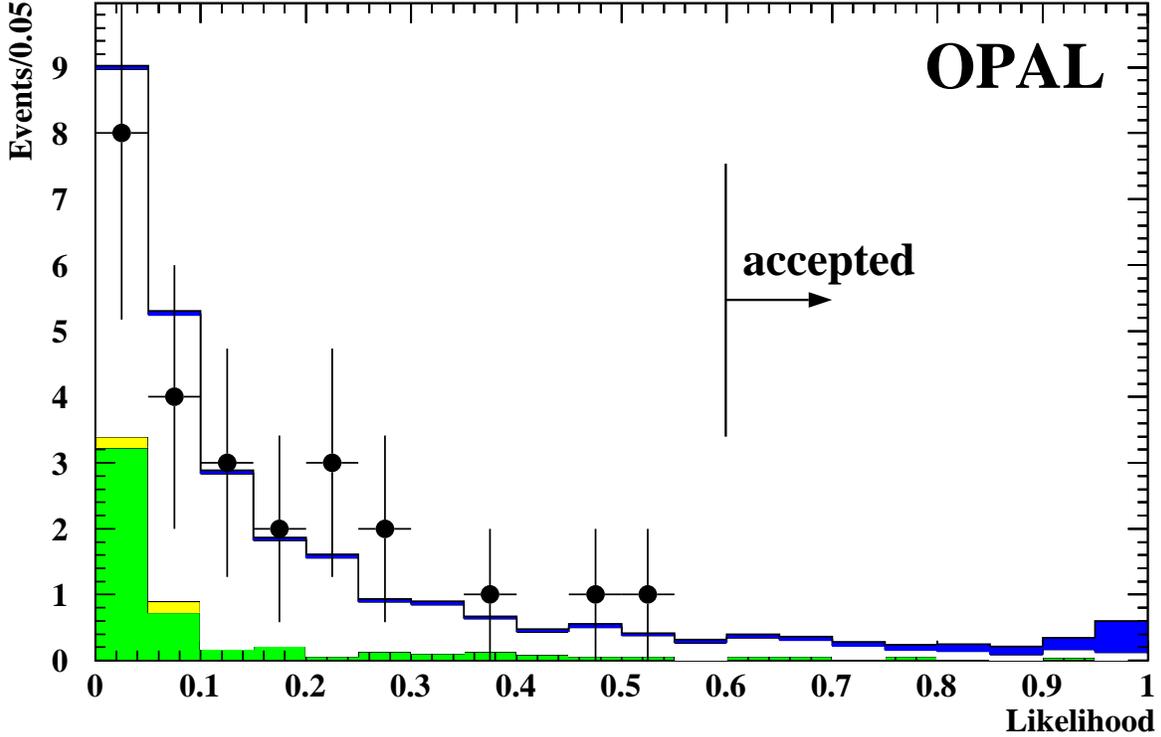,width=0.90\textwidth}}
\caption[]{\label{fig:miss4}\sl
           The likelihood distribution for the missing energy channel. 
           The points with error bars are OPAL data, the open histogram
           is the simulation of four-fermion processes and the grey
           (light grey) histogram is the simulation of 
           \Zgs\ra\qq --events (two-photon processes),
           all normalised to the recorded luminosity. The dark histogram
           is a simulated signal (\mH=85~GeV) added to the background
           expectation. The arrow indicates the position of the cut.
}
\end{figure}

Next, the b-tag (see Section~\ref{sect:btag}) as well as some 
other kinematic requirements are incorporated into the analysis
via a likelihood technique as described in~\cite{smpaper172}. 
Here, the information from
six quantities (described below) is combined.
The first set of variables
entering the likelihood are all
subject to loose cuts as part of the previously described preselection:
(1) $|\cos\theta_{\mathrm{miss}}|$, 
(2) $\mrm{max}\:|\cos\theta_{\mathrm{jet}}|$,
(3) $m_{\mathrm{miss}}$ and (4) the acoplanarity angle.
The remaining two variables are the b-tagging discriminants \Dbi 
of jets 1 and 2, as defined in Section~\ref{sect:btag},
ordered by decreasing jet energy.
Since the dominant remaining backgrounds are \qq$\ell\nu$ final
states where the charged lepton is included in one of the hadronic
jets, $p_t^\ell$ is not used in the calculation of \Dbi.  
The weight factors have been set to
$w_{\mathrm{b}} = w_{\mathrm{c}} = w_{\mathrm{uds}} = 1$.
The distributions of these input variables are shown in 
Figures~\ref{fig:miss2}(c)-(h).
In Figure~\ref{fig:miss4},
the resulting signal likelihood distributions are shown for the data and the
Monte Carlo simulations. The signal likelihood is required to be 
larger than 0.60.
\begin{table}[hbtp]
\begin{center}
\begin{tabular}{|c||r||r||r|r|r||c|} \hline
\multicolumn{1}{|c||}{Cut} &
\multicolumn{1}{c||}{Data} & 
\multicolumn{1}{c||}{Total bkg.} & 
\multicolumn{1}{c|}{\qq($\gamma$)} & 
\multicolumn{1}{c|}{4f} & 
\multicolumn{1}{c||}{$\gamma\gamma$} &
\multicolumn{1}{c|}{Efficiency (\%)} \\ 
    & 183 GeV &            &             &   &        & \mH~=~85 \Gc\   \\
\hline
(1) &  806 &  737.5 &  457.8 & 273.3  &   6.4  & 74.9 \\
(2) &  348 &  310.4 &   91.6 & 215.0  &   3.8  & 72.9 \\
(3) &  322 &  295.4 &   86.0 & 205.6  &   3.8  & 70.8 \\
(4) &  217 &  209.6 &   16.9 & 189.5  &   3.2  & 65.0 \\
(5) &   52 &   45.6 &    6.1 &  38.7  &   0.8  & 62.5 \\
(6) &   25 &   26.4  &   5.3 &  20.3  &   0.8  & 60.6 \\
\hline
${\cal L}^{HZ} > \mbox{0.6} $ &0 &1.56$\pm$0.13 &0.29 &1.27 &0.0 & 40.2 \\
\hline
\end{tabular}
\end{center}
\caption[]{\label{tab:nunu_t1}\sl
        The numbers of events after each cut for the
        data and the expected background for the missing energy channel.
        The background estimates are 
        normalised to 53.9~\pb. The quoted error is statistical.
        The last column shows the selection efficiencies
        for the \nn(\Ho\ra~all) final state
        for an 85~GeV Higgs boson. }
\end{table}
\vspace{-0.8cm}

The numbers of observed and expected events after each selection 
cut are given in Table~\ref{tab:nunu_t1}.
No events survive the selection, while
1.56 $\pm$0.13(stat.)$\pm$0.18(syst.) events
are expected from SM background processes.
The detection efficiencies as a function of the Higgs boson mass 
are listed in Table~\ref{tab:smsummary}.
In the calculation of the efficiencies and backgrounds 
a reduction by 3.7\% (relative)
has been applied in order to account for accidental vetos 
due to accelerator-related backgrounds in the forward detectors.

The systematic uncertainties due to modelling of the kinematic variables 
were estimated using $\mrm{W}\mrm{W}\rightarrow \qqp \ell\nu$ and 
$\ee\ra \qq\gamma$ events where the
identified isolated leptons or radiative photons were removed leaving a
system with kinematical properties similar to those of 
80--90 GeV Higgs bosons. The $\qqp \ell\nu$ events were also
used to estimate the uncertainties in the isolated lepton tag.
The detection efficiencies (number of expected background events)
have the following uncertainties:
modelling of the cut variables, 0.6\% (0.8\%);
and lepton tag, 0.7\% (0.8\%);
description of the tracking resolution, 0.2\% (9.7\%);
uncertainty in the knowledge of the true b-decay multiplicity 
and energy 2.2\% (5.8\%).
Adding the above systematic errors in quadrature,
the total systematic uncertainty in the signal efficiency (background)
is estimated to be 2.4\% (11.4\%). The additional error from Monte Carlo 
statistics is 0.8\% (10\%).

%
\subsection{The Tau Channels}
\label{sect:smbbtt}

The \qq\tautau final state accounts for approximately $9\%$ of the total 
\Zo\Ho\ production rate (both the (\Zo\ra\qq)(\Ho\ra\tautau) final state
and the (\Zo\ra\tautau)(\Ho\ra\qq) final state are considered)
and is characterised by a pair of tau leptons and a
pair of energetic hadronic jets.  The background is
suppressed by requiring that either the \tautau\ or the \qq\ pair yield a 
reconstructed invariant mass consistent with the \Zo\ mass.  The dominant 
backgrounds are the four-fermion processes \zzqqll\ and \wwln .
The process \ZZ\ra\tautau\bb\ is an irreducible background.

The selection begins by identifying tau leptons in the event using an
ANN.
The ANN is a track-based algorithm used to discriminate real tau decay tracks 
from tracks arising from the hadronic system.  The training process uses tracks
from high momentum tau leptons ($15$~GeV$ < p_{\tau} < 60$~GeV) in simulated 
$\qq\tautau$ events as signal and tracks in $\ee\ra\qq$ events as background.  
Any track with momentum greater than $2$~GeV and with no 
other good track within a cone of half-angle $10^{\circ}$ is considered a
one-prong tau candidate.  Any family of 
three charged tracks within a $10^{\circ}$ cone centred on any one of the 
tracks, having a total charge of $\pm 1$, and a total momentum greater than
$2$~GeV is considered as a three-prong tau
candidate.  Each candidate is then used as input to the ANN.

\begin{figure}[htbp]
\begin{center}
\centerline{\epsfig{file=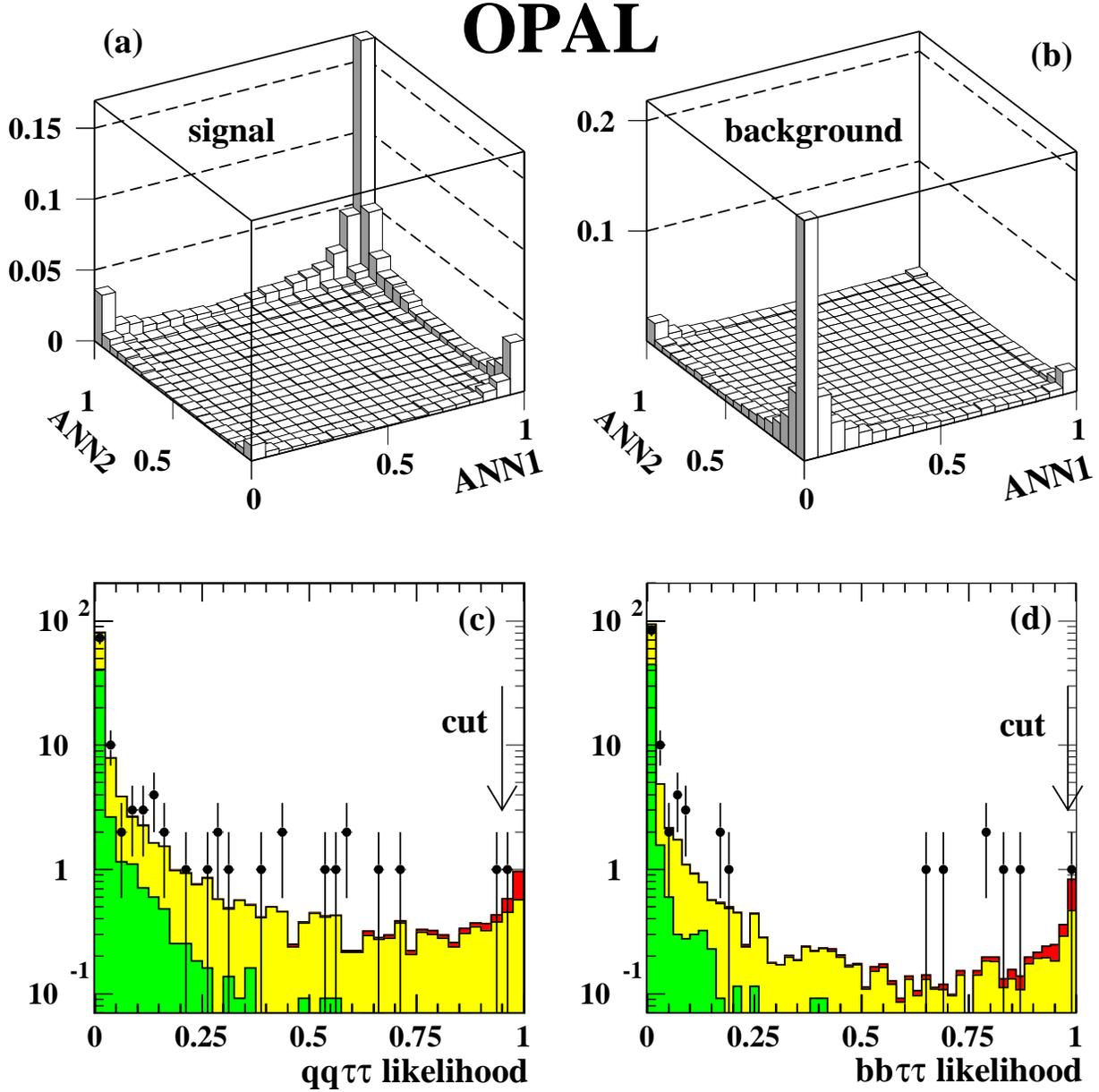,width=0.95\textwidth}}
\end{center}
\caption{\label{fig:qqtt}\sl
         Tau channel:
         the distributions of the two largest oppositely-charged tau ANN
         outputs for a Higgs \qq\tautau signal
         ($\mH=85$~GeV) (a) and the SM background (b). 
         The distributions of
         $\mathcal{L}( \qq\tautau )$ (c) and $\mathcal{L}( \bb\tautau )$ (d)
         for two-fermion background (dark grey),
         four-fermion background (light grey), OPAL data (points),
         and a simulated $85$~GeV Higgs signal
         added to the SM background (black).
}
\end{figure}

The ANN was trained separately for one-prong and three-prong tau decays.
Around each candidate an annular isolation cone of half-angle $30^{\circ}$
is constructed concentric with and excluding the narrow $10^{\circ}$ cone.
Both the one-prong and three-prong ANN use as inputs the
invariant mass of all tracks and neutral clusters in the $10^{\circ}$ cone, the
ratio of total energy contained in the isolation cone to that in the
$10^{\circ}$ cone, and the total number of tracks and neutral clusters with
energy greater than $750$~MeV in the isolation cone.  The one-prong net 
additionally takes as input the total energy in the $10^{\circ}$ cone, 
and the
track energy in the isolation cone.  The three-prong ANN
additionally uses the largest angle between the most energetic track
and any other track in the $10^{\circ}$ cone.  
Figure~\ref{fig:qqtt} demonstrates the power of the
ANN by comparing the two oppositely charged candidates with the 
largest ANN outputs in signal \qq\tautau events to those in SM 
background events. 

The modelling of the
fake rates is studied using high statistics $\ee\ra\qq$ data sets taken at
$\sqrt{s}\approx\mZ$.  The modelling of the signal inputs is studied using 
mixed events which are constructed by overlaying $\ee\ra\qq$ 
events with single hemispheres
of $\ee\ra\tautau$ events at $\sqrt{s}\approx\mZ$.
These mixed events are 
topologically and kinematically analogous to $\qq\tnu$ events at 
$\sqrt{s}\approx 183$~GeV.  The systematic errors estimated from these studies
are $\pm 10\%$ and $\pm 3\%$ for the fake rate and tau lepton efficiency,
respectively.

For each event, pairs of oppositely charged tau candidates 
are used to construct a two-tau likelihood, 
$\mathcal{L}_{\tau\tau} = \frac{\mrm{P}_{1}\mrm{P}_{2}}
{\mrm{P}_{1}\mrm{P}_{2} + (1-\mrm{P}_{1})(1-\mrm{P}_{2})}$, where 
${\mrm{P}}_{i}$ 
is the probability that the $i$th tau-candidate originates from a
real tau lepton.  This probability is calculated using the shape of the ANN 
output for signal and fake events and the 
accepted fake rate estimated from Monte Carlo.
The tau pair for the event is chosen to be that pair that maximises
$\mathcal{L}_{\tau\tau}$.

The selection uses a likelihood technique to 
discriminate between candidate $\Ho\Zo\ra\qq\tautau$ 
signal events and SM background 
processes.  Before constructing the likelihood, the following preselection is
made: the event is required to satisfy the high 
multiplicity selection described in~\cite{l2mh},
the number of charged tracks passing additional quality cuts must exceed six;
the total visible energy, divided by \sqrts , \rvis , must 
exceed 0.32;
the total missing momentum in the event divided by \sqrts\ 
must not exceed 0.40; the total scalar sum of momenta transverse to the beam
axis must exceed $40$~GeV; and the polar angle of the missing momentum vector,
$\theta_{\mrm{miss}}$, must satisfy 
$\left| \cos \theta_{\mrm{miss}} \right| < 0.95$.
After this preselection, the two-tau likelihood, $\mathcal{L}_{\tau\tau}$,
is required to exceed 0.10.  
A 2-C kinematic fit is then performed which requires energy and momentum 
conservation. The
directions of the tau lepton momenta 
is approximated by those of the
visible decay products while the energy is determined from the fit.  The 2-C
fit is required to yield a $\chi^2$-probability larger
than $10^{-5}$. The numbers of observed events 
passing these cuts and the numbers expected from the SM background 
processes are given in Table~\ref{tab:qqtt}.  The remaining events 
are used as input to two likelihood selections.

\begin{table}
\begin{center}
\begin{tabular}{|l||r||r||r|r||c||c|} \hline
\multicolumn{1}{|c||}{Cut} &
\multicolumn{1}{c||}{Data} & 
\multicolumn{1}{c||}{Total bkg.} & 
\multicolumn{1}{c|}{\qq($\gamma$)} & 
\multicolumn{1}{c||}{4f} & 
\multicolumn{1}{c||}{Efficiency (\%)} & 
\multicolumn{1}{c|}{Like-sign} \\
 & 183 GeV & & & & $\mH=85$~GeV & R(obs/exp) \\ \hline
 Presel. & 1596 & 1582 & 1007 & 575 & 77.6 & $0.98 \pm 0.03$ \\
 $\mathcal{L}_{\tau\tau}$ & 393 & 359 & 92 & 267 & 69.0 & $1.09 \pm 0.06$ \\
 2-C fit & 113 & 115 & 50 & 65 & 55.3 & $1.10 \pm 0.10$ \\ \hline
 $\mathcal{L}_{\mrm{HZ}}$ 
   & 1 & $1.3 \pm 0.1$ & 0.1 & 1.2 & 33.0 & - \\ \hline
\end{tabular}
\end{center}
\caption{\label{tab:qqtt}\sl
         The numbers of events sequentially surviving each cut as observed in
         the data compared with the total background expected from SM processes
         for the tau channel.
         The background estimates are normalised to 53.7~\pb.
         The errors are statistical only.
         Also shown is the signal efficiency for an 85~GeV Higgs boson
         (column 6) and the ratio of the number of observed events to the 
         number expected for events with a like-sign tau pair
         (column 7, see text for details).
}
\end{table}

Since roughly $50\%$ of
the $\Ho\Zo\ra\qq\tautau$ final state includes b-flavoured hadrons, one of the 
likelihoods uses b-tagging information, $\mathcal{L}( \bb\tautau )$, while a 
second likelihood ignores this information, $\mathcal{L}( \qq\tautau )$.  After
removing all tracks and clusters associated with the two tau candidates, the 
event is forced into two jets using the Durham algorithm.  A
3-C kinematic fit is performed which, in addition to energy and momentum
conservation, constrains
either the \qq\ or the \tautau\ system to the \Zo\ mass.  Both combinations are
tried and the one yielding the larger fit probability is retained.  This 
procedure correctly assigns the \qq\ pair in $93\%$ ($75\%$)
for a Higgs mass of $60$ ($80$)~GeV.  The
following variables are used as input to both likelihoods: \rvis , 
$\left| \cos \theta_{\mrm{miss}} \right|$, $\mathcal{L}_{\tau\tau}$, 
the logarithm of $y_{34}$ in the Durham scheme 
applied to the full event including the tau candidates,
the energy of the most energetic identified electron or muon in the event, 
the
angles between each tau candidate and the nearest jet
($\cos{\theta_{\mrm{nearest}}}$), the opening angle
between the most likely (largest $\mrm{P}_{i}$) tau candidate and the
missing momentum vector, and the logarithm of the fit probability for the more 
likely 3-C fit combination.  The $\mathcal{L}( \bb\tautau )$ likelihood
uses in addition the output of the b-tagging algorithm described in 
Section~\ref{sect:btag}. The weight factors
have been set to
$w_{\mathrm{b}} = w_{\mathrm{c}} = w_{\mathrm{uds}} = 1$.
 An event is retained if
$\mathcal{L}( \bb\tautau )$ exceeds 0.98 or $\mathcal{L}( \qq\tautau )$ exceeds
0.95.  For a 
Higgs mass of $85$~GeV, this selection has an efficiency of 33.0\%.
One event survives the likelihood cut compatible with the 
$\Zo ( \ra\qq ) \Ho ( \ra\tautau )$ signal hypothesis, consistent with the 
$1.3 \pm 0.1 (\mrm{stat}) \pm 0.2 (\mrm{syst})$ events expected from SM 
background processes. The fitted mass of the \tautau\ pair is
$22.7$~GeV.

A sample of like-sign tau pairs can be used to cross-check estimate for the 
dominant background in which at least one of the tau
candidates is a hadronic fake candidate. The last column of 
Table~\ref{tab:qqtt} shows the ratio of the number of observed to the number 
of expected events for this like-sign comparison for the first three cuts.  
The systematic uncertainty on the tau identification efficiency was estimated
to be 3\%
using the mixed event samples (as described above) at $\sqrts =\mZ$.
Further uncertainties on the signal efficiency arise from
the modelling of the b-hadron decay multiplicity, 1\%; the
modelling of b-fragmentation, 1\%; and
detector modelling, 1\%.  Adding these in quadrature yields a total
systematic error on the signal efficiency of 3\% (relative). The additional
error from Monte Carlo statistics is 2\%.
The total systematic error on the surviving background is $15\%$ (relative) 
not including the the Monte Carlo statistical error and
is dominated by uncertainties in the detector modelling of the fake tau rates 
and of the variables used to construct the final likelihood.

\label{sect:smqqtt}
An alternate selection also employs a likelihood
technique, but without using b-tagging information.
The identification of tau leptons is performed by considering different
sets of input variables targeted to three different types of tau lepton 
candidate, one such type consisting of explicitly identified electrons and
muons. Other input variables exploit kinematic differences of the ensemble 
of tracks and clusters not associated with the tau lepton candidate pair.
The efficiency of this analysis for \mH$=$ 85 GeV is 32.1\%
with total expected background similar to that of the main analysis.
No candidate events are observed in the data.
Of the selected simulated
signal events 60\%
are in common to both analyses. 
Of the background accepted
by the main analysis approximately 20\% is also 
accepted by the alternative analysis.

\subsection{The Electron and Muon Channels}
The $\ell^+\ell^-$\qq\ ($\ell =$ e or $\mu$) final state
arises mainly from the process
\ee\ra\Zo\Ho\ra$\ell^+\ell^-$\qq.
They amount to approximately 
6\% of the
Higgs boson production cross-section
with a small contribution (3.4\% (relative) for \mH=85~\Gc\@)
 from the \Zo\Zo\ fusion process
\ee\ra\ee\Ho\ra\ee\qq\@.
The analysis concentrates on those final states proceeding through the
first process which yield a clean experimental signature in the form of 
large visible energy, two energetic, isolated, oppositely-charged leptons of
the same species reconstructing to the \Zo\ 
boson mass, and two energetic hadronic jets carrying
b-flavour.
The dominant backgrounds are  \Zgs\ra\qq\ and four-fermion processes.
The selection is divided into two stages, a preselection and a likelihood
selection. 

The preselection is similar to
cuts (1) -- (3) described in~\cite{smpaper172} and proceeds as follows.
(1) The number of tracks must be at least six; 
$y_{\rm 34}$ in the Durham scheme 
has to be larger than $10^{-4}$;
$|P^z_{\mathrm{vis}}| < (E_{\rm vis}-0.5\sqrt{s})$ and
$E_{\rm vis} > 0.6\sqrt{s}$ are required.
(2) There must be at least one pair of oppositely charged,
same flavour leptons (e or $\mu$) as defined in~\cite{smpaper172}.
(3) The rest of the event, excluding the candidate lepton pair,
is reconstructed as two jets using the Durham algorithm;
for the muon channel, a 4-C kinematic fit is required to
yield a $\chi^2$ probability larger than 10$^{-5}$.
The invariant mass of the lepton pair should be larger than 40 GeV\@.

Next, a likelihood selection using the following input variables is applied:
$R_{\mathrm{vis}}=E_{\mathrm{vis}}/\sqrts$,
$\log_{10}(y_{34})$ in the Durham scheme,
the transverse momenta of the two leptons ordered by energy and
calculated with respect to
the nearest jet axis, and the invariant mass of the two leptons.
For the electron channel, electron identification variables
are used in addition to the previous five variables:
$(E/p)_{\mathrm{norm}} \equiv [(E/p)-1]/\sigma$
 of the two electron candidates\footnote{
$E$ and $p$ are cluster energies and track momenta, and $\sigma$ is
the error associated to $E/p$ obtained from the measurement errors of 
$E$ and $p$.},
and the normalised ionisation loss\footnote{%
  $(dE/dx)_{\rm norm}=\left[(dE/dx)-(dE/dx)_{\rm nominal}\right]/\sigma$
  where $(dE/dx)$ is the truncated ionisation loss in the jet chamber, 
  $(dE/dx)_{\rm nominal}$ is the nominal truncated ionisation 
  loss for an electron, and
  $\sigma$ is the error of $(dE/dx)$.
}, $(dE/dx)_{\rm norm}$ of the two electron candidates.
>From these variables the likelihood ${\cal K}$
is calculated as explained in~\cite{smpaper172}.

The b-flavour requirement is taken into account by combining
${\cal K}$ and the b-tagging discriminant
${\cal B_{\mathrm{2jet}}}$ from the two hadronic
jets:
\begin{eqnarray*}
  {\cal B}_{\rm 2jet} = \frac{w_{\rm b}\cdot p_{\rm b}^{(1)}\cdot p_{\rm b}^{(2)}}
  {w_{\rm b}\cdot p_{\rm b}^{(1)}\cdot p_{\rm b}^{(2)} +
    w_{\rm c}\cdot p_{\rm c}^{(1)}\cdot p_{\rm c}^{(2)} +
    w_{\rm uds}\cdot p_{\rm uds}^{(1)} \cdot  p_{\rm uds}^{(2)}} \ \ \ ,
\end{eqnarray*}
where $p_{\rm q}^{(i)} = f_{\rm q}^{\tau}\cdot f_{\rm q}^{\ell} 
\cdot f_{\rm q}^{s}$ with
$q =$ b, c, uds (see Eq.~\ref{fun-lhc}).
The weight factors have been set to $w_{\rm b} = 0.22$, $w_{\rm c} = 0.17$ and
$w_{\rm uds} = 0.61$, corresponding to the branching fractions for \Zo\ decays.
This is motivated by the fact that the dominant background
arises from \ZZ\ production.

The signal likelihood is given by:
\[ {\cal L} = \frac{{\cal K}\cdot \cal B_{\mathrm{2jet}}}
{{\cal K}\cdot {\cal B}_{\rm 2jet} + 
 (1 - {\cal K})(1 - {\cal B}_{\rm 2jet})}. \]

Candidate events are required to have a likelihood ${\cal L}>$0.9 (0.4)
for the electron (muon) channel. 
The different cut values are the result of an optimisation
which maximises the sensitivity of the two channels separately.
The signal selection efficiency for an
85 GeV Higgs boson is 57.9 \% (62.7 \%) for the electron (muon) channel. 

\begin{figure}[htbp]
  \begin{center}
    \epsfig{file=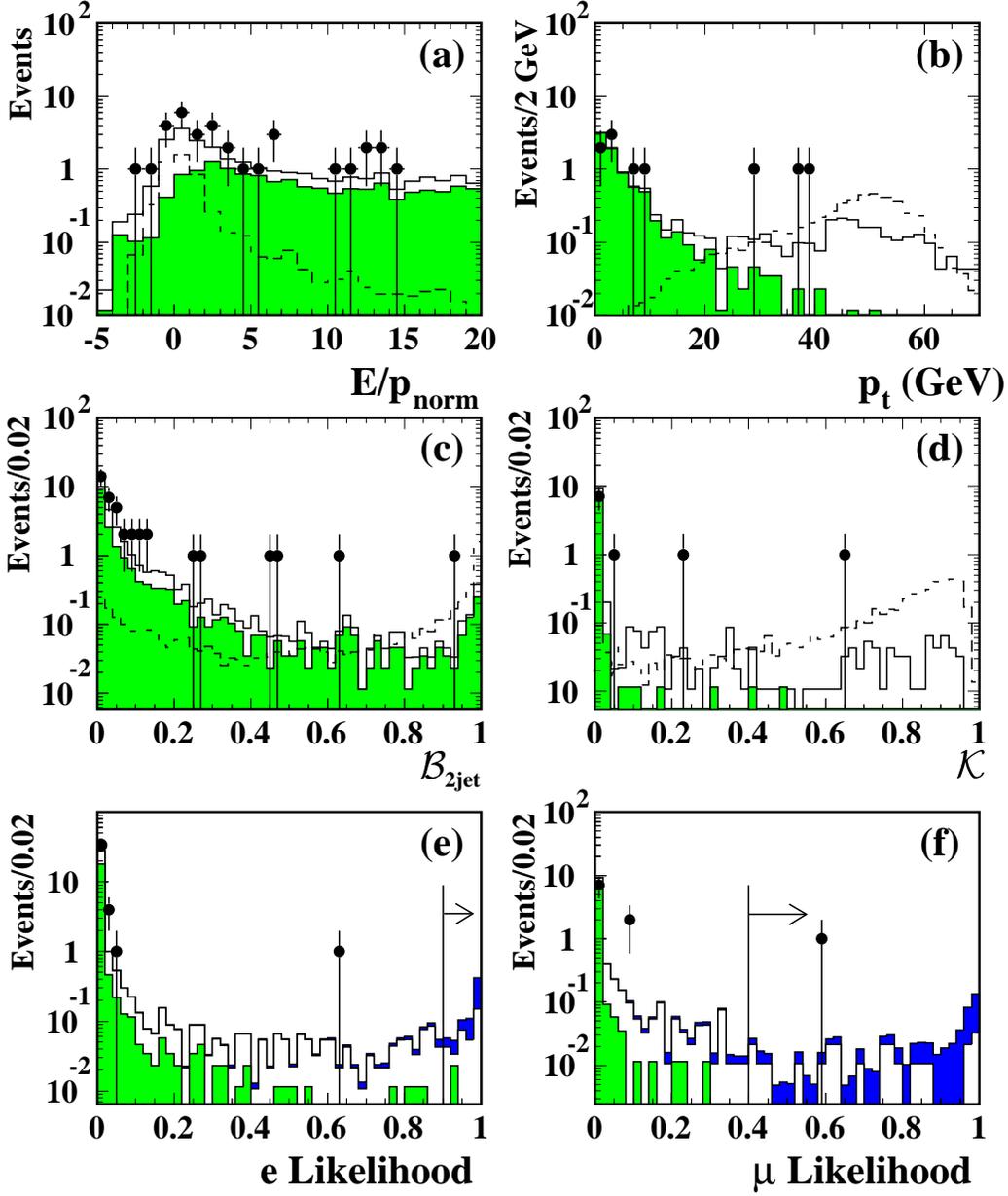,width=0.80\textwidth}
    \caption{\label{fig:lepton}\sl
      Electron and muon channels:
      distributions used in the likelihood selection for preselected events:
      (a) $(E/p)_{\mathrm{norm}}$ for the higher energy electron candidate;
      (b) $p_t$ of the higher momentum muon candidate;
      (c) ${\cal B}_{\rm 2jet}$ for the electron channel;
      (d) ${\cal K}$ for the muon channel;
      (e) final likelihood for the electron channel;
      (f) final likelihood for the muon channel.
      Data are shown as 
      points with error bars. Background simulations, 
      normalised to the integrated luminosity of 53.7~\pb,
      are represented by the
      open (grey) histograms for four-fermion (q\=q)
      events. Dashed lines show the expected signal distribution
      for an 85 \Gc\  Higgs boson. In (e) and (f), the signal simulation
      is added to the expected background (black area). 
      The arrows indicate the position
      of the cuts. In (a)-(d) the simulated signal
      is scaled by a factor of 10 for better visibility.}
  \end{center}
\end{figure}
Distributions of some variables used in the likelihood selection are
shown in Figure~\ref{fig:lepton}.

\begin{table}[htb]
\begin{center}
\begin{tabular}{|c||r|r||r|r||r|r|r|r||c|c|} \hline
\multicolumn{1}{|c||}{Cut} &
\multicolumn{2}{c||}{Data} & 
\multicolumn{2}{c||}{Total bkg.} & 
\multicolumn{2}{c|}{\qq($\gamma$)} & 
\multicolumn{2}{c||}{4f} & 
\multicolumn{2}{c|}{Efficiency $\mH = 85$ GeV} \\ \cline{2-11}
 & 
\multicolumn{1}{c|}{e} &
\multicolumn{1}{c||}{$\mu$} &
\multicolumn{1}{c|}{e} &
\multicolumn{1}{c||}{$\mu$} &
\multicolumn{1}{c|}{e} &
\multicolumn{1}{c|}{$\mu$} &
\multicolumn{1}{c|}{e} &
\multicolumn{1}{c||}{$\mu$} &
\hspace{0.48cm}e (\%)\hspace{0.48cm} & $\mu$ (\%)\\\hline\hline
(1) &  \multicolumn{2}{c||}{2732} &  
       \multicolumn{2}{c||}{2987} &  
       \multicolumn{2}{c|}{2254} & 
       \multicolumn{2}{c||}{733}  & 92.5 & 87.0 \\ \hline
(2) & 53 & 27 & 39.7 & 33.5 & 23.4 & 25.7 & 16.3 & 7.8 & 67.4 & 76.6 \\
(3) & 40 & 10 & 34.0 & 11.5 &  20.3 & 8.3 & 13.7 & 3.2 & 66.9 & 75.7 \\\hline
Likelihood & 0 & 1 & 0.37 & 0.27 & 0.02 & 0.0 & 0.35 & 0.27 & 57.9 & 62.7 \\\hline
\end{tabular}

\caption{\label{tab_lepton}\sl
  The numbers of events after each preselection cut
  and the likelihood cut for the data and the expected background in
  the electron and muon channels.
  Background estimates are normalised to $53.7\:\:\mrm{pb}^{-1}$. 
  The last two columns show the detection efficiencies for the
  processes \ee\ra(\ee\ or \mm)\Ho\ for an 85~GeV Higgs boson.}
\end{center}
\end{table}
The numbers of observed and expected events after each stage of the selection
are given in Table~\ref{tab_lepton}, together with 
the detection efficiency for an 85~GeV  Higgs boson. 
The selection retains one event in the muon channel. The total 
background
expectation is 0.64$\pm$0.08(stat.)$\pm$0.20(syst.) events
(0.37$\pm$0.07 events in the electron channel, 0.27$\pm$0.06
 events in the muon channel).
The candidate event
has a di-lepton mass of 65.5$\pm$3.7~GeV.
The Higgs mass, obtained from a 4-C kinematic fit, is
108.7$\pm$2.7~GeV for the candidate event.

The signal selection efficiencies as a function of the Higgs boson mass are
given in Table \,\ref{tab:smsummary}.
These are affected by the following systematic uncertainties for
the electron (muon) channel:
uncertainties in the lepton identification, 0.5\% (0.4\%);
uncertainties in modelling the likelihood variables
0.8\% (0.3\%); tracking resolution in the b-tagging, 0.9\% (0.9\%).
Taking these uncertainties as independent and adding them in quadrature
results in a total systematic uncertainty of 1.3\% (1.0\%) (relative errors).
The additional error from Monte Carlo statistics is 1.2\% (1.0\%).

The residual background has the following  systematic uncertainties:
uncertainty in the lepton identification,
3.5\%; uncertainties in modelling the likelihood variables, 7.5\%;
uncertainties in the generation of four-fermion processes, 25.2\%;
tracking resolution of 9.8\%. The total systematic uncertainty
on the background estimate is 28.3\%. The additional error from Monte Carlo 
statistics is 12.5\%.
%
\subsection{Search for ${\protect \boldmath \Zo\ho}$ with ${\protect \boldmath \ho\ra\Ao\Ao}$ \label{sect:hzaa}} 
All of the above searches are also sensitive to the
process \ee\ra\ho\Zo\ followed by \ho\ra\Ao\Ao\ and \Ao\ra\bb\
which appears in the 2HDM
and the MSSM if kinematically allowed.

The selection in the four-jet channel described in Section~\ref{sect:sm4jet} 
has been re-optimised for the \Zo\Ao\Ao\ final state.
The preselection cuts are kept,
variables 3, 5 and 6 are dropped, and two variables sensitive to the
six quark final state are added
to the likelihood inputs: the logarithm of $y_{56}$ in the Durham scheme and 
the number of good charged tracks in the event.
Finally, the signal likelihood ${\cal L}^{HZ}$ is required to be larger
than 0.98. The efficiency for $\mh=60$~GeV and $\mA=30$~GeV
is 38.4$\pm$2.2(stat.)$\pm$3.1(syst.)\%. 
The expected background is 1.8 events from \Zgs\ and 
2.6 events from four-fermion processes. Other sources of background 
are negligible. Four candidate events are selected, consistent
with a total expected background of 4.4$\pm$0.3$\pm$0.9 events. 
Two of the candidate events 
selected in this analysis are the same as for the four-jet analysis of
Section~\ref{sect:sm4jet}
(event 2 and event 6 in Table~\ref{fourj_t2}). The other two candidate events
have reconstructed \mh (${\cal L}^{HZ}$) of 59.5~\Gc\ (0.999) and
77.9~\Gc\ (0.983). 
\begin{table}[htbp]
\begin{center}
\begin{tabular}{|l|l|c|}
\hline
SM search & applied to the process & Efficiency (\%) \\  
\hline\hline
four jet        &(\Ao\Ao\ra\bb\bb)(\Zo\ra\qq)     & 38 \\
missing energy  &(\Ao\Ao\ra\bb\bb)(\Zo\ra\nn)     & 26 \\
electron        &(\Ao\Ao\ra\bb\bb)(\Zo\ra\ee)     & 75 \\
muon            &(\Ao\Ao\ra\bb\bb)(\Zo\ra\mm)     & 64 \\
tau lepton      &(\Ao\Ao\ra\bb\bb)(\Zo\ra\tautau) & 29 \\
\hline
\end{tabular}
\caption[]{\label{tab:hAA}\sl
         Signal detection efficiencies for the searches for the SM Higgs boson, 
         applied to the processes with \ho\ra\Ao\Ao\ followed by \Ao\ra\bb. 
         The efficiencies are quoted for $\mh=60$~GeV and
         $\mA=30$~GeV, with typical statistical errors of 1--4\%.
}
\end{center}
\end{table}

For the selections in the missing energy channel and the charged lepton 
channels, Monte Carlo simulations have demonstrated that the
detection efficiencies for the two-stage process involving \ho\ra\Ao\Ao\
followed by \Ao\ra\bb\ are close to those of the \ho\ra\bb\ decay.
For example, the detection efficiencies for $\mh=60$~GeV
and $\mA=30$~GeV, a point close to the kinematical boundary of the process
\ho\ra\Ao\Ao, are shown in Table~\ref{tab:hAA}. 
By construction, the candidate events selected are the same as for 
the corresponding \Ho\ra\bb\ analyses.
\section{The ${\protect \boldmath \Ao\ho}$ search channels}
\label{sect:hasearches}
The process \ee\ra\Ao\ho\ which appears in the 2HDM and the MSSM 
is searched for in the final
states \Ao\ho\ra\bb\bb\ and \Ao\ho\ra\bb\tautau. The case \ho\ra\Ao\Ao\
as also treated in searching for the process \Ao\ho\ra\Ao\Ao\Ao\ra\bb\bb\bb .

\subsection{The ${\protect \boldmath \Ao\ho\ra\bb\bb}$ Final State}

The signature for events
from the process \Ao\ho\ra\bb\bb\ 
is four energetic jets containing b-hadron decays
and a visible energy close to the centre-of-mass energy. 
The dominant background processes
are \Zgs\ra\qq , with or without initial state radiation accompanied by hard
gluon radiation, and four-fermion processes, in particular 
hadronic \WW\ final states. \ZZ\ production
with both \Zo\ bosons decaying into \bb\, 
constitutes an irreducible background;
however, its cross-section
is small at a centre-of-mass energy of 183 GeV.
  
The event preselection proceeds through similar cuts as in the
four jet channel described in Section~\ref{sect:sm4jet}; 
however, the cut value on the $C$--parameter (cut (4))
is 0.45. Cut (5) is replaced by the requirement that
for each of the four jets, the sum of the reconstructed charged tracks
and unassociated electromagnetic calorimeter clusters remaining after the
energy-flow calculation~\cite{lep2neutralino} be larger than six.
No 5-C fit is performed in cut (6).

For events passing the preselection, a likelihood technique
is applied to classify the events as belonging to one of the three
classes: \Zgs, four-fermion processes, or \Ao\ho\ra\bb\bb . 
Seven input variables are used. Four variables are the 
b-tagging discriminants \Dbi\ described in Section 3 (the index $i$
denotes the jet number). 
In the calculation of \Dbi , the weight factors have been
optimised for this search, $w_{\mathrm{b}} = w_{\mathrm{c}} = 0.2$ 
and $w_{\mathrm{uds}} = 0.6$.
The four jets are ordered with decreasing jet energy. 
These variables are supplemented by $y_{34}$ in the Durham scheme,
the event thrust $T$, and
the mean $|\cos{\theta}_{\mathrm{jet}}|$ of the four jets.

\begin{figure}[htbp]
\centerline{\hbox{
\epsfig{figure=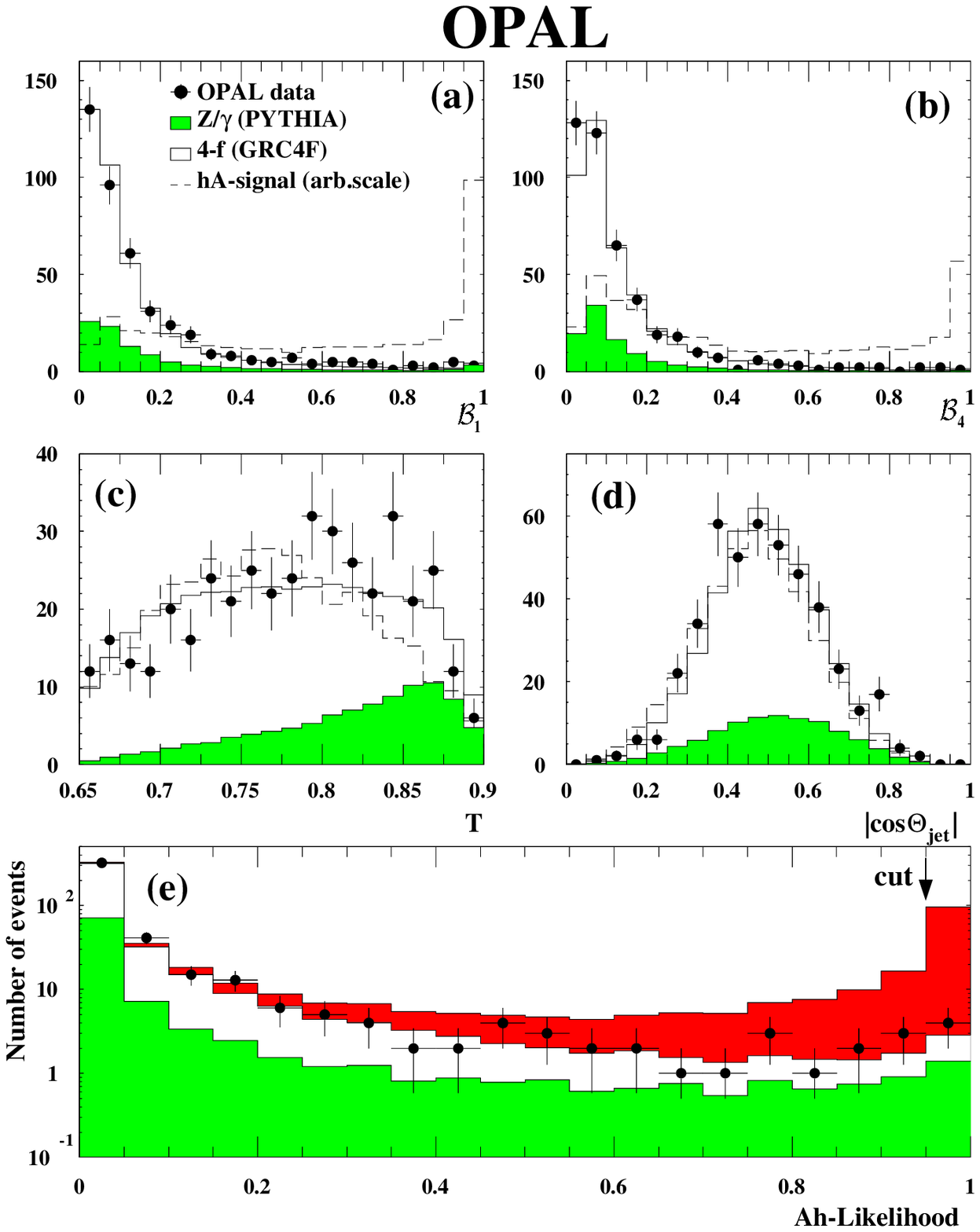,width=0.95\textwidth}
}}
\vspace{-0.5cm}
\caption[1]{\label{fig1-ah}\sl
            \Ao\ho\ra\bb\bb -channel: 
            (a)--(d) distribution of input variables
            to the likelihood selection
            (as described in the text) for data compared to Monte Carlo
            expectations. (e) distribution of the likelihood output
            variable ${\cal L}^{\mathrm{Ah}}$.
           The points with error bars are data, the shaded (open)
           histogram is the simulation of \Zgs\ra\qq\ (four-fermion) events,
           normalised to the recorded luminosity. The dashed line in
           (a)--(d) and the dark area in (e) 
           represent a simulated signal (\mh=\mA=70~GeV) 
           scaled arbitrarily for better visibility.
           The arrow in (e) indicates the position of the cut.
}
\end{figure}
The distributions of four of the seven input variables are shown in
Figure~\ref{fig1-ah}. The final likelihood discriminant
${\cal L}^{\mathrm{Ah}}$ is also shown.

\begin{table}[htbp]
\begin{center}
\begin{tabular}{|c||r||r||r|r|r||c|} \hline
\multicolumn{1}{|c||}{Cut} & 
\multicolumn{1}{c||}{Data} & 
\multicolumn{1}{c||}{Total bkg.} & 
\multicolumn{1}{c|}{\qq($\gamma$)} & 
\multicolumn{1}{c|}{4f} & 
\multicolumn{1}{c||}{$\gamma\gamma$} &
\multicolumn{1}{c|}{Efficiency (\%)} \\
 & 183 GeV & & & & & $\mA = \mh =70$ GeV \\
\hline
(1) & 6131 &  6153 &  5096 &   950 &    108 & 99.8 \\
(2) & 1956 &  1959 &  1405 &   549 &     5.2 & 96.5 \\
(3) &  711 &   677 &   254 &   421 &     2.1 & 87.8 \\
(4) &  562 &   540 &   140 &   398 &     1.6 & 85.8 \\
(5) &  447 &   434 &   106 &   328 &     0.7 & 83.2 \\
(6) &  433 &   418 &    99 &   319 &     0.1 & 80.7 \\ \hline
${\cal L}^{Ah}>0.95$ &   4 &     2.92$\pm$0.18 & 1.43 & 1.49 & -- & 50.3 \\
\hline
\end{tabular} 
\caption[]{\label{tab-ah1}\sl
         Effect of the cuts on data,
         background (normalised to the integrated luminosity of 54.1~\pb) 
         and signal simulation (\mh=\mA=70~GeV)
         for \ho\Ao\ra\bb\bb . The quoted error on the
         background is statistical.
}
\end{center}
\end{table}

Candidate events are selected by requiring ${\cal L}^{\mathrm{Ah}} >
\mbox{0.95}$. Four candidate events are observed in the data, consistent with
2.9$\pm$0.2(stat.)$\pm$0.5(syst.) events expected 
from SM background processes.
Two of the four candidate events are common to those found in the 
four jet channel of Section~\ref{sect:sm4jet},
and one is in common with the \Zo\ho\ra\Zo\Ao\Ao\ra\qq\bb\bb\ search of
Section~\ref{sect:hzaa}.

Table~\ref{tab-ah1} shows the number of selected events together
with the expectation from background processes and the signal
selection efficiency for \mA = \mh = 70~GeV,
after each cut in the
preselection and after the final cut on ${\cal L}^{\mathrm{Ah}}$.

\begin{table}[htbp]
\begin{center}
\begin{tabular}{|c|c|c||c|c||c|c|}
\hline
\multicolumn{1}{|c|}{} & \multicolumn{2}{|c||}{combination 1} &
\multicolumn{2}{|c||}{combination 2} &
\multicolumn{2}{|c|}{combination 3} \\ \cline{2-7}
\multicolumn{1}{|c|}{Candidate} & \rule[-3mm]{0mm}{10mm}
 $m_{1}$ & $m_{2}$ &
 $m_{1}$ & $m_{2}$ &
 $m_{1}$ & $m_{2}$ \\
\multicolumn{1}{|c|}{}& (GeV) & (GeV) & (GeV) & (GeV) & (GeV) & (GeV) \\ \hline
 1 & 34.2 & 141.3 & 32.4 & 71.3 & 41.9 & 75.5 \\ \hline
 2 & 25.1 & 114.0 & 43.3 & 99.3 & 36.3 & 87.0 \\ \hline
 3 & 30.4 &  70.1 & 69.4 & 106.2& 78.1 & 87.6 \\ \hline
 4 & 34.3 & 83.3  & 68.8 & 110.5& 70.7 & 86.0 \\ \hline

\end{tabular} 
\caption[]{\label{tab-ah2}\sl
           Reconstructed di-jet mass combinations for the four candidate
           events in the search for \Ao\ho\ra\bb\bb . The last two
           events are also selected by the four jet selection.
}
\end{center}
\end{table}

The systematic uncertainties on the
signal selection efficiencies and background estimates 
were determined using the same methods as described
in Section~\ref{sect:sm4jet}.
The overall systematic uncertainty is 4\% on the selection efficiencies
and 17\% on the expected number of background events.

Candidate Higgs masses are calculated
from the measured jet momenta using a 4-C fit.
Since the four jets can be combined in three ways, and since
\ho\, and \Ao\, cannot be distinguished, each candidate event enters at
six points in the (\mA ,\mh) plane. 
The resolution on the mass sum, $M = \mA + \mh$, is estimated to be
approximately 3 GeV for 
$M =$ 150~GeV~\cite{mssmpaper172}. 
For $\mA = \mh$, 68\% of the events
have a reconstructed mass difference $|\mA^{\mathrm{rec}} -
\mh^{\mathrm{rec}}|$ of less than 13~GeV. 
The di-jet masses of the four candidate events are given
in Table~\ref{tab-ah2}.

\subsection{The ${\protect \boldmath \Ao\ho\ra\bb\tautau}$ Final State}

The \Ao\ho\ra\bb\tautau final state, where either \Ao\ or
\ho\ decays into the tau pair, is topologically similar to
the $\Ho\Zo\ra\qq\tautau$ final state described in Section~\ref{sect:smbbtt},
the main difference being the loss of the \Zo\ mass constraint.
Therefore the selection proceeds in exactly the same manner as
described in Section~\ref{sect:smqqtt}, with only a minor modification in the
final likelihood selection.  Here only the $\mathcal{L}( \bb\tautau )$ 
likelihood is used, calculated without the 3-C
fit probability and the opening angle between the most likely tau and the
missing momentum vector.  With a cut on this modified likelihood,
($\mathcal{L}_{\mrm{hA}}$) at 0.9, an efficiency of 
44.7$\pm$1.6(stat.)$\pm$1.8(syst.)\,\% for
$\mh =\mA = 70$~GeV is obtained.  Three candidates are observed in the data,
one of which is the \qq\tautau candidate reported in Section~\ref{sect:smbbtt}.
This is consistent
with the $1.5 \pm 0.1 (\mrm{stat.}) \pm 0.2 (\mrm{syst.})$ events expected from
SM background processes. The invariant masses $m_{\tau\tau}$ ($m_{\mrm{had}}$)
of the three candidate events are 38.6~GeV (79.4~GeV), 20.6~GeV (94.1~GeV), 
and 84.9~GeV (46.0~GeV). Since the \Ao\ and \ho\ cannot be distinguished,
each event enters two times in the (\mA ,\mh) plane.
Systematic uncertainties on backgrounds and
efficiencies were evaluated as in Section~\ref{sect:smqqtt}.

\subsection{The ${\protect \boldmath 
\Ao\ho\ra\Ao\Ao\Ao\ra\bb\bb\bb}$ Final State}
\label{section:ah6b}
When $2\mA\leq\mh$, the decay  
\ho\ra\Ao\Ao\ is kinematically allowed and may be the dominant decay in
parts of the 2HDM and MSSM parameter space. In
this case the process \ee\ra\ho\Ao\ra\Ao\Ao\Ao\
can have a large branching ratio to the final state \bb\bb\bb.
The events are characterised by a
large number of jets and a large charged track multiplicity.
To reduce backgrounds, b-tagging plays a crucial role.
At 183~GeV, backgrounds from \Zgs\ra{\bb}g($\gamma$) 
with hard gluon emission and four-fermion processes contribute
approximately equally. Backgrounds from two-photon
processes are reduced to a negligible level in the 
course of the event selection.

The selection is identical to that described in~\cite{mssmpaper172}, 
consisting of five cuts:
(1)~requirement of a hadronic final state~\cite{l2mh};
(2)~at least five jets with $y_{\mathrm{cut}}=0.0015$ using the Durham
algorithm; 
(3)~$\sqrtsp>110$~GeV;
(4)~more than 35 charged particle tracks;
(5)~three or more jets with evidence for b flavour using the b-tagging
algorithm described in~\cite{mssmpaper172}.
Distributions of the variables relevant for the selection were
compared with Monte Carlo simulations and found to agree reasonably
well within the limited statistics of the data.

The numbers of events passing each requirement, compared with estimates from
the background simulations, are shown
in Table~\ref{table:ah6b_cuts}.
Also shown are the detection efficiencies for a simulated signal
sample with $\mh=60$~GeV and
$\mA=30$~GeV. Two events pass the selection
requirements, consistent with the
background expectation of 2.3$\pm$0.2 events.

\begin{table}[htbp]
\begin{center}
\begin{tabular}{|c||r||r||r|c||c|}
\hline
\multicolumn{1}{|c||}{Cut} & 
\multicolumn{1}{c||}{Data} & 
\multicolumn{1}{c||}{Total bkg.} & 
\multicolumn{1}{c|}{\qq($\gamma$)} & 
\multicolumn{1}{c||}{4f} & 
\multicolumn{1}{c|}{Efficiency (\%)} \\
 & 183 GeV & & & & $(\mh,\mA) =$ (60,30) GeV \\
\hline
(1) & 6131  & 6047              & 5097     & 950    & 99.5 \\
(2) &  997  &  840              &  517     & 322    & 88.2 \\
(3) &  622  &  538              &  234     & 304    & 81.2 \\
(4) &  198  &  181              &   53     & 128    & 67.0 \\ \hline
(5) &    2  &  2.3$\pm$0.2      &    1.2   &   1.1  & 36.0 \\
\hline
\end{tabular}
\caption[]{\label{table:ah6b_cuts}\sl
         Effect of the selection criteria on data,
         background (normalised to the integrated luminosity of 54.1~\pb) 
         and signal simulation ($\mh=60$~GeV, $\mA=30$~GeV) for the signal 
         channel \ho\Ao\ra\bb\bb\bb. 
         The quoted error on the background is statistical.
}
\end{center}
\end{table}

The systematic errors on the signal detection efficiencies 
(background estimates) are:
jet reconstruction, 1.3\% (4.3\%); 
requirement on \sqrtsp , 1.3\% (1.6\%); 
tracking resolution, 0.8\% (11.8\%);
uncertainty in the b-hadron decay multiplicity, 1.3\%;
mismodelling of detector effects on the multiplicity, 4.2\% (9.6\%). 
Different Monte Carlo generators to simulate the SM background processes
(HERWIG instead of PYTHIA for \Zgs -events and EXCALIBUR instead of grc4f
for four-fermion events) were found to be statistically consistent.
The total systematic error on the detection efficiency (background estimate)
is 4.8\% (15.9\%).  The additional error from Monte Carlo statistics is
6\% (7\%).

An alternative search for \ee\ra\ho\Ao\ra\Ao\Ao\Ao\ra\bb\bb\bb\ has
also been performed.
Selection of candidates is done through a neural network analysis which
combines kinematic and topological variables  with heavy flavour tagging.
The sensitivity of this analysis is similar to the main analysis. 
For example, the efficiency of the ANN 
analysis (the main analysis) is 47.2\% (41.6\%) for $\mh =$ 60 GeV
and $\mA =$ 30 GeV and 26.6\% (28.2\%) for $\mh =$ 70 GeV and
$\mA =$ 20 GeV with similar background levels for the two analyses.
Of the selected simulated
signal events approximately 60\% (depending on the masses) 
are in common to both analyses. 
Of the accepted background
cross-section for the main analysis 20\% is also 
accepted by the ANN analysis.
One of the two selected
candidate events of the ANN analysis is in common with the main analysis.

\section{\boldmath The \Hpm\ search channels}\label{sect:hpmsearches}

In this search
we consider leptonic and hadronic decays of charged Higgs bosons. 
The charged Higgs production process \ee\ra\Hp\Hm\ is searched for
in the three final states \Hp\Hm\ra\tpnu\tmnu\ (leptonic final state),
\Hp\Hm\ra\tnt\qqp\ (semileptonic final state), and \Hp\Hm\ra\qqp\qqp\
(hadronic final state).

\subsection{The Leptonic Final State}

A search at $\sqrts$~=~161,~172 and~183~GeV for pair-produced charged Higgs
bosons in the leptonic channel, \Hp\Hm\ra\tpnu\tmnu , has been
described in detail in~\cite{llpaper} within the context of a general
search for the anomalous production of di-lepton events with
missing transverse momentum. 
A likelihood technique is employed to combine information from the
various discriminating variables.
A cut is made on the relative likelihood of an event being
consistent with the charged Higgs signal hypothesis as opposed to the 
Standard Model background hypothesis.
The cut value is adjusted such that the {\it a priori}\,
average value
of the 95\% CL upper limit on the cross-section for \Hp\Hm\
is minimised using Monte Carlo simulation only.
The optimisation is performed  separately at each value of \sqrts\ and
for each value of \mHpm\ in 5~GeV steps.

The results of the analysis at 183~GeV 
are summarised in Table~\ref{tab:hpm-lepres}. 
The numbers of selected candidates are in agreement with the Standard Model 
expectation.
The dominant Standard Model background  results from
\WW\ production, which is well understood and for which the available  high
statistics Monte Carlo samples  describe well the OPAL data
\cite{wwpapers}.  The systematic error on the
expected background was estimated to be 5\%.
In addition to the uncertainty due to the limited Monte Carlo statistics
for \Hp\Hm, the systematic error on the
selection efficiency was estimated to be 5\% 
taking into account deficiencies in the
Monte Carlo generators and the detector simulation. 
\begin{table}[htbp]
\centering
{
\begin{tabular}{|c|c|c|c|}
\hline
Selection for & Data & Exp.~background & Efficiency (\%) \\ 
$\mHpm$ (GeV) & 183 GeV & & \\ \hline
50 & 4 & 6.58$\pm$0.31 & 38.9 \\ \hline
60 & 5 & 7.48$\pm$0.32 & 42.9 \\ \hline
70 & 5 & 9.17$\pm$0.36 & 48.6 \\ \hline
80 & 8 & 9.65$\pm$0.36 & 51.4 \\ \hline
90 & 4 & 6.35$\pm$0.27 & 45.1 \\ \hline
\end{tabular}
}
\caption[]{\label{tab:hpm-lepres} \sl
          Leptonic charged Higgs channel:
          the number of selected  and  expected events  
          together with selection efficiencies  at $\sqrts=$
          183 GeV  for different values of \mHpm.
          Monte Carlo statistical errors are given.
          Note that there is significant overlap between the various
          \mHpm-dependent selections. The background expectations are
          normalised to the integrated luminosity of 55.8~\pb.
}
\end{table}
\subsection{The Semileptonic Final State}

The  semileptonic channel \Hp\Hm\ra\tpnu\qqp\ (or the charge-conjugate
decay)  is characterised by  an
isolated tau lepton, a pair of acoplanar jets and sizable missing
energy and momentum. 
The main background comes from the 
\WW\ra\qqp$\ell^+\nu_\ell$ process which has 
a similar topology to the signal,
particularly if the charged Higgs boson mass is close to the W$^\pm$ mass.

The analysis proceeds in two steps.  
First, events consistent with the final state
topology are preselected. These events are then 
categorised into different classes
using a likelihood method.  

The preselection consists of the following cuts:
(1) the event must qualify as a hadronic final state~\cite{l2mh} with
(2) significant missing energy,
$R_{\mathrm{vis}} = E_{\mathrm{vis}}/\sqrts<0.85$.
(3) The total missing momentum transverse to the beam direction
($P^T_{\mathrm{vis}}$) has to be larger than 10~GeV.
The polar angle of the missing momentum is required to satisfy
$|\cos{\theta_{\mathrm{miss}}}|<0.9$.
The sum of the energies in the forward detector, gamma
catcher and silicon tungsten calorimeter is required
to be less than 20~GeV.
(4) There must be at least one tau lepton identified by the
track-based ANN algorithm, described in Section~\ref{sect:smbbtt},
with output larger than 0.5. If there is more than one tau lepton candidate in
the event, the one with the largest ANN output is retained.
(5) The two hadronic jets of the event are defined 
using the Durham algorithm
after removing the decay products of the tau lepton.
Both jets should contain at least one charged track.

The likelihood selection uses 12 input variables 
to further exploit the
differences between the signal and the background events.
Three event classes are defined: two-fermion events, four-fermion events,
and \Hp\Hm\ra\tpnu\qqp.  
The input variables are:
the transverse momentum of the event ($p_T$),
the scalar sum of the charged track momenta ($\Sigma p$), 
the number of charged tracks in a 30$^\circ$ cone around
the tau direction excluding the tracks within the 10$^\circ$ tau cone
the cosine of the angle between
the tau and the nearest jet, ($\cos{\theta_{\mathrm{nearest}}}$),
the tau lepton ANN output,
the number of charged tracks within the tau cone ($N_{\tau}^{\mrm{CT}}$), 
the highest track momentum ($p_{\mrm{max}}$),
the highest electromagnetic cluster energy ($E_{\mrm{max}}$),
the polar angle of the hadronic system multiplied by the sign
of the tau lepton charge ($Q_{\tau} \cos{\theta_{\mrm{hadr}}}$),
the polar angle of the tau lepton
in the rest-frame of the hadronic system multiplied by the sign of the
tau lepton charge ($Q_{\tau} \cos{\theta^*_{\tau}}$),
the $C$-parameter,
and the Durham scheme jet resolution parameter $y_{12}^{\mathrm{hadr}}$,
calculated from the hadronic system after removing the tau lepton candidate.

Candidate events are selected if their likelihood output ${\cal L}$
is greater than 0.85.

\begin{figure}[htbp] 
\centering
\epsfig{file=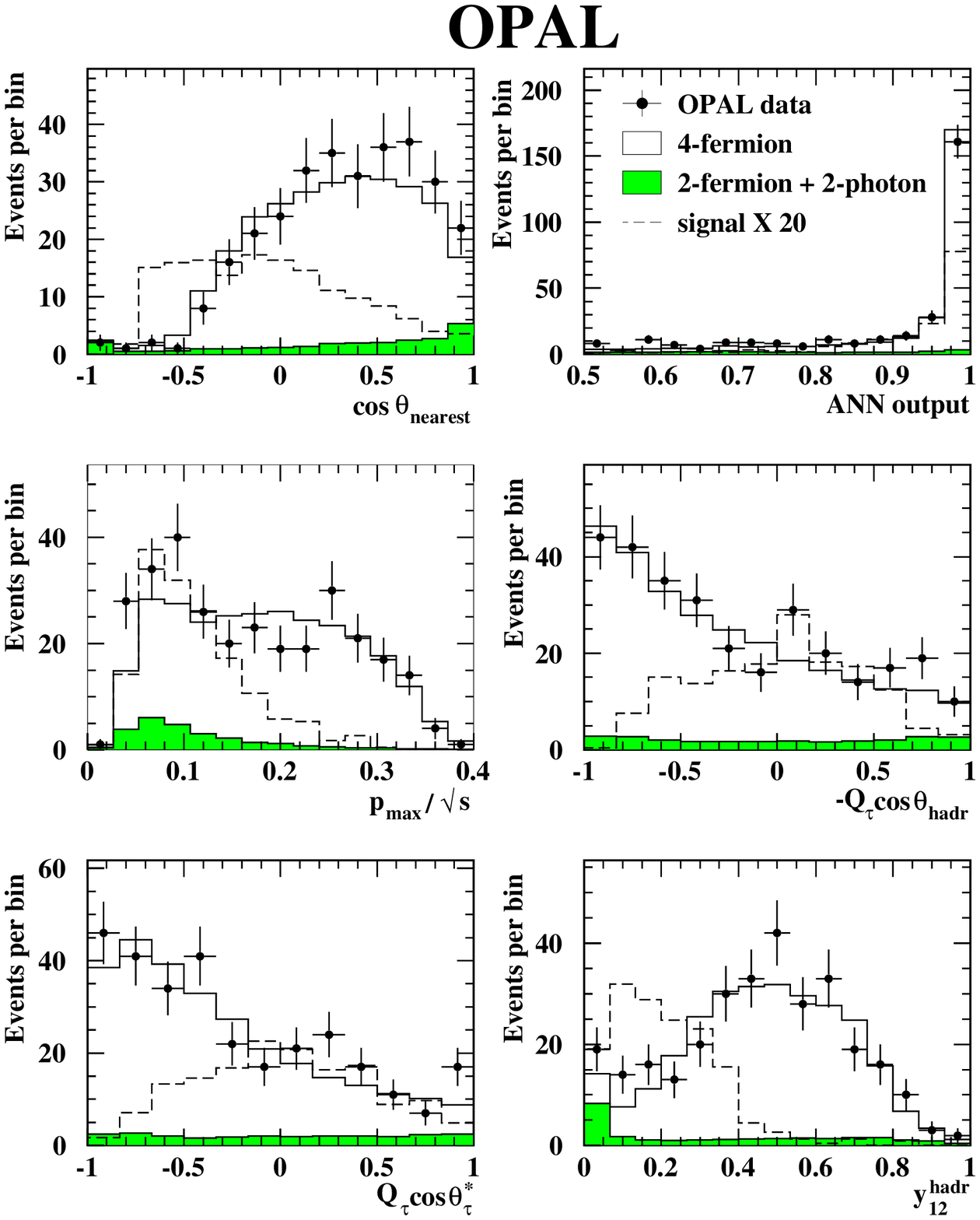,width=0.9\textwidth}
\caption{ \sl
Semileptonic charged Higgs boson channel:
distributions used in the likelihood selection (see text for explanation).  
The points with error bars are data, the shaded (open)
histogram is the simulation of two-fermion (four-fermion) events,
normalised to the recorded luminosity. The dashed line
is a simulated signal (\mHpm\ = 60~GeV) assuming  BR($\Hp\to\tpnu$) = 0.5 and
scaled by a factor of 20 for better visibility.}
\label{fig:semivars}
\end{figure}

\begin{figure}[htbp] 
\centering
\epsfig{file=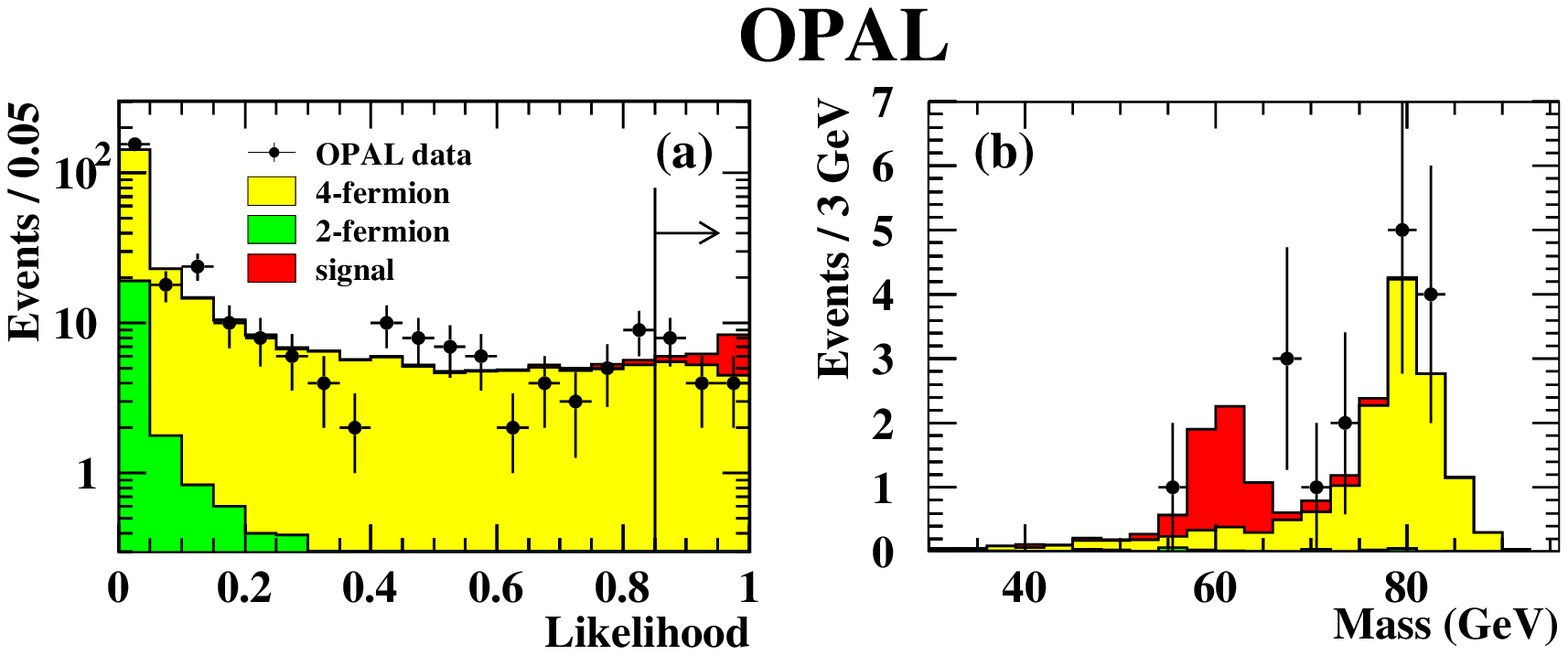,width=\textwidth}
\caption{\sl Semileptonic charged Higgs boson channel:
(a) the likelihood output and (b) the
invariant mass distribution for 183~GeV data.
The points with error bars are data, the grey (light grey)
histogram is the simulation of two-fermion (four-fermion) events,
normalised to the recorded luminosity. The dark grey histogram
is a simulated signal (\mHpm\ = 60~GeV) assuming  BR($\Hp\to\tpnu$) = 0.5
added to the background expectation.
The arrow indicates the cut on the likelihood output.
}
\label{fig:semimass}
\end{figure}

In Figure~\ref{fig:semivars},
the distributions of six likelihood input variables
are shown. The resulting likelihood distributions are shown in 
Figure~\ref{fig:semimass}(a).

Table~\ref{tab:semicuts} shows the  number
of selected data events at 183~GeV, the expected background 
and the signal efficiency for \mHpm\ = 60 GeV after each cut.
After all requirements,
16 events are selected in the data
sample, while 15.3$\pm$0.4(stat.)$\pm$1.8(syst.) events are expected
from Standard Model processes. Of these,
the four-fermion processes account for 98\%. 

\begin{table}[htb]
\centering
\begin{tabular}{|c||r||r||r|r|r||c|}
\hline
\multicolumn{1}{|c||}{Cut} & 
\multicolumn{1}{c||}{Data} & 
\multicolumn{1}{c||}{Total bkg.} & 
\multicolumn{1}{c|}{\qq($\gamma$)} & 
\multicolumn{1}{c|}{4f} & 
\multicolumn{1}{c||}{$\gamma\gamma$} & 
\multicolumn{1}{c|}{Efficiency (\%)} \\
 & 183 GeV & & & & & $\mHpm = 60$ GeV \\ \hline
(1)
&  6333 & 6405 & 5304 & 987 & 114 & 94.4
\\
(2)
&  3642 & 3466 & 2889 & 502 & 75.1  & 89.8
\\
(3)
& 536   & 478 & 158 & 320 & 0.8   & 85.2
\\
(4)
&  304  & 285 & 29.2  & 256 & 0.8   & 71.0
\\
(5)
& 298   & 279 & 24.8  & 253 &  0.8  & 69.6
\\ \hline
${\cal L} > 0.85$ 
& 16    & 15.3$\pm$0.4  & 0.3$\pm$0.1   & 15.0$\pm$0.4  & ---   & 48.6$\pm$2.2 
\\
\hline
\end{tabular}
\caption{ \sl
Semileptonic charged Higgs boson channel:
comparison of the number of observed events
and expected background (normalised to 56.2~\pb) together with the selected
fraction of simulated signal  events (\mHpm\ = 60 GeV)  after each 
cut. The errors are statistical.
}
\label{tab:semicuts}
\end{table}

The signal detection efficiencies are listed in Table~\ref{tab:semilepton}.
A decrease of the efficiency is observed with increasing Higgs mass, 
since the signal topology becomes more and more background-like.
In the calculation of the efficiencies and backgrounds 
a reduction by 1.8\% (relative)
has been applied in order to account for accidental vetos 
due to accelerator-related backgrounds in the forward detectors.

\begin{table}[htb]
\begin{center}
{\small
\begin{tabular}{|c|c|c|c|c|c|c|c|c|}
\hline  
\multicolumn{9}{|c|}{Signal selection efficiencies ($\%$)} 
\\ \cline{1-9}
50 GeV  & 55 GeV & 60 GeV & 65 GeV & 70 GeV 
& 75 GeV & 80 GeV & 85 GeV & 90 GeV \\ 
\hline
47.8$\pm$2.2    & 50.4$\pm$2.2  & 48.6$\pm$2.2  &
46.4$\pm$2.2    & 35.0$\pm$2.1  & 30.6$\pm$2.1  &
17.4$\pm$1.7    & 7.0$\pm$1.1   & 3.2$\pm$0.8 \\ 
\hline
\end{tabular}
}
\end{center} 
\caption{\sl Semileptonic charged Higgs boson channel:
signal selection efficiencies (in \%) for 
various charged Higgs masses. The
errors are statistical.
} 
\label{tab:semilepton}
\end{table} 

The Higgs mass is reconstructed from
the hadronic system  with 2.0 -- 2.5 GeV resolution using a 
one-constraint
kinematic fit requiring energy and momentum conservation
and the decay of two equal mass objects.
If the fit has a $\chi^2$ probability of less than
10$^{-5}$, the mass is calculated, instead, from the measured jet
four-momenta using the angular information and scaling 
the total  energy of the hadronic system
to the beam energy. 
The resulting mass distributions are shown in  Figure~\ref{fig:semimass}(b).

The signal selection efficiencies are affected by systematic
uncertainties on the tau lepton identification (3.0\%) and   
the modelling of the kinematic variables (4.8\%). The total systematic
uncertainty is 5.7\%.
The additional 
statistical error of the background Monte Carlo samples is 2.5\%.
The background estimate is affected by the following 
systematic uncertainties:  
modelling of the hadronisation process estimated by comparing different event
generators (8\%), uncertainty on the tau lepton identification
(3\%), and modelling of the kinematic variables (9\%). The total systematic
error on the background estimate is 12\%.
The additional error from Monte Carlo statistics is typically 5\%.
%
\subsection{The Hadronic Final State}

The hadronic channel, $\HH\ra\qqp\qqp$,  is
characterised by an event topology of four well separated hadron
jets and large visible energy.
The background comes from \qq($\gamma$) events with hard gluon emission
and from four-fermion processes, predominantly 
$\WW\ra\qqp\qqp$.

First, well-defined four-jet events are 
preselected; then a set of variables are combined using a likelihood technique.

The preselection consists of the following cuts:
(1) The event must qualify as a hadronic final state~\cite{l2mh}.
(2) The effective centre-of-mass energy~\cite{l2mh} 
($\sqrt{s^\prime}$) is required to be at least 150 GeV and
the visible energy ($E_{vis}$) is required to be at least $0.7\sqrt{s}$.
(3) The events are reconstructed into four jets using the Durham
algorithm.
The jet resolution parameter $y_{34}$ has to be larger than 0.0025.
Each jet must contain at least one charged track.
(4) A 4-C fit
requiring energy and momentum conservation is required to 
yield a $\chi^2$ probability larger than 10$^{-5}$, and
a 5-C fit requiring equal di-jet invariant
masses in addition is required to converge and yield a 
$\chi^2$ probability larger than 10$^{-5}$ for at least one of the three  
jet pair combinations.
(5) The $C$-parameter must be larger than 0.45.
\begin{table}[htp]
\centering
\begin{tabular}{|c||r||r||r|r|r||c|}
\hline
\multicolumn{1}{|c||}{Cut} & 
\multicolumn{1}{c||}{Data} & 
\multicolumn{1}{c||}{Total bkg.} & 
\multicolumn{1}{c|}{\qq($\gamma$)} & 
\multicolumn{1}{c|}{4f} & 
\multicolumn{1}{c||}{$\gamma\gamma$} & 
\multicolumn{1}{c|}{Efficiency (\%)} \\
 & 183 GeV & & & & & $\mHpm = 60$ GeV \\ \hline
(1) & 6333 & 6405 & 5304 & 987 & 114 & 100. \\
(2) & 1939  & 1980 & 1457 & 519 & 4.3   & 97.0 \\
(3) & 707   & 703 & 280 & 422 & 1.3   & 89.0 \\
(4) &  534  & 542 & 183 & 358 & 0.4   & 78.0 \\
(5) &  454  & 445 & 104 & 341 & 0.3   & 76.6 \\ \hline
${\cal L} > 0.6$
&  50   & 48.8$\pm$0.7
                & 9.0$\pm$0.3   
                        & 39.8$\pm$0.7   
                                & ---   
                                        & 42.8$\pm$2.2
\\
\hline
\end{tabular}
\caption{\sl 
Hadronic charged Higgs boson channel:
Comparison of the number of observed events and expected
background (normalised to 56.2~\pb) together with the 
signal efficiency for $\mHpm = $ 60~GeV after each cut. 
The errors are statistical.
}
\label{tab:hadcuts}
\end{table}

To separate the signal from the background events surviving the preselection,
a likelihood technique is applied. Three event classes are defined:
two-fermion, four-fermion, and $\HH \to \qqp \qqp$.
The following five variables are used as input: 
the cosine of the smallest jet-jet angle 
($\cos{\alpha_{\mrm{min}}}$);
the difference between the largest and smallest jet energy
($E_{\mrm{max}}-E_{\mrm{min}}$) after the 4-C fit;
the cosine of the polar angle of the thrust axis
($\cos{\theta_{\mrm{thrust}}}$);
the cosine of the
di-jet production angle ($\cos{\theta_{\mrm{di-jet}}}$)  multiplied by
the di-jet charge\footnote{If there is more than one charged track in a
jet, its charge is calculated as $\Sigma q^{(i)} \sqrt{p_{\mathrm{L}}^{(i)}}
 / \Sigma \sqrt{p_{\mathrm{L}}^{(i)}}$, 
where the sum goes over each track within the jet, $q^{(i)}$ is
the charge of the track and $p_{\mathrm{L}}^{(i)}$ 
is its momentum parallel to the jet
direction. A charge of +1 is assigned to the di-jet system with the larger
sum of the two individual jet charges, and a charge of -1 to the other.}
($Q_{\mrm{di-jet}}$) for the 
combination with the highest probability given by the 5-C fit;
and the smallest di-jet mass difference ($\Delta M_{\mrm{min}}$) 
after the 4-C fit.
An event is selected if its likelihood output ${\cal L}$ is greater than 0.6.

\begin{table}[htbp]
\begin{center}
\null
{\small  
\begin{tabular}{|c|c|c|c|c|c|c|c|c|}
\hline
\multicolumn{9}{|c|}{Signal selection efficiencies ($\%$)} 
\\ \cline{1-9}
50 GeV  & 55 GeV & 60 GeV & 65 GeV & 70 GeV 
& 75 GeV & 80 GeV & 85 GeV & 90 GeV 
\\ 
\hline
 36.8$\pm$2.2     & 
 42.0$\pm$2.2     &               
 42.8$\pm$2.2     &               
 33.0$\pm$2.1     &               
 26.0$\pm$2.0     &
 16.4$\pm$1.7     &
 12.4$\pm$1.5     &
 12.2$\pm$1.5     &
 11.6$\pm$1.4                       
\\ 
\hline 
\end{tabular}
}
\end{center}
\caption{\sl 
Hadronic charged Higgs boson channel:
Signal selection efficiencies (in \%)
for various charged Higgs masses. 
The errors are statistical. 
}
\label{tab:hadr}
\end{table}
\begin{figure}[htbp] 
\centering
\epsfig{file=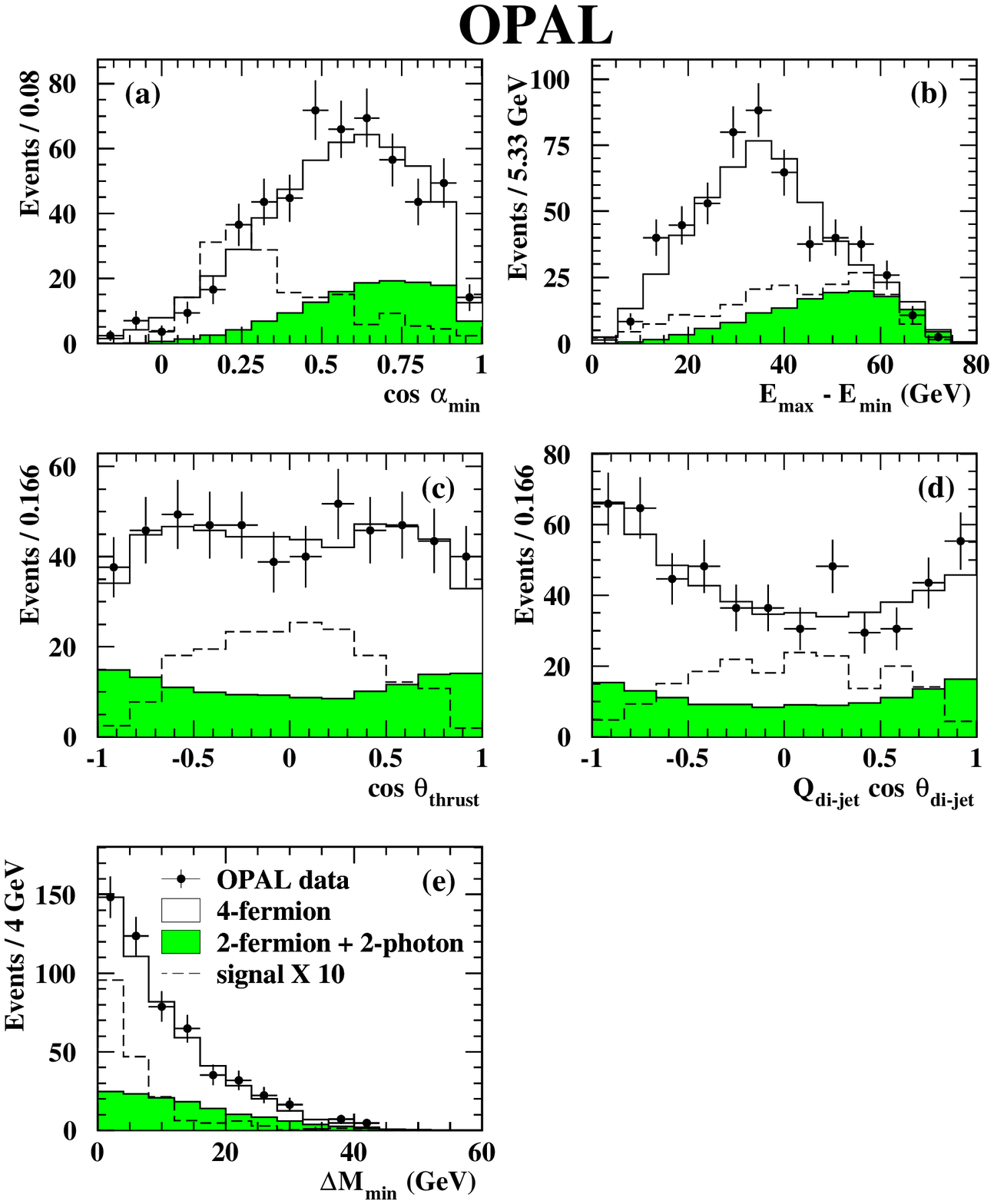,width=16cm}
\caption{\sl
Hadronic charged Higgs boson channel:
distributions used in the likelihood selection.  
The points with error bars are data, the shaded (open)
histogram is the simulation of two-fermion (four-fermion) events,
normalised to the recorded luminosity. The dashed line
is a simulated signal (\mHpm\ = 60~GeV) assuming  BR($\Hp\qqp$) = 1 and
scaled by a factor of 10 for better visibility.}
\label{fig:hadrvars}
\end{figure}

\begin{figure}[htbp]
\centering
\epsfig{file=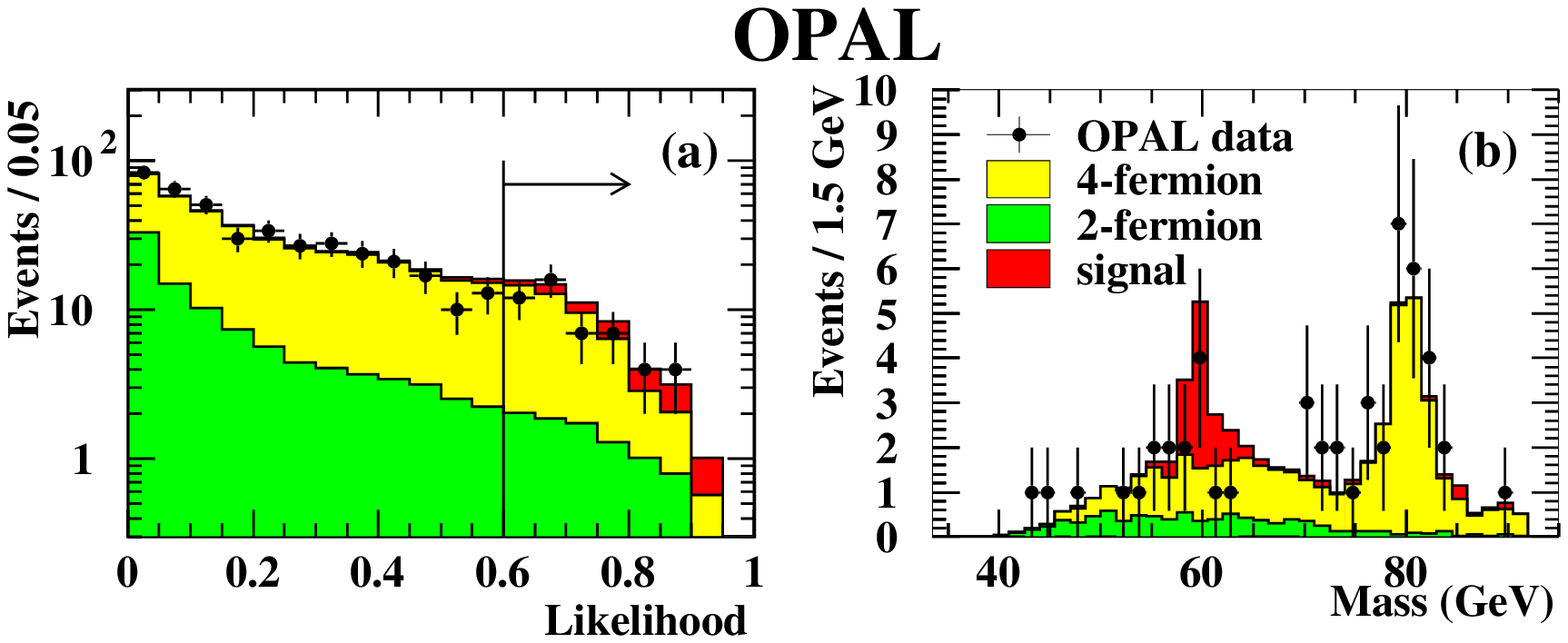,width=\textwidth}
\caption{ \sl
Hadronic charged Higgs boson channel:
(a) the likelihood output and (b) the invariant mass
distribution.
The points with error bars are data, the grey (light grey)
histogram is the simulation of two-fermion (four-fermion) events,
normalised to the recorded luminosity. The dark grey histogram
is a simulated signal (\mHpm\ = 60~GeV) assuming  BR($\Hp\to\qqp$) = 1
added to the background expectation.
The arrow indicates the cut on the likelihood output.
}
\label{fig:hadmres}
\end{figure}

In Figure~\ref{fig:hadrvars} the distributions of the 
input variables to the likelihood selection are shown. 
The likelihood distribution is shown in Figure~\ref{fig:hadmres}(a).
Table~\ref{tab:hadcuts} shows the  number of selected events, the
estimated background, and the fraction of signal events retained for 
\mHpm\ = 60 GeV after each cut.
In total, 50 events
are selected in the data, while 48.8$\pm$0.7 (statistical error) 
events are expected from Standard Model processes. 
The four-fermion processes account for 82\% of
the expected background, and result in
a large peak centred at the W$^\pm$ mass.

For the selected events, the jet pair association giving 
the highest $\chi^2$ probability in the 5-C fit
is retained. The resulting mass resolution ranges from 1.0~GeV to 1.5~GeV.
Figure~\ref{fig:hadmres}(b) shows the 
invariant mass distribution of the selected
events together with the Standard Model background expectation and 
a simulated signal of \mHpm\ = 60 GeV.

The uncertainties on the signal selection efficiency 
include typically 6\%
from the limited Monte Carlo statistics and 3\% systematic 
uncertainty from the modelling of the cut variables.

Systematic uncertainties arise from
modelling of the hadronisation process (2.0\%), estimated
by comparing different event generators and from modelling
of the cut variables (4.9\%), yielding a total systematic uncertainty of
5.3\%. The additional error from Monte Carlo statistics is 1.6\%.

\section{Interpretation of the Search Results}
\label{sect:interpret}

None of the searches presented in the previous sections revealed a 
significant excess over the expectation from SM background processes.
This negative result is used to derive limits at the 95\% confidence level (CL)
on neutral Higgs boson masses in the SM, in 2HDM and in the MSSM under various
assumptions for the values of the free parameters of the models. A
limit on the charged Higgs boson mass is also given.

The search channels are combined using the method described in Section~5
of~\cite{mssmpaper172}. This method takes into account the experimental
mass resolution, including tails, in all search channels. The expected
background is reduced by its systematic error in each channel and then 
subtracted.

\subsection{Mass Limit for the Standard Model Higgs Boson}
\label{sect:smlimit}
\begin{figure}[htbp]
\centerline{\epsfig{file=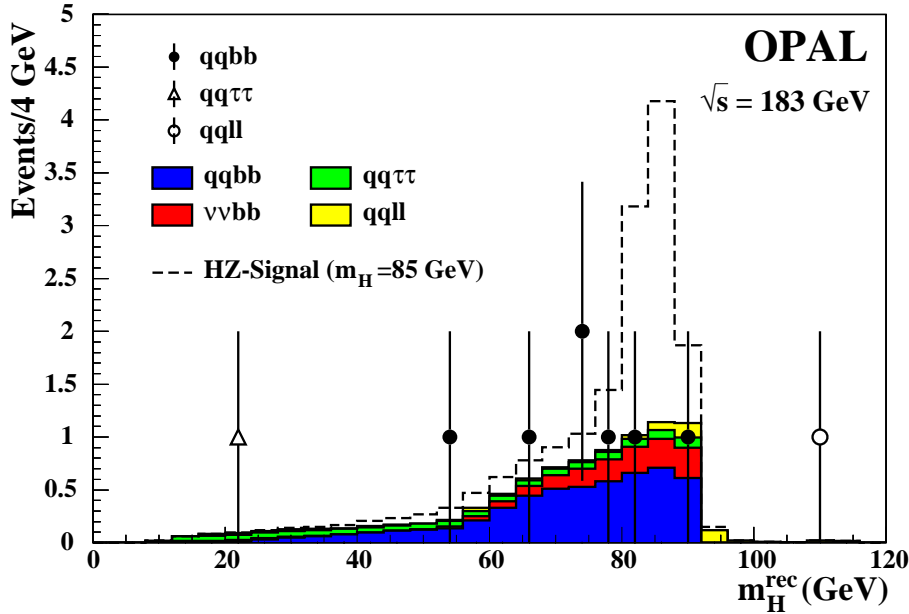,width=12cm}}
\caption[]{\label{fig:massplot}\sl
Distribution of the reconstructed Higgs boson candidate masses, 
$m_{\mathrm{H}}^{\mathrm{rec}}$,
for all SM channels.
The data (points with error bars) are 
compared to the Monte Carlo expectations for the backgrounds
from the various processes for the different selection channels (full
histograms). 
A simulated signal for $\mH = $ 85~GeV (dashed line) 
is also shown, added to the background expectation.
}
\end{figure}

Table~\ref{tab:smsummary} lists the efficiencies and 
expected signal event rates
for all search channels relevant for the SM Higgs boson as a function of
the Higgs boson mass. The total expected
event rate from all channels combined is also shown.
In Figure~\ref{fig:massplot}
the masses of the nine candidate events are shown together with the expected
background and a simulated signal at $\mH =$ 85 GeV. Only the data taken at
$\sqrts\approx$183~GeV are considered.
\begin{table}[hbtp]
\begin{center}
\begin{footnotesize}
\begin{tabular}{|c||c|c|c|c|c||c|} \hline
 \mH &\qq\Ho   &\nn\Ho&\tautau\qq&\ee\Ho&\mm\Ho&expected \\
(\Gc)&\Ho\ra\bb&      &         &      &      &events (total) \\
\hline\hline
 70&30.2 (8.1)&41.7 (3.9)&31.2 (1.2)&57.3 (0.9)&69.0 (0.9)&15.0 \\ \hline
 75&33.9 (7.5)&43.8 (3.4)&32.5 (1.1)&58.5 (0.8)&60.7 (0.8)&13.5 \\ \hline
 80&37.1 (6.4)&43.7 (2.7)&33.1 (0.9)&58.7 (0.6)&62.0 (0.6)&11.2 \\ \hline
 85&39.2 (4.7)&40.2 (1.7)&33.0 (0.6)&57.9 (0.4)&62.7 (0.5)& 7.9 \\ \hline
 90&39.4 (2.1)&34.6 (0.7)&32.0 (0.3)&55.2 (0.2)&62.1 (0.2)& 3.4 \\ \hline
 95&36.6 (0.30)&28.7 (0.13)&29.9 (0.04)&47.0 (0.03)&57.7 (0.03)& 0.53 \\ \hline
100&29.9 (0.10)&26.4 (0.07)&26.6 (0.01)&32.3 (0.01)&47.2 (0.01)& 0.20 \\ \hline
\hline
Background & 5.0$\pm$0.2 &1.6$\pm$0.1 & 1.3$\pm$0.1 &
\multicolumn{2}{|c||}{0.6$\pm$0.1} & 8.5$\pm$0.4 \\\hline
Systematics&     $\pm$0.6 &     $\pm$0.2&     $\pm$0.2 &
\multicolumn{2}{|c||}{    $\pm$0.1} &      $\pm$ 0.7 \\\hline
Observed   & 7 &      0    &      1 &\multicolumn{2}{|c||}{1}& 9 \\
\hline
\end{tabular}
\end{footnotesize}
\caption[]{\label{tab:smsummary}\sl
        Detection efficiencies (in \%) and numbers of expected Higgs 
        boson events 
         (in parentheses) at $\sqrt{s}$= 183~\Gc\ 
          for each search channel separately,
        as a function of the Higgs boson mass. 
        The last column shows the total numbers of  
        expected events in the present search at $\sqrt{s}$= 183~\Gc\@ .
}
\end{center}
\end{table}

\begin{figure}[htbp]
\centerline{\epsfig{file=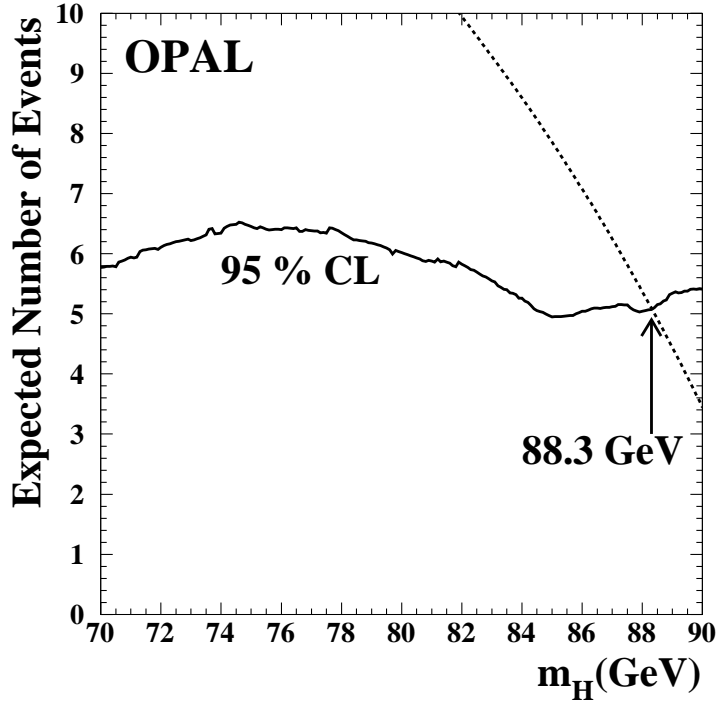,width=0.55\textwidth}}
\caption[]{\label{fig:smresults1}\sl
           Upper limit on the production rate for SM Higgs bosons
           at 95\% CL (solid line) and the expected event rate (dashed line)
           as a function of the Higgs boson mass. }
\end{figure}

\begin{figure}[htbp]
\centerline{\epsfig{file=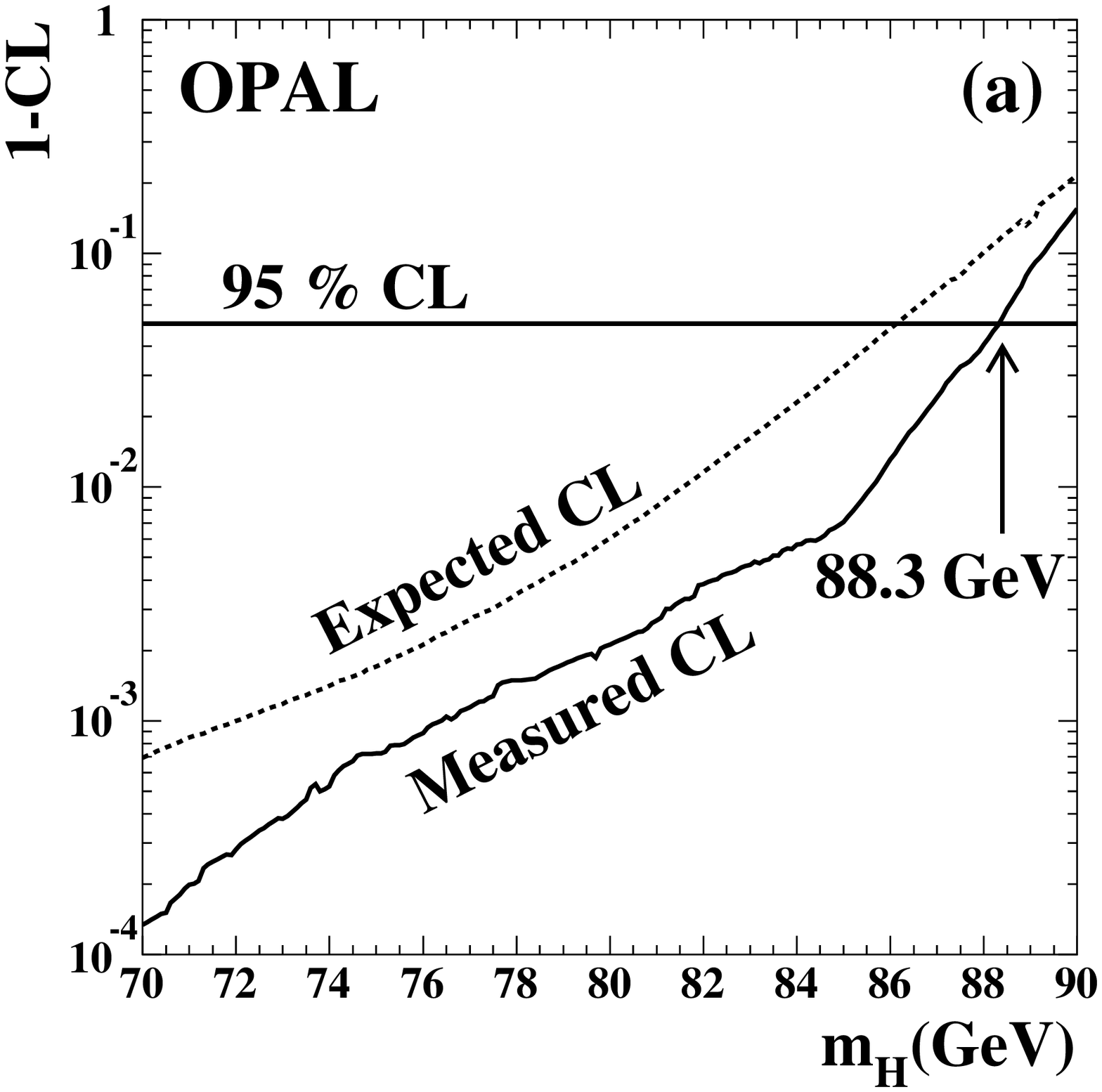,width=0.4\textwidth}
            \epsfig{file=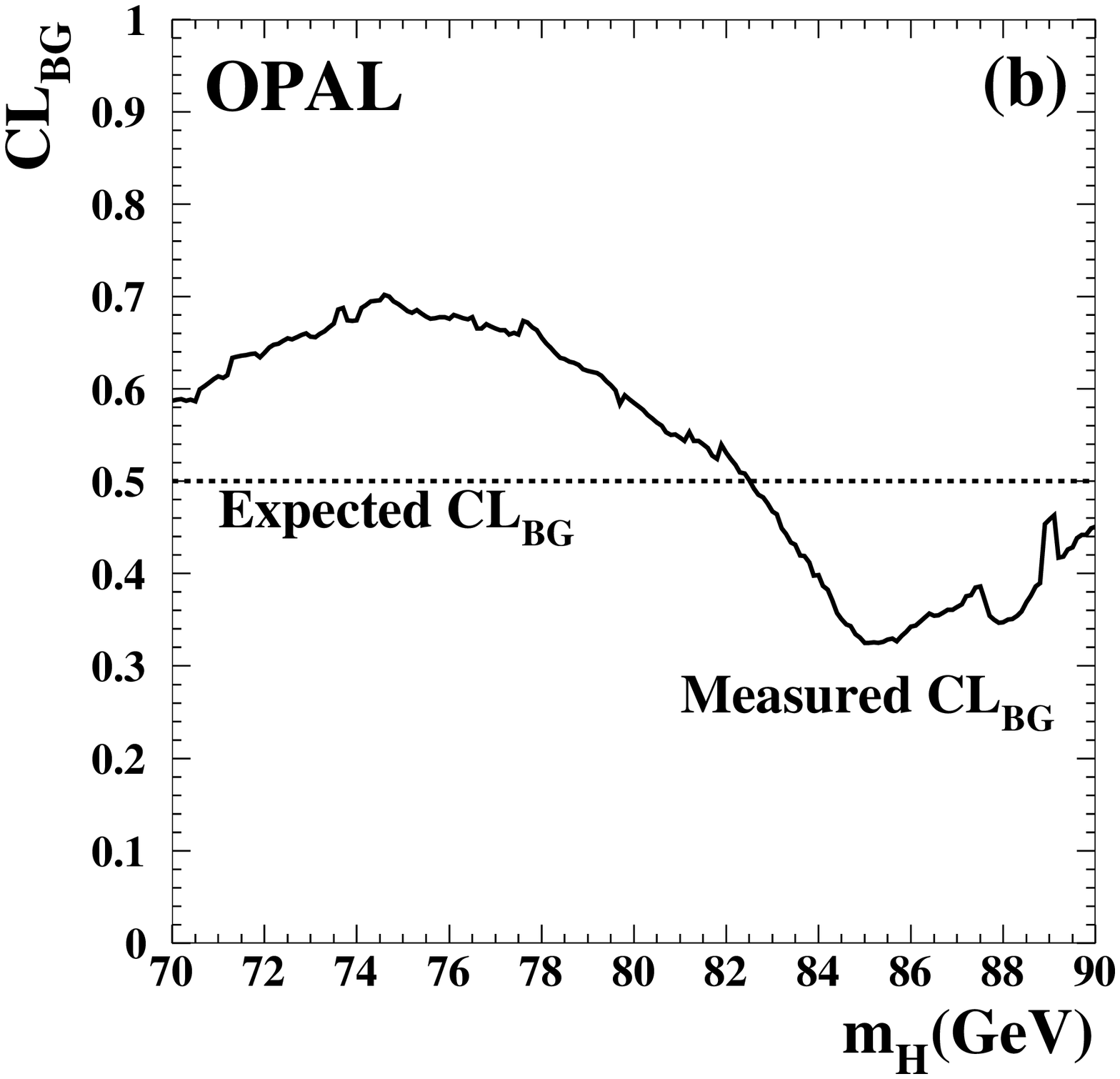,width=0.4\textwidth}}
\caption[]{\label{fig:smresults2}\sl
           Search for the SM Higgs boson:
           (a) Measured (solid line) and average expected (dashed line)
           confidence levels for the signal hypothesis
           as a function of the Higgs boson mass. 
           (b) Measured (solid line) and average expected (dashed line)
           confidence levels for the background hypothesis
           as a function of the Higgs boson mass.
}
\end{figure}

Figures~\ref{fig:smresults1} and~\ref{fig:smresults2}
show the results for signal event rates and
confidence levels for the signal and background hypotheses.
At 95\% CL the derived observed lower limit for the SM Higgs boson mass 
is found to be $\mH>88.3$~GeV, while the average 
expected limit from simulated background-only experiments is $\mH>86.1$~GeV.
From Figure~\ref{fig:smresults2}(b) it can be seen that this observation
is quite compatible with the SM background for Higgs boson mass 
hypotheses between 70 and 90~GeV.
The probability for obtaining a limit larger than 88.3~GeV 
was found to be 40\% if no signal is present.

%
\subsection{Model--independent and 2HDM Interpretation}
\label{sect:modindep}
Model-independent limits are determined for the cross-section for the
generic processes \ee\ra~S$^0$\Zo\ and \ee\ra~S$^0$P$^0$, where S$^0$ and
P$^0$ denote scalar and pseudo-scalar neutral bosons 
which decay into a pair of leptons or quarks, respectively.
This is achieved by combining the searches presented in this publication
with previous OPAL Higgs 
searches~\cite{smpaper172,mssmpaper172,smhiggs91,mssmhiggs91}
at \sqrts\ values between \mZ\ and 172~GeV.
The limits are conveniently expressed in terms of 
scale factors, $s^2$ and $c^2$, which relate the cross-sections of these
generic processes to those of the SM cross-sections
(c.f.~Eqs.~(\ref{equation:xsec_zh}),~(\ref{equation:xsec_ah})):
\begin{equation}
\sigma_{\mathrm{SZ}}=s^2~\sigma^{\mathrm{SM}}_{\mathrm{HZ}},
\label{eq:s}
\end{equation}
\begin{equation}
\sigma_{\mathrm{SP}}=c^2~\bar{\lambda}~\sigma^{\mathrm{SM}}_{\nn}.
\label{eq:c}
\end{equation}

\begin{figure}[tbp]
\centerline{ \epsfig{file=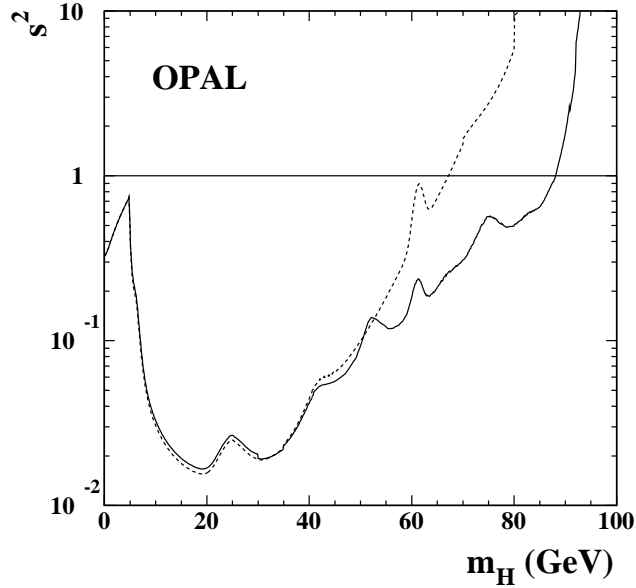,width=8.3cm} }
\caption[]{\label{modindepZh}\sl
         Upper limits at 95\% CL on $s^2$ (as defined by Eq.~(\ref{eq:s}))
         using all SM search channels and assuming the SM Higgs boson branching
         ratios for the S$^0$ (solid line). The dashed line is from a previous
         OPAL search~\cite{mssmpaper172} and includes only channels
         that do not use b-tagging. A hadronic branching ratio of 
         the S$^0$ of 100\% is assumed.
}
\end{figure}
Figure~\ref{modindepZh}
shows the 95\% CL upper bound for $s^2$
as a function of the S$^0$ mass, obtained using:
\vspace*{-1mm}
\[s^2=
  \frac{N_{95}^{\mathrm{SZ}}}{
       \sum~(\epsilon~{\cal L}~\sigma^{\mathrm{SM}}_{\mathrm{HZ}})},
\vspace*{-1mm}
\]
where $N^{\mathrm{SZ}}_{95}$ is the 95\% CL upper limit
for the number of possible
signal events in the data, $\epsilon$ is the signal detection efficiency,
${\cal L}$ is the integrated luminosity, and
the sum runs over the different centre-of-mass energies of the data.
The solid line is computed using all SM search channels and assumes
SM Higgs branching ratios for the S$^0$.
The dashed line (from a previous OPAL search~\cite{mssmpaper172} 
is computed assuming 100\% hadronic branching ratio for the
S$^0$ and uses only search channels that do
not employ b-tagging (see~\cite{mssmpaper172} for a list of the search
channels) and is therefore more general.
At low masses, the searches lose sensitivity
rapidly, and the limit for $s^2$ is determined from
the decay width of the \Zo\ boson only, as described in~\cite{mssmpaper172}.

\begin{figure}[tbp]
\centerline{
\mbox{\epsfig{file=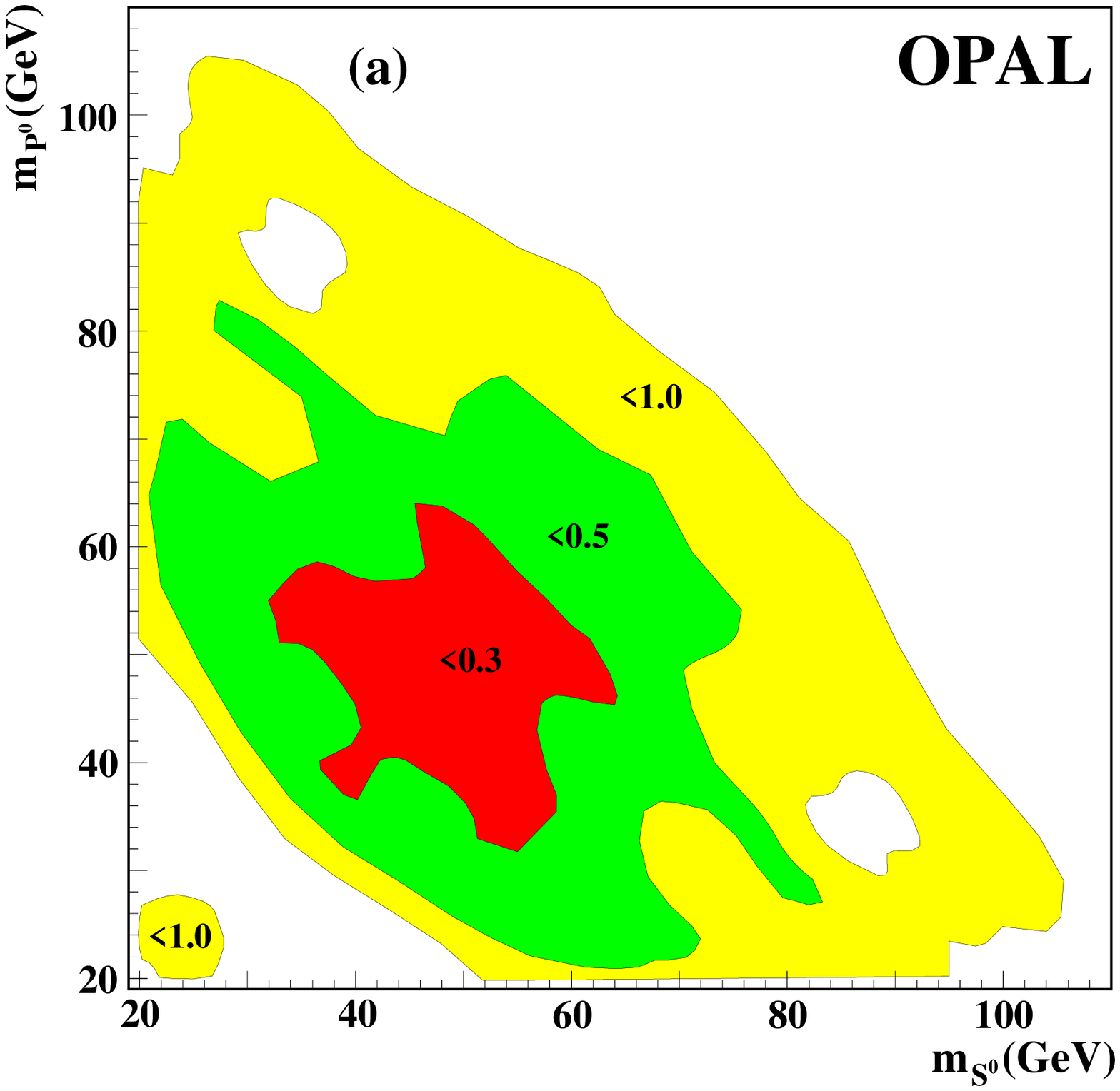,width=8.3cm}}\hfill
\mbox{\epsfig{file=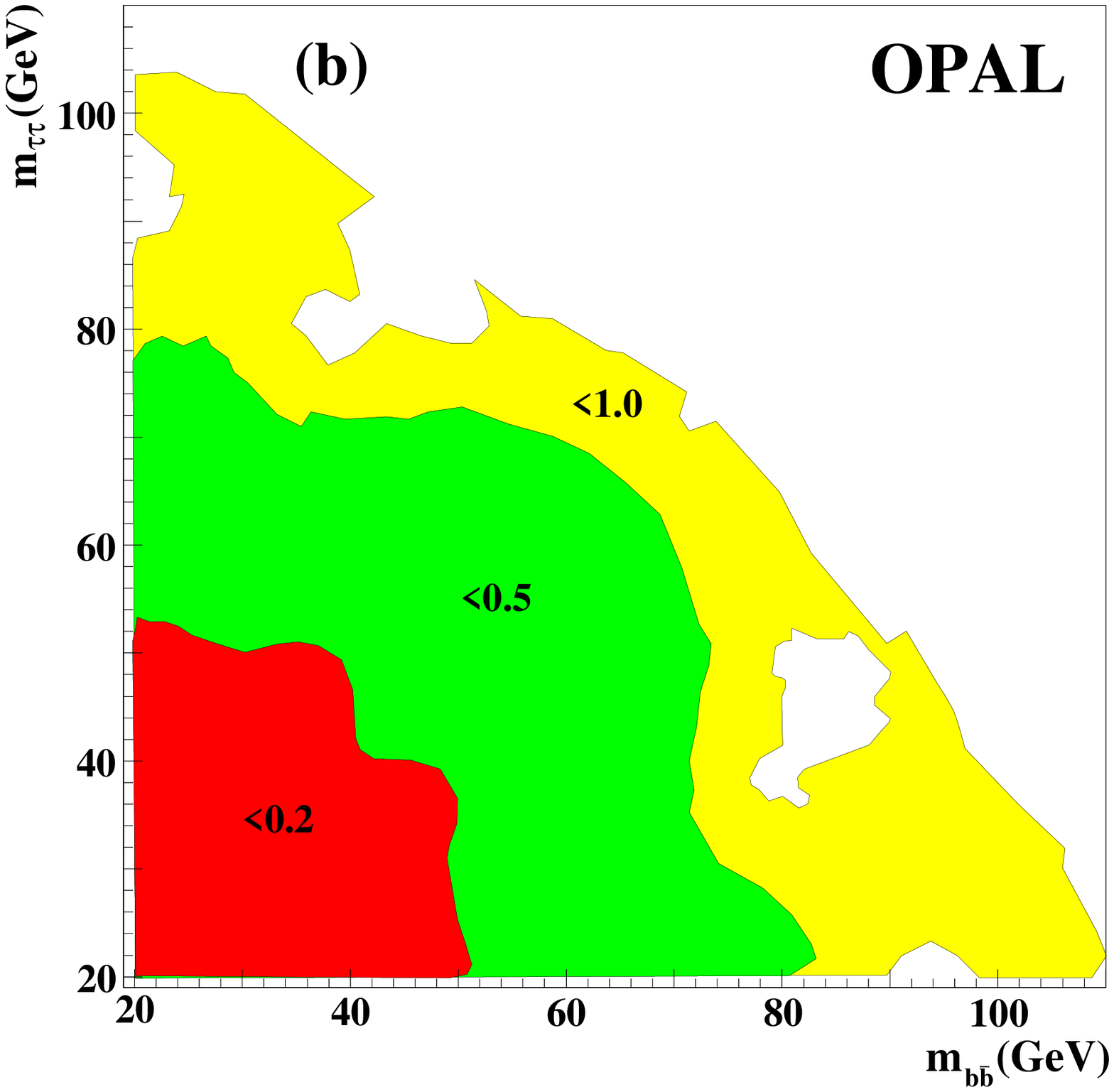,width=8.3cm}}
}
\caption[]{\label{modindephA}\sl
         Upper limits at 95\% CL on $c^2$ (see Eq.~(\ref{eq:c}))
         for:
         (a) the S$^0$P$^0$\ra\bb\bb\ search channel assuming the \bb\ branching         ratio for both S$^0$ and P$^0$ to be 100\%, and
         (b) the S$^0$P$^0$\ra\bb\tautau\ search channel assuming a
         100\% branching ratio for this final state.
         The invariant masses of the tau lepton pair and hadron jet pair
         are denoted $m_{\tau\tau}$ and $m_{\mathrm{b\bar{b}}}$, respectively.
}
\end{figure}
Figure~\ref{modindephA} shows contours of 95\% CL
upper limits for $c^2$ in the S$^0$ and P$^0$ mass plane,
for the processes \ee\ra~S$^0$P$^0$\ra\bb\bb\ and \bb\tautau, respectively.
In both cases a 100\% branching ratio into the specified final state is
assumed.
The contours are obtained from:
\vspace*{-2mm}
\[c^2=
  \frac{N_{95}^{\mathrm{SP}}}{
       \sum~(\epsilon~{\cal L}~\bar{\lambda}~\sigma^{\mathrm{SM}}_{\nn})},
\vspace*{-1.5mm}
\]
with $N_{95}^{\mathrm{SP}}$ being the 95\% CL upper limit for the number of
signal events in the data.
The results obtained for \bb\bb\ (Figure~\ref{modindephA}(a)) are
symmetric with respect to interchanging S$^0$ and P$^0$.
In Figure~\ref{modindephA}(b), the results for the
\tautau\bb\ final state are shown with the
mass of the particle decaying into \tautau\ along the abscissa 
and that of the particle
decaying into \bb\ along the ordinate.
The irregularities of the contours are due to the presence
of candidate events that affect $N_{95}^{\mathrm{SP}}$.

\begin{figure}[tbp]
\centerline{
\epsfig{file=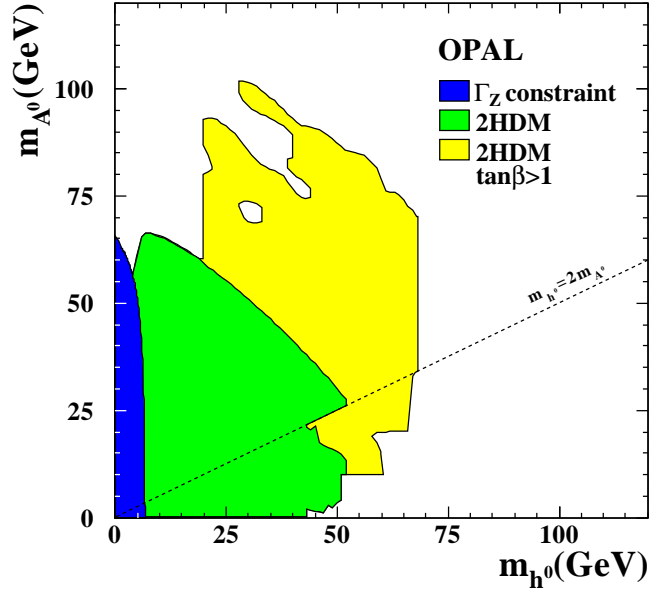,width=8.5cm}
}
\caption[]{\label{modind_gz}\sl
  Regions excluded at 95\% CL in the Type II 2HDM.
  The black region is excluded using constraints from $\Gamma_{\mathrm{Z}}$
  only.
  The dark grey region uses the direct searches for the SM Higgs in addition,
  but discarding the search channels that use b-tagging, assuming a hadronic
  branching ratio of the \ho\ of 92\%.
  The light grey region is excluded for $\tanb>1$ in the 2HDM, assuming
  SM Higgs branching ratios for \ho\ and \Ao.
}
\end{figure}
In the 2HDM the bosons S$^0$ and P$^0$ are identified with \ho\ and \Ao,
and the couplings $s^2$ and $c^2$ are identified with \sba\ and \cba,
respectively.
The assignment of the possible excess width in $\Gamma_{\mathrm{Z}}$
to the process \Zo\ra{\ho}\Zs\ yields an upper bound for
$s^2$ which depends only on the mass of \ho\ whereas the assignment to
\Zo\ra\ho\Ao\ yields an upper bound for $c^2$ which depends on the
masses of both \ho\ and \Ao.
Combining these limits, the black region shown in
Figure~\ref{modind_gz} is excluded at 95\% CL
regardless of the \ho\ and \Ao\ decay modes.
In the 2HDM, the most important
final states of the decays of the \ho\ and \Ao\ bosons are \bb, \cc\ and
\tautau\ but \ho\ra\Ao\Ao\ is also possible.
The branching ratios depend on \tanb, but the hadronic
branching fraction always exceeds 92\%~\cite{hdecay}.
For $\tanb{\geq}1$ the \bb\ channel dominates
while for $\tanb<1$ the \cc\ contribution may become the largest.

In Figure~\ref{modind_gz} the excluded area in the $(\mh,\mA)$ plane
is shown when the limits on $c^2$ and $s^2$ are combined.
Below the dotted line, where the \ho\ra\Ao\Ao\ decay is kinematically allowed
and competes with the \ho\ra\ff\ decay,
the smaller of the detection efficiencies is used.
The excluded area is therefore
valid regardless of the \ho\ra\Ao\Ao\ branching ratio.
The dark grey area is excluded at 95\% CL when BR(\ho\ra\qq)$\ge$92\%
and is most generally valid in the 2HDM. This 95\% CL limit is 
obtained using only search channels that do not employ b-tagging.
The limit in the 2HDM for equal \ho\ and \Ao\ masses is at
$\mh=\mA=41.0$~GeV.
The light grey area is excluded when either SM Higgs branching ratios
or BR(\ho\ra\Ao\Ao)=100\% is assumed for \ho\ (whatever yields a more
conservative result) and SM Higgs branching ratios are assumed for \Ao.
This assumption provides conservative results in the 2HDM for $\tanb>1$.
In that case, the 95\% CL limit
for equal \ho\ and \Ao\ masses is at $\mh=\mA=68.0$~GeV.
The hole in the exclusion of the light grey area
is caused by a candidate event in the \ho\Ao\ra\bb\bb\ search.


\subsection{Interpretation of the Search Results within the MSSM}
\label{sect:mssm}

We consider a constrained MSSM with the following
free parameters in addition to those of the SM. 
The model assumes
unification of the scalar-fermion masses, $m_0$,
at the grand unification (GUT) scale, and unification of the 
gaugino masses (parametrised using $M_2$, the SU(2) gaugino mass 
term at the electroweak scale)
and unification of the scalar-fermion tri-linear
couplings, $A$, at the electroweak scale.
The remaining parameters are chosen to be the supersymmetric
Higgs mass parameter $\mu$, the ratio of the vacuum expectation values 
of the two Higgs field doublets, $\tanb=v_2/v_1$,
and the mass of the CP--odd Higgs boson, \mA .
The above simplifications have practically no impact on the MSSM Higgs
phenomenology; in particular, common
scalar-fermion mass and tri-linear couplings
are justified since only the scalar top (\sctop) sector gives important
contributions to Higgs boson masses and couplings.

Those six parameters were scanned within ranges motivated by theory.
The details of the MSSM parameter scans are described in~\cite{mssmpaper172}.
Since the precise value of the top quark mass, $m_{\rm t}$, has a strong
impact through loop corrections (on \mh\ in particular), it was considered
in the more general scans as a supplementary parameter, with values
$m_{\rm t} =$ 165, 175, and 185 GeV.

In this paper we consider the same three MSSM parameter scans (A, B and C)
already used in~\cite{mssmpaper172}. 

{\em Scan (A)}, 
proposed in~\cite{lep2higgs}, is the least general since, of the
seven parameters (including $m_{\rm t}$), only \mA\ and \tanb\ are varied
while $m_0$ and $M_2$ are fixed at 1~TeV and $\mu$ is chosen to be -100~GeV.
The top quark mass is fixed at 175~GeV.
Two sub-cases are considered, with
the tri-linear coupling fixed at $A=0$~TeV or $\sqrt{6}$~TeV, corresponding to
{\em no mixing} or {\em maximal mixing} in the scalar-top sector. 
In {\em Scan (B)}, $m_0$, $M_2$, \mA , \tanb , and $m_{\rm t}$ are varied 
independently while $\mu$ and $A$ are linked by relations which, in each
case, correspond to either {\em minimal} or {\em maximal mixing} in the
scalar-top sector.
In {\em Scan (C)}, the most general, all seven parameters were varied 
independently.
In each of these scans, the parameter
sets were used as input to the HZHA program~\cite{hzha} which calculates
the Higgs masses, cross--sections~\cite{remt,gkw} and branching 
ratios~\cite{hdecay}.  SUSYGEN~\cite{susygen}
was used to calculate scalar fermion masses at the electroweak scale.

Parameter sets giving rise to chargino or neutralino
masses~\cite{neutralino183}, or stop masses~\cite{lep2stop}, excluded by
OPAL searches, or to \Zo\ra\ho\Zs,\ho\Ao\ cross-sections incompatible
with the \Zo\ decay width (see~\cite{mssmpaper172}), have been discarded.
In the case of scan (C), they were also tested against 
criteria~\cite{frere,casas,kusenko} that exclude parameter sets
leading to charge- or colour-breaking (CCB) minima of the MSSM Lagrangian.

The searches presented in this publication are combined 
with previous OPAL Higgs 
searches~\cite{smpaper172,mssmpaper172,smhiggs91,mssmhiggs91} at \sqrts\
between \mZ\ and 172~GeV.

The results are presented, separately for each scan, in four sub-figures:
(a) in the (\mh,~\mA) plane for $\tanb>1$,
(b) in the same plane for $\tanb>0.7$,
(c) in the (\mh,~\tanb) plane, and (d) in the (\mA,~\tanb) plane.
For scans (A) and (B) the experimental 
lower limits for the minimal and maximal mixing
cases differ only marginally; therefore only the weaker of the two
exclusion limits is given.
The theoretically accessible area corresponds to the larger one,
for maximal scalar top mixing.
The theoretically inaccessible areas are shown in the figures in grey.

\begin{figure}[p]
\centerline{
\epsfig{file=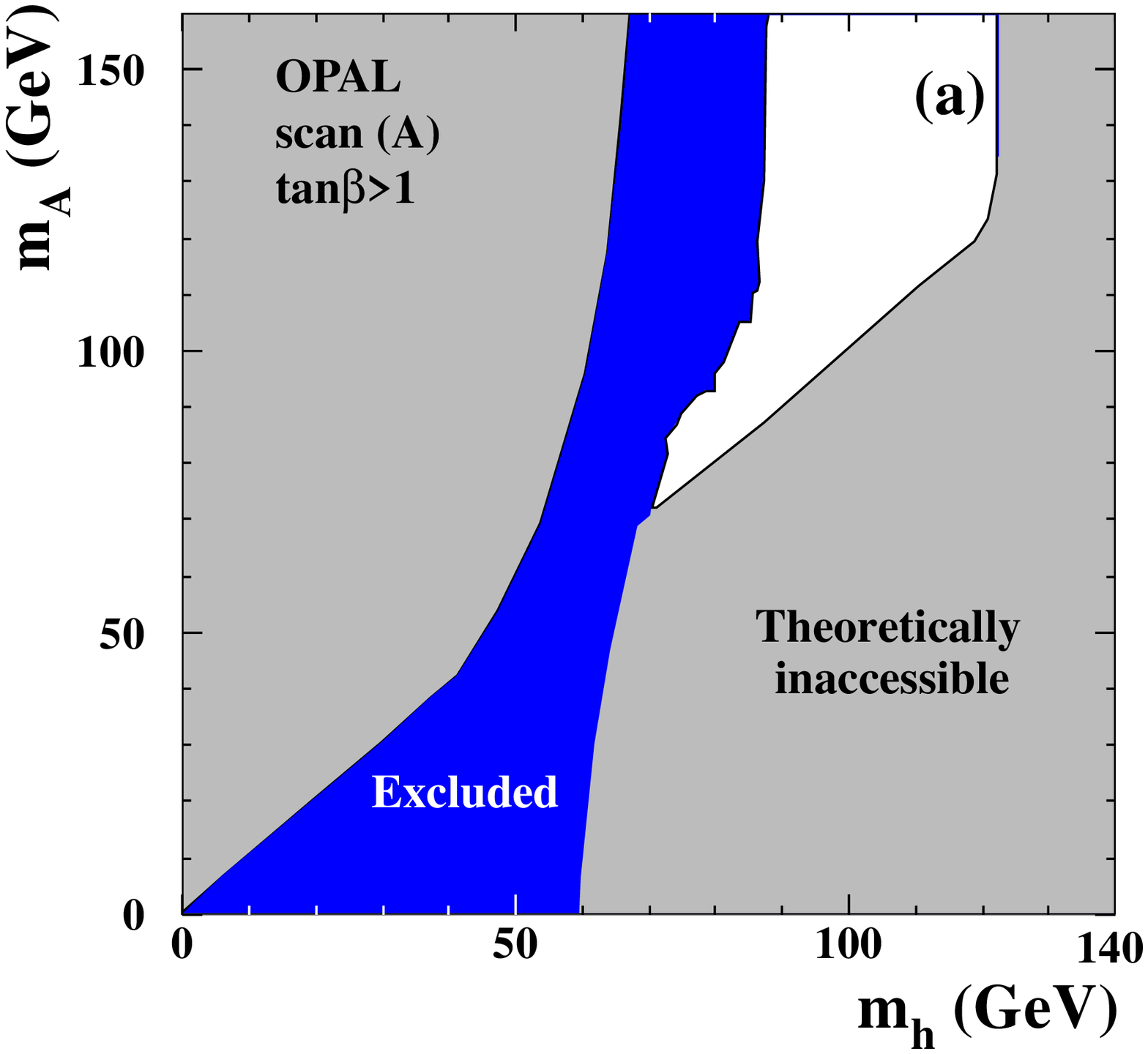,width=8.5cm}\hfill
\epsfig{file=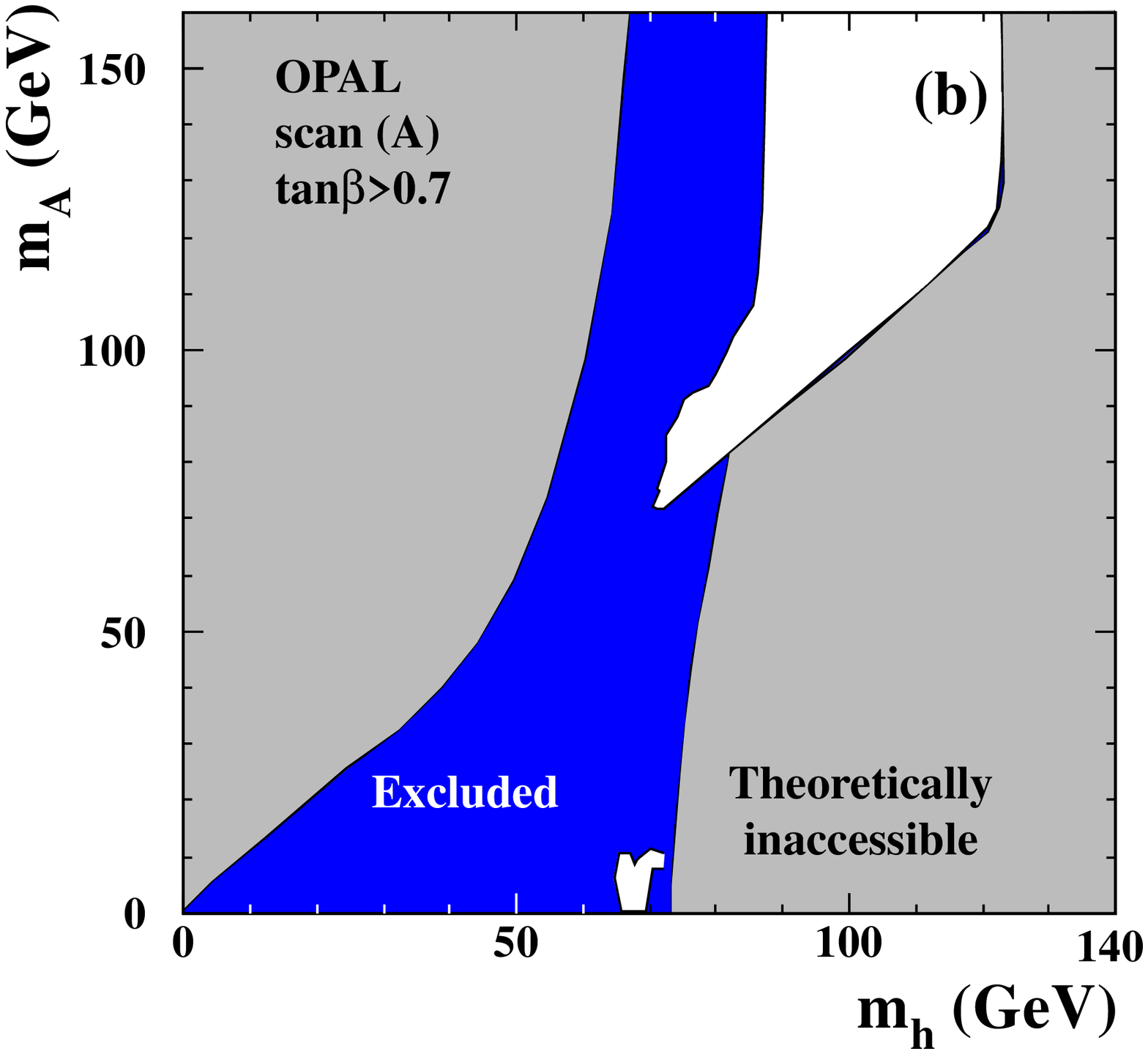,width=8.5cm}}
\centerline{
\epsfig{file=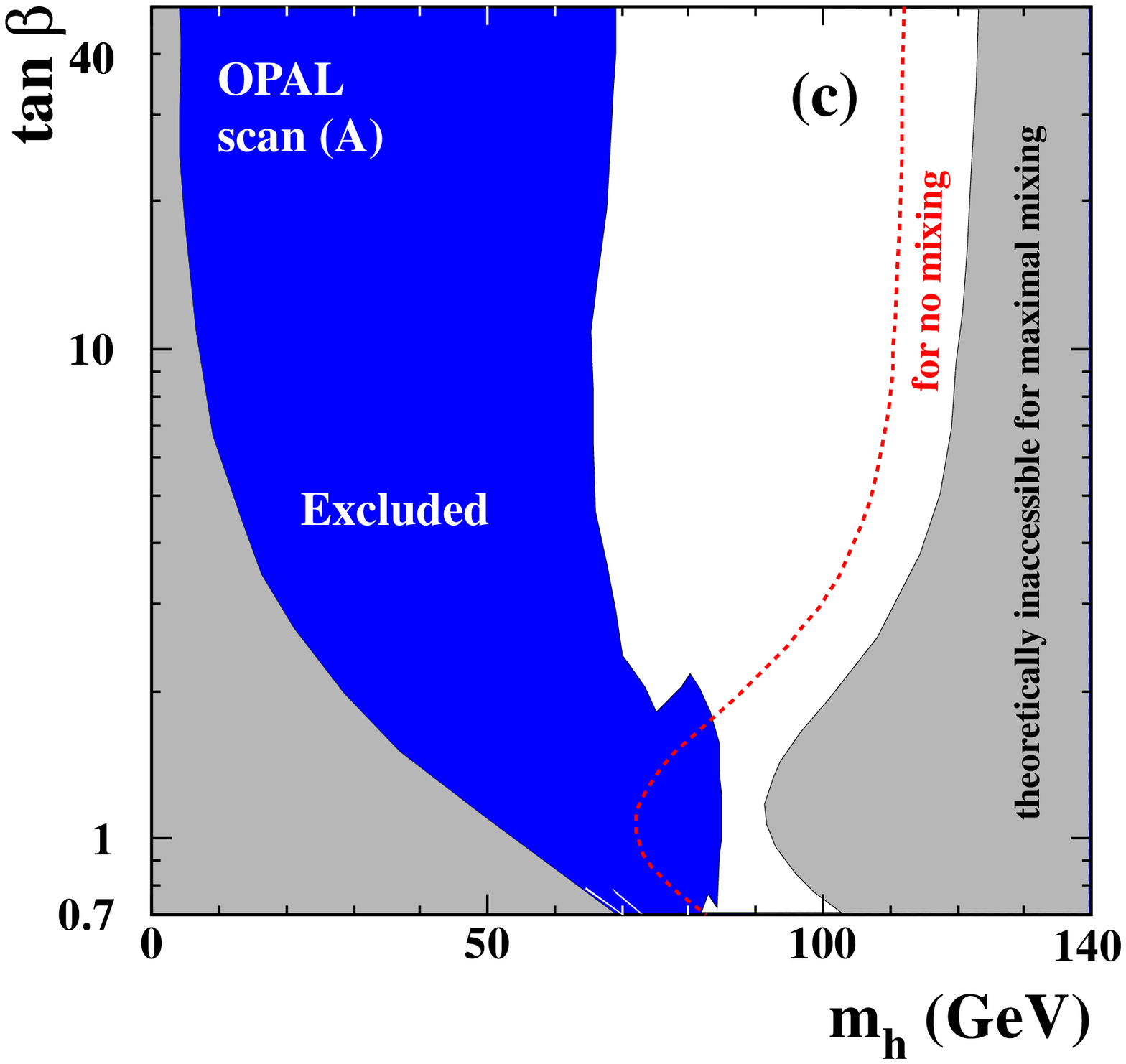,width=8.5cm}\hfill\hfill
\epsfig{file=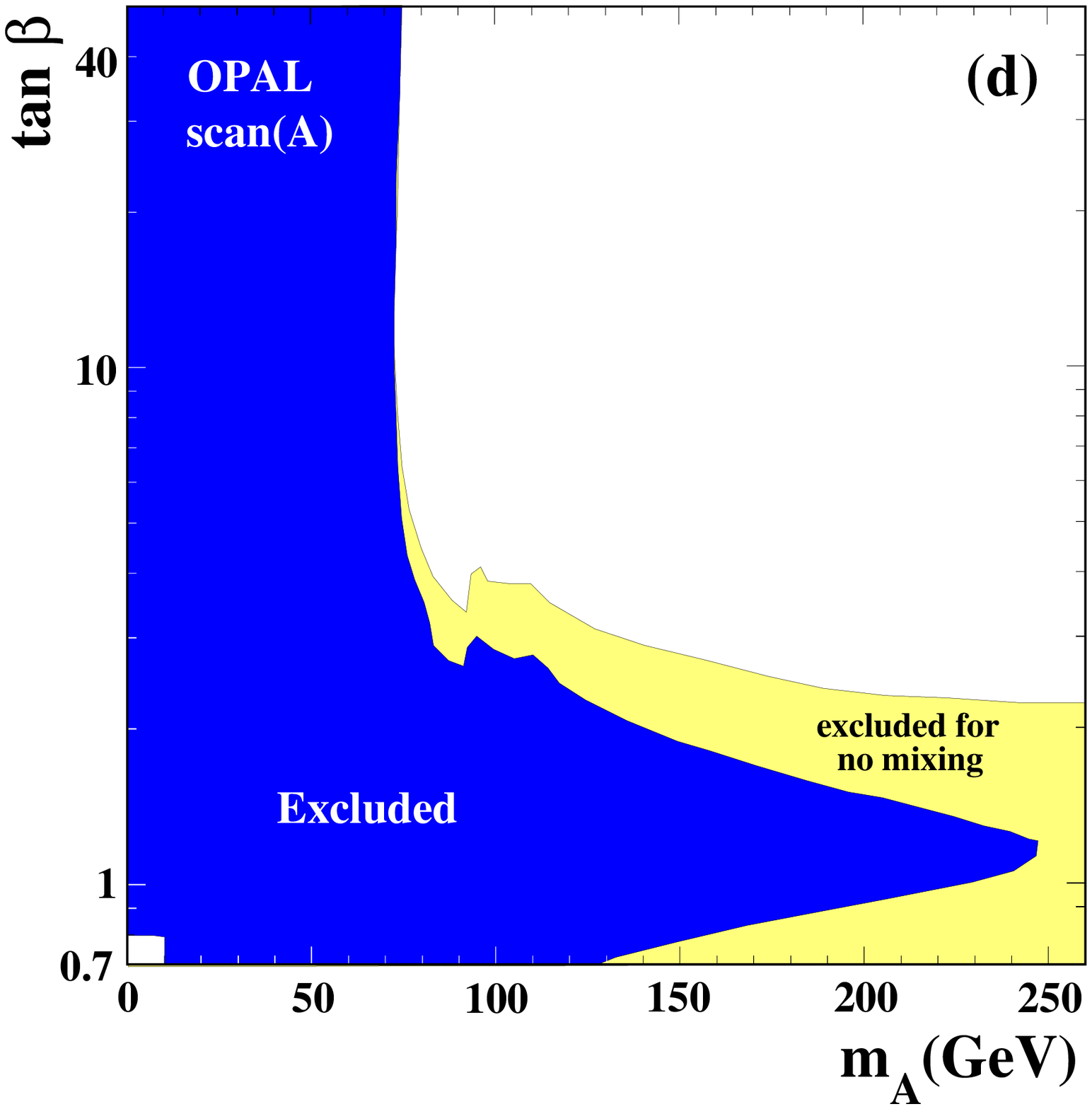,width=8.0cm}}
\caption[]{\label{fig:fig1-mssm}\sl
         The MSSM exclusion for scan (A) described in the text of
         Section~\ref{sect:mssm}.
         Excluded regions are shown for
         (a) the (\mh,~\mA) plane for $\tanb>1$,
         (b) the (\mh,~\mA) plane for $\tanb>0.7$,
         (c) the (\mh,~\tanb) plane, and
         (d) the (\mA,~\tanb) plane.
         The black area is excluded at 95\% CL.
         The grey areas in (a), (b) and (c) are theoretically inaccessible.
         The light grey area in (d) is excluded only for no scalar-top
         mixing.
}
\end{figure}
\begin{figure}[p]
\centerline{
\epsfig{file=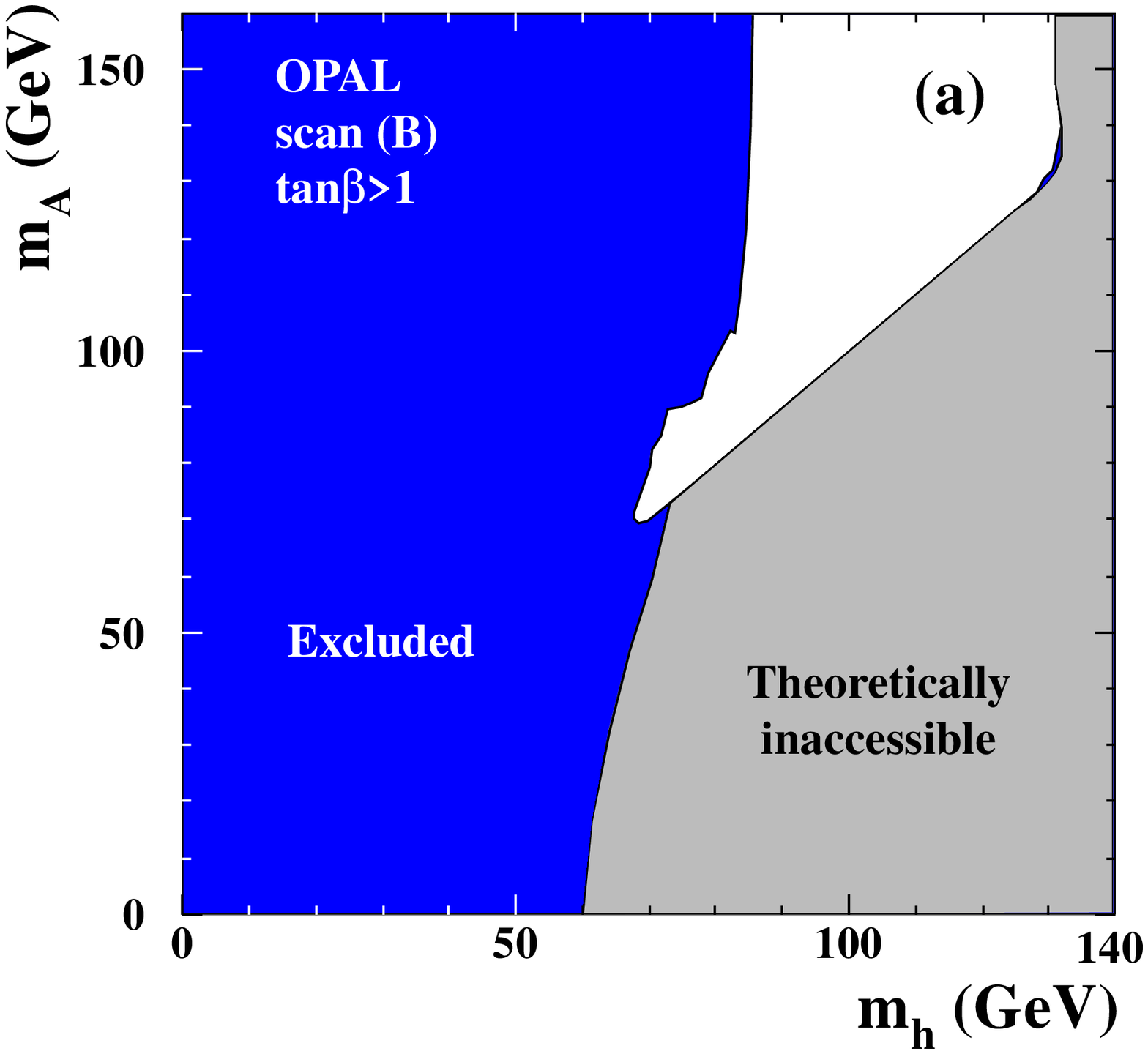,width=8.5cm}\hfill
\epsfig{file=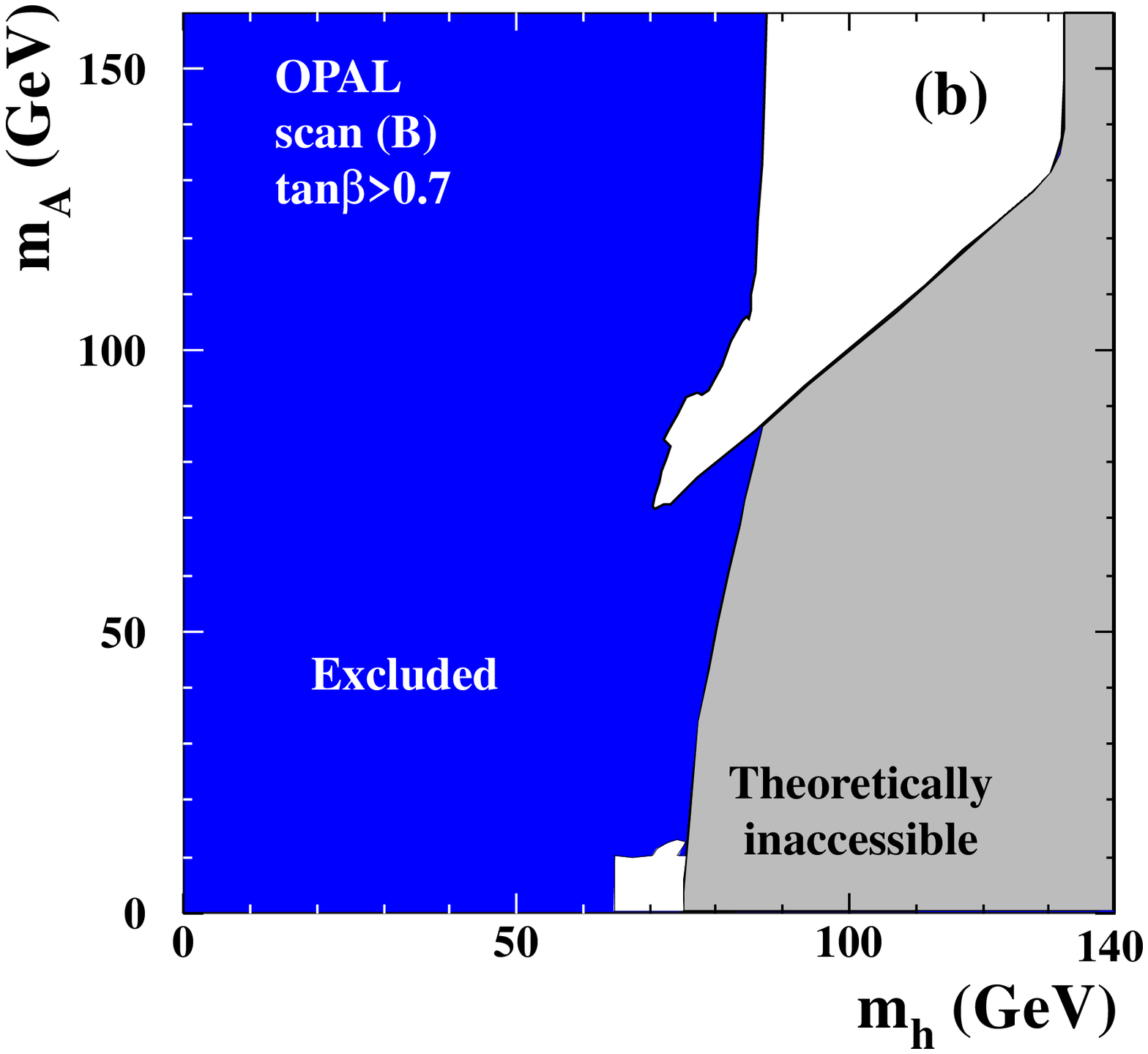,width=8.5cm}}
\centerline{
\epsfig{file=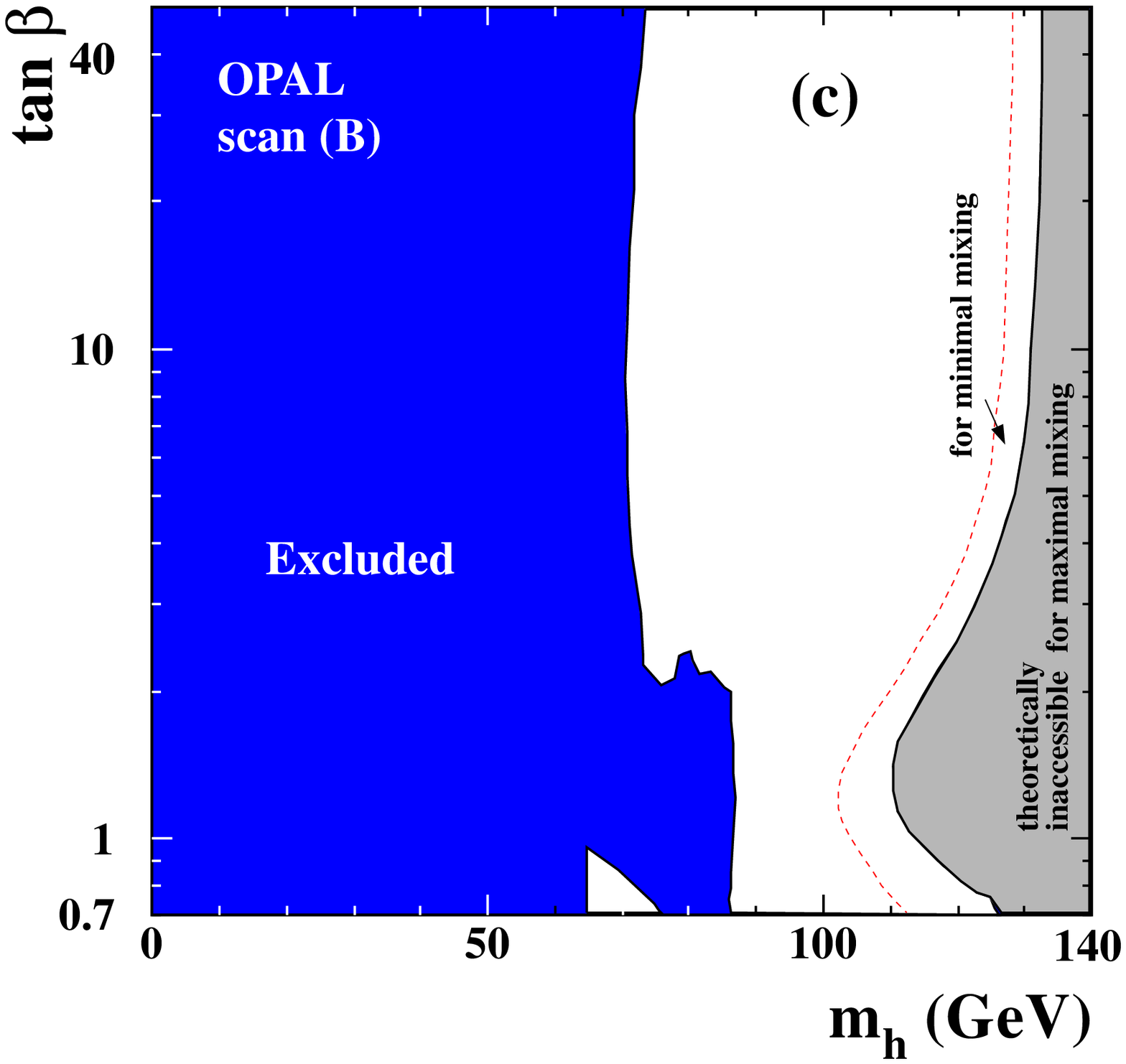,width=8.0cm}\hfill
\epsfig{file=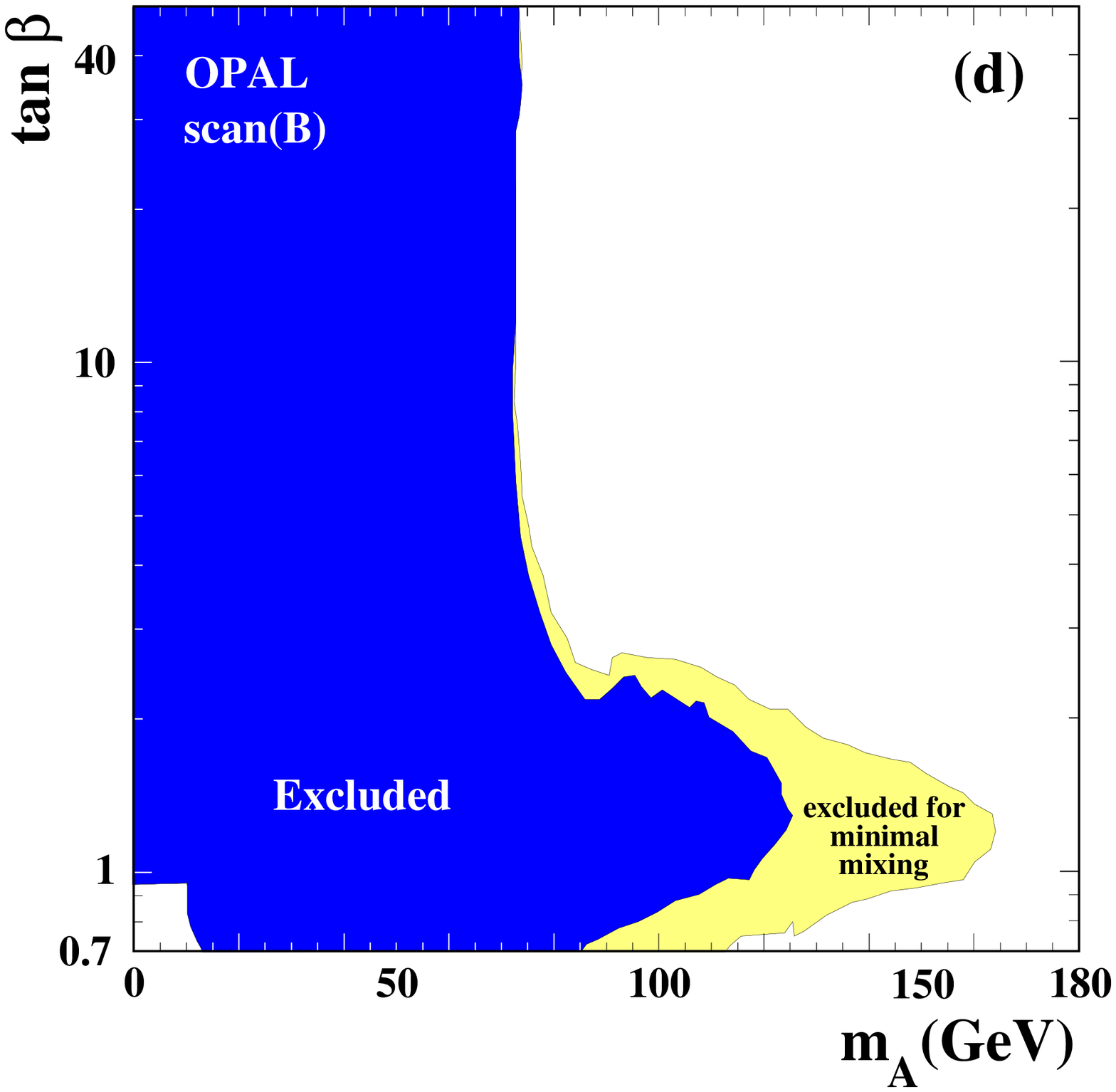,width=8.0cm}}
\caption[]{\label{fig:fig2-mssm}\sl
         The MSSM exclusion for the scan (B) described in the text of
         Section~\ref{sect:mssm}.
         Excluded regions are shown for
         (a) the (\mh,~\mA) plane for $\tanb>1$,
         (b) the (\mh,~\mA) plane for $\tanb>0.7$,
         (c) the (\mh,~\tanb) plane, and
         (d) the (\mA,~\tanb) plane.
         The black area is excluded at 95\% CL.
         The grey
         areas in (a), (b) and (c) are theoretically inaccessible.
         The light grey area in (d) is excluded only for minimal scalar-top
         mixing.
}
\end{figure}
\begin{figure}[p]
\centerline{
\epsfig{file=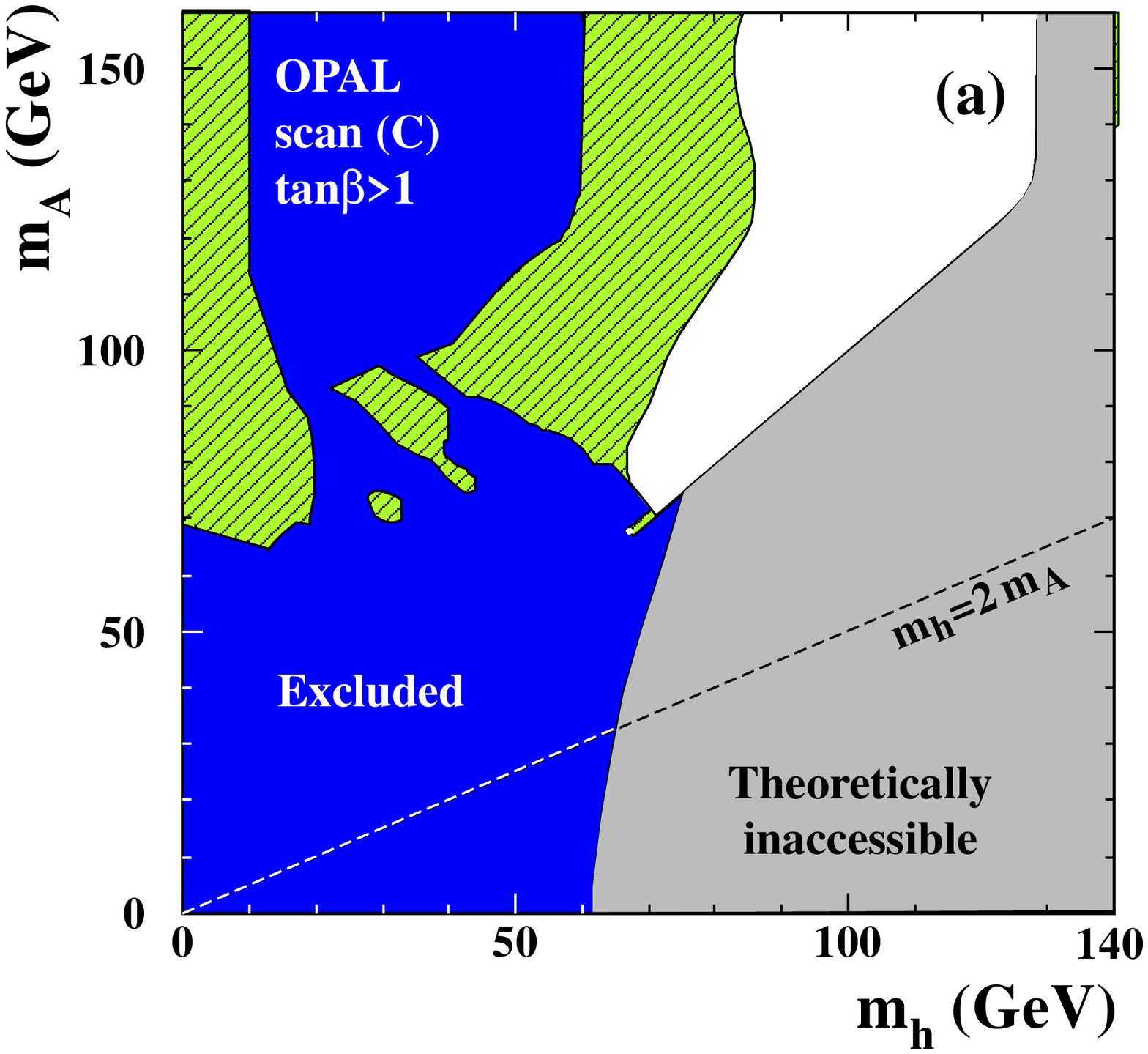,width=8.5cm}\hfill
\epsfig{file=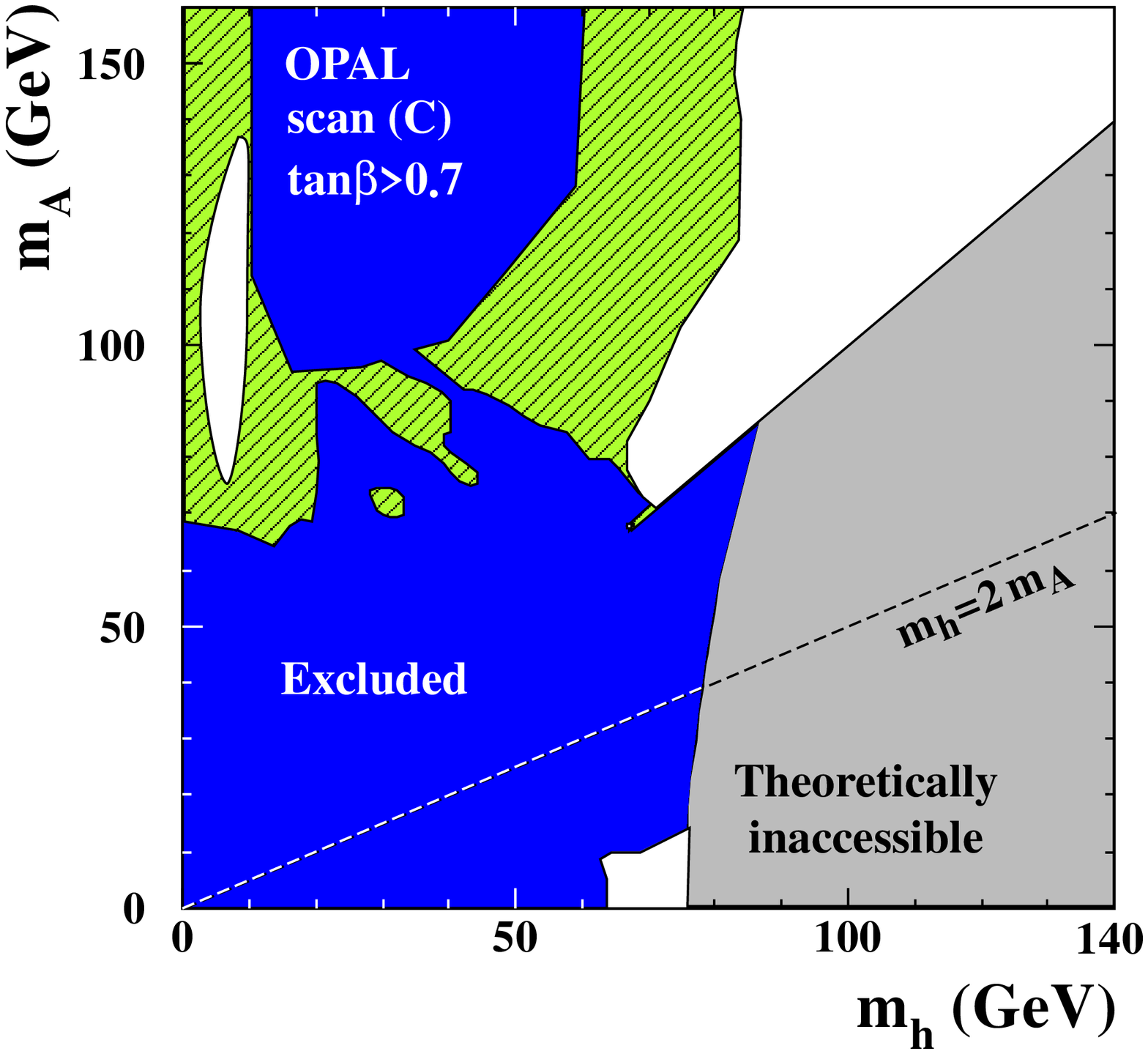,width=8.5cm}
}
\centerline{
\epsfig{file=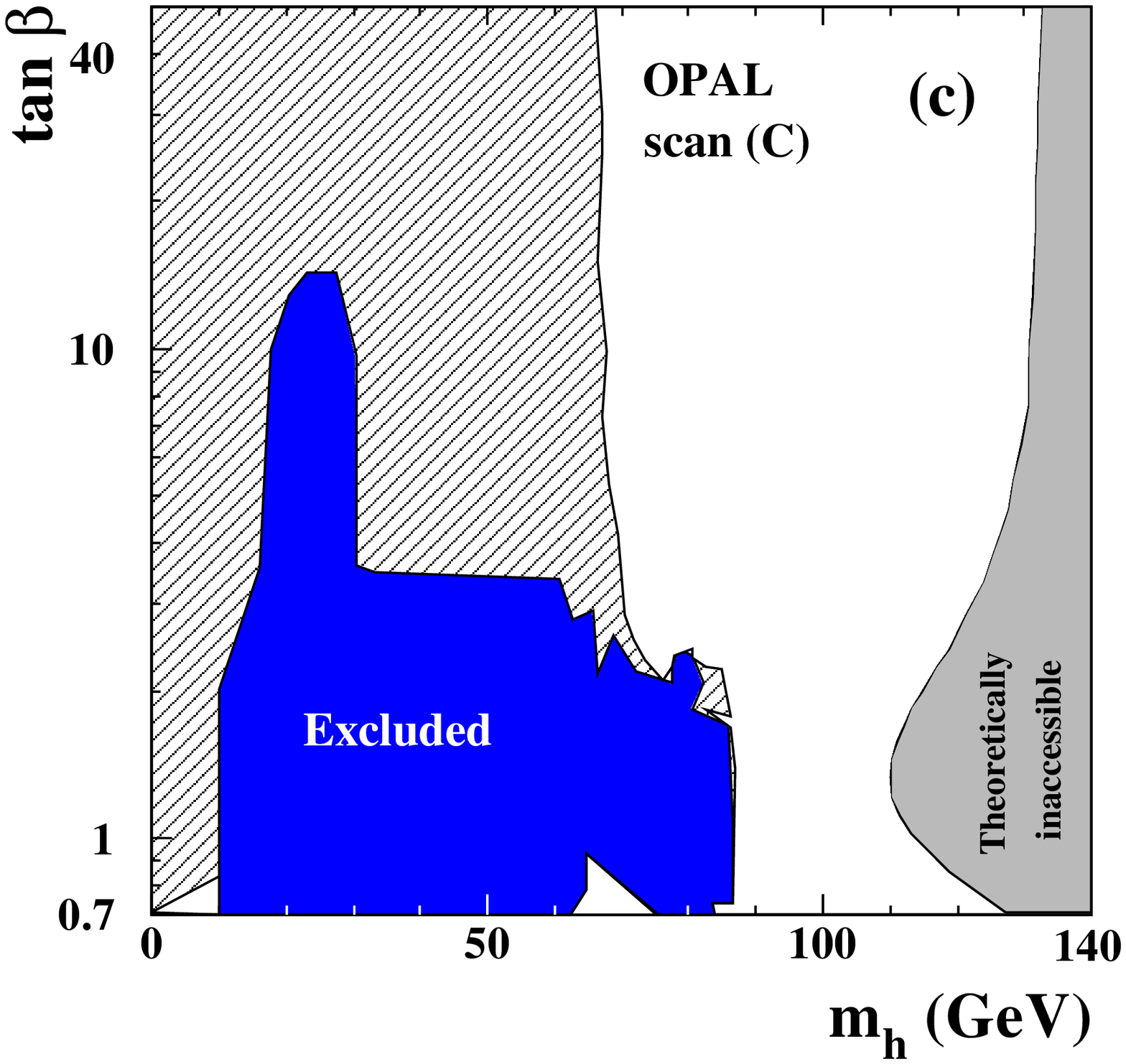,width=8.5cm}\hfill
\epsfig{file=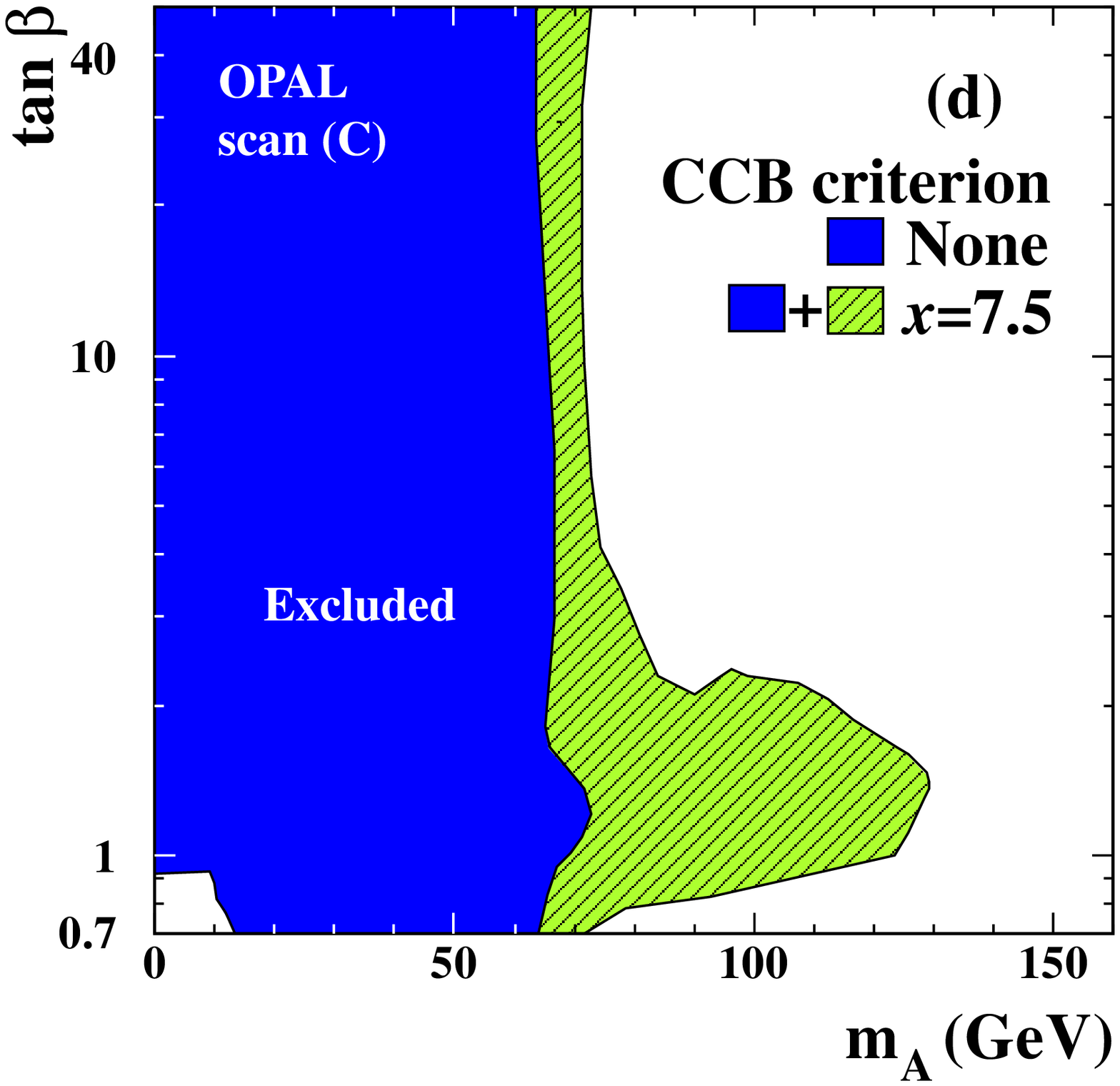,width=8.5cm}
}
\caption[]{\label{fig:fig3-mssm}\sl
         The MSSM exclusion for scan (C) described in the text of
         Section~\ref{sect:mssm}.
         Excluded regions are shown for
         (a) the (\mh,~\mA) plane for $\tanb>1$,
         (b) the (\mh,~\mA) plane for $\tanb>0.7$,
         (c) the (\mh,~\tanb) plane, and
         (d) the (\mA,~\tanb) plane.
         All exclusion limits are at 95\% CL.
         The black areas are excluded without applying any CCB criterion.
         When the CCB criterion is applied with $x=7.5$
         the grey hatched areas are excluded in addition.
         The grey areas in (a), (b) and (c) are theoretically
         inaccessible.
}
\end{figure}

The results for scan (A) are shown in Figure~\ref{fig:fig1-mssm}.
For $\tanb \ge 1$, the 95\% CL lower limits obtained 
are $\mh>70.5$~GeV and $\mA>72.0$~GeV (Figure~\ref{fig:fig1-mssm}(a)).
When the \tanb\ range is enlarged to $\tanb>0.7$
(Figure~\ref{fig:fig1-mssm}(b)),
the lower limits on \mh\ and \mA\ are not affected, except for a small
unexcluded region at $\mA < 10$ GeV and 65 GeV $< \mh <$ 72 GeV. 
In this region the searches for \ho\ra\Ao\Ao\ are not sensitive.
For a detailed discussion of the region $\mA<5$~GeV see~\cite{mssmpaper172}.
Figure~\ref{fig:fig1-mssm}(c) shows the projection onto the (\mh,~\tanb)
plane. For the specific parameter choices of scan (A), a region
0.8 $< \tanb < $ 1.9 can be excluded at 95\% CL for the case of
no scalar-top mixing. Note, however, that
this applies only for $m_{\mathrm{top}} \le 175$~GeV. Since for larger
top quark masses the theoretically allowed area widens significantly, 
no exclusion can be made in \tanb\ e.g.~for $\mtop = 185$ GeV.
In Figure~\ref{fig:fig1-mssm}(d) the (\mA,~\tanb) projection is shown.

Figure~\ref{fig:fig2-mssm} shows the results for scan (B).
Differences with respect to scan (A) are due to
the possibility of having lower $\mstop$ values.
This leads in general to modified couplings and in particular, for some
parameter sets, to a strongly enhanced branching ratio for $\ho\ra\mathrm{gg}$.
The wider range of $\mstop$ in conjunction
with $m_{\mathrm{t}}=185$~GeV leads to larger theoretically accessible
regions.
Despite these modifications, many essential features such as the
limit on \mh\ and \mA\ for $\tanb > 1$ 
(Figure~\ref{fig:fig2-mssm}(a)) remain unchanged. For $\tanb > 0.7$ 
(Figure~\ref{fig:fig2-mssm}(b)) the unexcluded
region at low \mA\ becomes slightly larger, extending up to $\mA \approx 
13$ GeV. From Figures~\ref{fig:fig2-mssm}(c) and (d) it can be seen that
an exclusion in \tanb\ is no longer possible because of the larger 
theoretically allowed area.

The results for scan (C) are shown in Figure~\ref{fig:fig3-mssm}.
The dark area is excluded at 95\% CL.
The grey hatched area is excluded if, in addition, 
a soft CCB criterion with $x=7.5$
is applied as discussed in~\cite{mssmpaper172}.
Lower values for $x$ do not extend the exclusion.
The exclusion in the low \tanb\ region, $\tanb<3$, is obtained by
applying the SM search analysis also to \Zo\Ho\ production,
where \Ho\ is the heavy CP-even Higgs boson.
For $\tanb<3$, the combination of $\mh<60$~GeV, $\mA>80$~GeV and
very small $\sin^2(\beta-\alpha)$ typically
leads to a heavy CP-even Higgs boson mass $\mH<90$~GeV, while 
\Zo\Ho\ production is enhanced by the large $\cos^2(\beta-\alpha)$ value.
As a consequence, the area of low $\tanb<3$ and $10<\mh<60$~GeV is
excluded.
However, as a side effect, an unexcluded region at
$\mA\approx\mh\approx 67.5$~GeV and \mH\ close to 90~GeV
appears for large \tanb\ due to the presence of candidates.
%
%
The unexcluded region at $\mh<10$~GeV and $75$~GeV$<\mA<140$~GeV for low
\tanb\ is a result of the limited sensitivity for \Zo\ho\ production
for these \ho\ masses (see Figure~\ref{modindepZh}).

For $\tanb>1$ an absolute lower limit of $\mA>64.5$~GeV can be derived
in the general scan at 95\% CL. For $\tanb>0.7$, the region
$13$~GeV~$<\mA<64.5$~GeV is excluded at 95\% CL, with no CCB criterion
applied.
When a soft CCB criterion is applied ($x=7.5$) the mass limits at 95\% CL are
$\mA>67.5$~GeV and $\mh>67.5$~GeV, while for the latter also a region
$\mh<10$~GeV is allowed if $\tanb<0.85$.
\vspace{-0.3cm}

\subsection{Results of the Searches for Charged Higgs Bosons}
%

\begin{figure}[htbp]
\centering
\epsfig{file=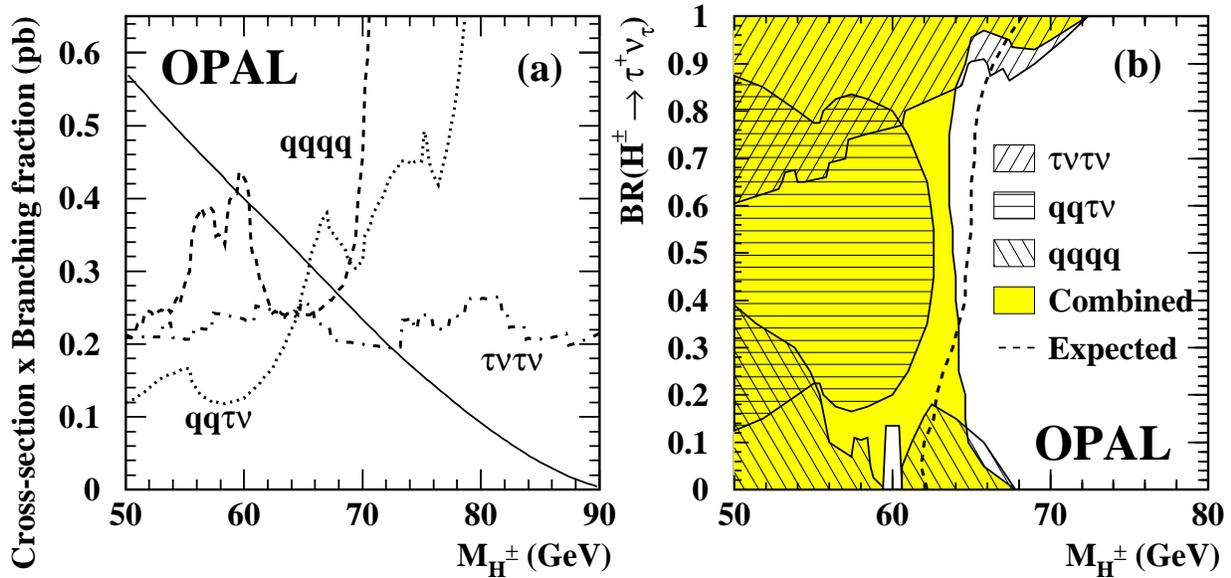,width=0.95\textwidth}
\caption{\sl
(a) Upper limits at 95\% CL, 
scaled to  $\sqrts = $ 183~GeV, on the production
cross-section times branching fraction of the decay for the process
\ee\ra\Hp\Hm\ for the three final states considered. Different
centre-of-mass energies are combined,
using the predicted $s$-dependence of the 
charged Higgs boson production cross-section.
The charged Higgs boson production cross-section 
at $\sqrts = $ 183~GeV is shown as a solid line. Note that the maximum 
branching fraction for the \tpnu\qqp\ final state is 0.5.
(b) Excluded areas at 95\% CL
 in the $[M_{{\rm H}^\pm}$, 
BR$(\Hp\to\tpnu)]$ plane. The 
results from each of the channels separately are indicated by
different hatch styles, and the combined exclusion by the shaded area.
The dashed line shows the expected 95\% CL limit from simulated
background experiments. }
\label{fig:hphmres}
\end{figure}

Upper limits for the production cross-section 
times branching fraction of the decay into a given final state
are presented in Figure~\ref{fig:hphmres}(a).
The results from various centre-of-mass energies are 
scaled to $\sqrts = $ 183~GeV,
assuming the predicted $s$-dependence of the charged Higgs boson production
cross-section.

Lower bounds on the mass of the charged Higgs boson
are presented in Figure~\ref{fig:hphmres}(b) as a function
of the $\Hp\to\tpnu$ branching ratio. The expected mass limit from
simulated background experiments (assuming no signals) is also shown.
Charged Higgs bosons are excluded up
to a mass of 59.5~GeV
at 95\% CL,
independently of the $\Hp\to\tpnu$ branching ratio.
For $BR(\Hp\to\tpnu) > 0.15$, a limit of 63.6~GeV
can be set at 95\% CL.
Some regions are excluded by the searches in individual channels
but not in their combination. This is mainly due to 
three candidate events observed in the semileptonic channel
around 66-68 GeV. All three events are consistent with \WW\ production.

\section{Summary}

The searches for Higgs bosons presented here and based on data collected
by OPAL at \sqrts = 183 GeV,
have not
revealed any significant excess over the expected backgrounds. 
In combination with previous search results,
new limits on the masses of neutral and charged Higgs bosons have been
set at 95\% CL. 
In particular, the Standard Model Higgs boson is excluded for masses below
88.3~GeV. In the MSSM, for parameter sets corresponding
to minimal and maximal scalar top mixing, masses of \mA\ (\mh ) below
72.0~GeV (70.5~GeV) are excluded for $\tanb > 1$.
For minimal scalar top 
mixing, soft SUSY breaking masses of 1~TeV, and
$m_{\mathrm{top}} \le 175$~GeV, the range
$0.8<\tanb<1.9$ is excluded.
If the MSSM parameters are varied in a general scan, masses of
\Ao\ and \ho\ below 67.5~GeV are excluded for $\tanb>1$.
Charged Higgs bosons are excluded below 59.5~GeV.

\section*{Acknowledgements}

\noindent
We particularly wish to thank the SL Division for the efficient operation
of the LEP accelerator at all energies
 and for their continuing close cooperation with
our experimental group.  We thank our colleagues from CEA, DAPNIA/SPP,
CE-Saclay for their efforts over the years on the time-of-flight and trigger
systems which we continue to use.  In addition to the support staff at our own
institutions we are pleased to acknowledge the  \\
Department of Energy, USA, \\
National Science Foundation, USA, \\
Particle Physics and Astronomy Research Council, UK, \\
Natural Sciences and Engineering Research Council, Canada, \\
Israel Science Foundation, administered by the Israel
Academy of Science and Humanities, \\
Minerva Gesellschaft, \\
Benoziyo Center for High Energy Physics,\\
Japanese Ministry of Education, Science and Culture (the
Monbusho) and a grant under the Monbusho International
Science Research Program,\\
Japanese Society for the Promotion of Science (JSPS),\\
German Israeli Bi-national Science Foundation (GIF), \\
Bundesministerium f\"ur Bildung, Wissenschaft,
Forschung und Technologie, Germany, \\
National Research Council of Canada, \\
Research Corporation, USA,\\
Hungarian Foundation for Scientific Research, OTKA T-016660, 
T023793 and OTKA F-023259.\\

\begin{appendix}

\section*{Appendix: Lifetime Tag}
  The five quantities which are input to an artificial neural network (ANN)
  to form the lifetime tag for b-flavour, 
  $\beta_\tau$, are described here.

  The first three of these quantities rely on the reconstruction of secondary
  vertices. Within a jet, sub-jets are formed using a cone 
  algorithm~\cite{cone} with a cone half angle of 0.5 radians and 
  a minimum sub-jet energy of 7~GeV. In each of these sub-jets, a secondary
  vertex is reconstructed using the method described in~\cite{btag1}.
  In order to compensate for the loss in b-tagging efficiency due to 
  the requirement of secondary
  vertex reconstruction, the last two of the inputs to the ANN are based on 
  track impact parameters only.

  The quantities are the following:
\begin{itemize}
\item[(1)] Secondary vertex likelihood, ${\cal L}_{S}$:
      a vertex-multiplicity-dependent likelihood ${\cal L}_{S}$ is formed
      using
      the decay length significance, $S$ (the
      decay length divided by its error).
      ${\cal L}_{S}$
      is calculated from the probability density function (p.d.f.) of $S$
      for b, c and uds flavours, $f_{\mathrm{b}}$, $f_{\mathrm{c}}$, 
      $f_{\mathrm{uds}}$. 
      If more than one sub-jet is formed, the secondary vertex with the
      largest ${\cal L}_{S}$ in a given jet is selected for this and the 
      following two quantities.

\item[(2)] Reduced secondary vertex likelihood, ${\cal L}_{R}$:
      the reduced decay length is obtained from a vertex fit using
      all tracks in the secondary vertex, except the one with the
      largest impact parameter significance, i.e., the impact parameter
      with respect to the primary vertex divided by its error. While for
      b-flavoured hadron decays, the reduced decay length often coincides
      with the decay length, randomly formed vertices are less
      robust against removing the most significant track.
      The reduced decay length significance $R$ is given by the reduced
      decay length divided by its error. From $R$ a 
      multiplicity-dependent 
      likelihood ${\cal L}_{R}$ is calculated. If a secondary
      vertex consists of only two tracks, $R$ is not defined. In that
      case, ${\cal L}_{R}$ is set to the value corresponding to the 
      likelihood for b-flavour to form a reconstructed 
      two track vertex relative to all flavours.
\item[(3)] Critical track discriminator, $T_{\mathrm{crit}}$:
      An auxiliary ANN is trained to discriminate between
      tracks originating from the b-flavoured hadron decay and from tracks due
      to fragmentation or decays of light-flavoured hadrons. 
      The inputs to this ANN are the impact parameter
      of the track with respect to the primary vertex,
      the impact parameter with respect to
      the secondary vertex,
      the momentum of the track, and 
      its transverse momentum with respect to the corresponding sub-jet axis.
      The tracks belonging to the sub-jet are then sorted according
      to the output of the auxiliary ANN in a descending order. Tracks are
      added one by one to a `cluster' of tracks whose invariant
      mass is calculated, assuming that all tracks have the pion mass.
      $T_{\mathrm{crit}}$ is the auxiliary ANN output of that track
      which causes the cluster invariant mass to exceed 1.9 GeV. 
      This algorithm exploits the higher mass of b-flavoured hadrons compared
      to charmed and lighter hadrons.
      The algorithm is described in detail in~\cite{tcrit}.

\item[(4)] Two-dimensional impact parameter joint probability, 
  ($P_{\mathrm{join}}$):
  The impact parameter distribution for tracks with negative 
  impact parameter significance\footnote{
  The impact parameter is taken to be
  positive if, in the two-dimensional projection, the track path, starting
  from the point of closest approach to the primary vertex,
  crosses the jet axis in the flight direction; otherwise it is negative.
  }
  is assumed to represent the class of tracks 
  from the primary vertex and thus provides
  an estimate of the detector resolution function.
  This resolution function is then used to ``weight'' the tracks, and the
  joint probability for the tracks in a jet to come from the primary
  vertex is given by

  $$ P_{\mathrm{join}} = y \sum_{m=0}^{N-1} \frac{(-\ln y)^{m}}{m!},$$

  where $y$ is the product of the probabilities of all $N$ tracks with
  positive impact parameters in the jet~\cite{aleph_btag}.
  Only tracks that pass stringent track quality criteria
  are used in the calculation of $P_{\mathrm{join}}$.

\item[(5)]  Impact parameter mass tag ($P_{\mathrm{mass}}$):
  Tracks in each sub-jet are sorted in descending order of the impact parameter
  significance and iteratively clustered. $P_{\mathrm{mass}}$ is defined as the
  impact parameter significance of that track which causes the invariant 
  mass of the cluster to exceed 1.2 GeV. When more than one sub-jet is
  reconstructed in a given jet, the highest $P_{\mathrm{mass}}$ value is used.
  Only tracks that pass stringent track quality criteria
  are used in the calculation of $P_{\mathrm{mass}}$.
\end{itemize}

  The five variables ${\cal L}_{S}$, ${\cal L}_{R}$,
  $T_{\mathrm{crit}}$, $P_{\mathrm{join}}$ and $P_{\mathrm{mass}}$ 
  are then input to an ANN. 
  Monte Carlo samples at $\sqrts = \mZ$ are used to train the ANN.
  The program
  JETNET 3.4~\cite{jetnet34} is used with five input nodes, one hidden layer
  with 10 nodes and one output node, the lifetime-tag $\beta_{\tau}$.
  Since the vertex tagging performance depends on the jet polar angle,
  three separate ANN's are trained for jets with
  $| \mathrm{cos}\theta_{\mathrm{jet}} | \leq 0.75$, 
  $0.75 < | \mathrm{cos}\theta_{\mathrm{jet}} | \leq 0.9$, and  
  $| \mathrm{cos}\theta_{\mathrm{jet}} | > 0.9$.
\end{appendix}


\end{document}